%% file: hoogerwerf.tex
\documentclass{aa}
\usepackage{psfig}
\usepackage{graphicx,lscape}
\usepackage{relsize}
\input hoogerwerf_figures.tex
\def\mathv{ v}

\def\spose#1{\hbox to 0pt{#1\hss}}
\def\lta{\mathrel{\spose{\lower 3pt\hbox{$\sim$}}
    \raise 2.0pt\hbox{$<$}}}
\def\gta{\mathrel{\spose{\lower 3pt\hbox{$\sim$}}
    \raise 2.0pt\hbox{$>$}}}
\begin{document}
%
%
\thesaurus{05(04.01.2; 05.01.1; 05.18.1)}
%
%
\title{On~the~origin~of~the~O~and~B-type~stars~with~high~velocities~II}
\subtitle{Runaway stars and pulsars ejected from the nearby young stellar groups}
\author{R.\ Hoogerwerf \and J.H.J.\ de Bruijne \and P.T.\ de Zeeuw}
\institute{Sterrewacht Leiden, Postbus 9513, 2300 RA Leiden, the Netherlands}
\offprints{P.T.\ de Zeeuw}
\mail{tim@strw.leidenuniv.nl}
\date{August 2000}
%
%
\maketitle
%
%
\begin{abstract}
We use milli-arcsecond accuracy astrometry (proper motions and
parallaxes) from Hipparcos and from radio observations to retrace the
orbits of 56 runaway stars and nine compact objects with distances
less than 700 pc, to identify the parent stellar group. It is possible
to deduce the specific formation scenario with near certainty for two
cases. (i) We find that the runaway star $\zeta$~Ophiuchi and the
pulsar PSR~J1932$+$1059 originated about 1~Myr ago in a supernova
explosion in a binary in the Upper Scorpius subgroup of the Sco OB2
association. The pulsar received a kick velocity of $\sim$350 km
s$^{-1}$ in this event, which dissociated the binary, and gave
$\zeta$~Oph its large space velocity. (ii) Blaauw \& Morgan and Gies
\& Bolton already postulated a common origin for the runaway-pair
AE~Aur and $\mu$~Col, possibly involving the massive highly-eccentric
binary $\iota$~Ori, based on their equal and opposite velocities. We
demonstrate that these three objects indeed occupied a very small
volume $\sim$2.5~Myr ago, and show that they were ejected from the
nascent Trapezium cluster.

We identify the parent group for two more pulsars: both likely
originate in the $\sim 50$~Myr old association Per OB3, which
contains the open cluster $\alpha$ Persei. At least 21 of the 56
runaway stars in our sample can be linked to the nearby associations
and young open clusters. These include the classical runaways
53~Arietis (Ori~OB1), $\xi$~Persei (Per~OB2), and $\lambda$~Cephei
(Cep~OB3), and fifteen new identifications, amongst which a pair of
stars running away in opposite directions from the region containing
the $\lambda$ Ori cluster. Other currently nearby runaways and pulsars
originated beyond 700 pc, where our knowledge of the parent groups is
very incomplete.
\end{abstract}
\section{Introduction}\label{sec:intro}
About 10--30\% of the O stars and 5--10\% of the B stars (Gies
\cite{gie1987}; Stone \cite{sto1991}) have large peculiar velocities
(up to 200~km~s$^{-1}$), and are often found in isolated locations;
these are the so-called `runaway stars' (Blaauw \cite{bla1961},
hereafter Paper I). The velocity dispersion of the population of
runaway stars, $\sigma_\mathv \sim 30$~km~s$^{-1}$ (e.g., Stone
\cite{sto1991}), is much larger than that of the `normal' early-type
stars, $\sigma_\mathv \sim 10$~km~s$^{-1}$.  Besides their peculiar
kinematics, runaway stars are also distinguished from the normal
early-type stars by an almost complete absence of multiplicity (cf.\
the binary fraction of normal early-type stars is $>$50\% [e.g., Mason
et al.\ \cite{mas1998}]). Furthermore, over 50\% of the (massive)
runaways have large rotational velocities and enhanced surface helium
abundances (Blaauw \cite{bla1993}).

Several mechanisms have been suggested for the origin of runaway stars
(Zwicky \cite{zwi1957}; Paper~I; Poveda, Ruiz \& Allen \cite{pra1967};
Carrasco et al.\ \cite{car1980}; Gies \& Bolton \cite{gb1986}), two of
which are still viable: the binary-supernova scenario (Paper~I) and
the dynamical ejection scenario (Poveda et al.\ \cite{pra1967}). We
summarize them in turn.

\paragraph{Binary-supernova scenario (BSS):} In this scenario a runaway 
star receives its velocity when the primary component of a massive
binary system explodes as a supernova. When the supernova shell passes
the secondary the gravitational attraction of the primary reduces
considerably, and the secondary starts to move through space with a
velocity comparable to its original orbital velocity
(30--150~km~s$^{-1}$).  What remains of the primary after the
explosion is a compact object, either a neutron star or a black
hole. Depending on the details of the preceding binary evolution, the
eccentricity of the orbit, and the kick velocity
$\vec{\mathv}_\mathrm{kick}$ due to the asymmetry of the supernova
explosion (e.g., Burrows, Hayes \& Fryxell \cite{bhf1995}), the
compact object may or may not remain bound to the runaway star (e.g.,
Hills \cite{hil1983}). In most cases less than half of the total mass
of the binary is expelled in the explosion. As this is insufficient to
break up the binary (Paper I), most BSS runaways are
expected to remain double.  Examples of such systems are provided by
the high-mass-X-ray binaries (e.g., van den Heuvel et al.\
\cite{heu2000}). Their typical velocities of $\sim$50 km s$^{-1}$
(Kaper et al.\ \cite{kap1997}; Chevalier \& Ilovaisky \cite{ci1998})
are the natural result of the recoil velocity acquired when the
supernova shell leaves the binary system.

Several searches failed to find compact companions of classical
runaway stars (Gies \& Bolton \cite{gb1986}; Philp et al.\
\cite{phi1996}; Sayer, Nice \& Kaspi \cite{snk1996}), suggesting that
for these systems $\vec{\mathv}_\mathrm{kick}$ of the neutron star was
large enough (several 100~km~s$^{-1}$) to unbind the binary (e.g.,
Frail \& Kulkarni \cite{fk1991}; Cordes, Romani \& Lundgren
\cite{crl1993}; Lai \cite{lai1999}). The typical magnitude of this
`threshold' kick velocity is uncertain (e.g., Hills \cite{hil1983};
Lorimer, Bailes \& Harrison \cite{lbh1997}; Hansen \& Phinney
\cite{hp1997}; Hartman \cite{har1997}).

Single BSS runaways must originate in close binaries because these
systems have the largest orbital velocities, and therefore they have
experienced close binary evolution before being ejected as a
runaway. This leads to the following observable characteristics:
\newcounter{it61}
\begin{list}{\arabic{it61}:}{
     \usecounter{it61}
     \setlength{\topsep}{1mm}
     \setlength{\labelwidth}{3mm}
     \setlength{\labelsep}{2mm}
     \setlength{\itemindent}{0mm}
     \setlength{\parsep}{0mm}
     \setlength{\itemsep}{0mm}
     \setlength{\leftmargin}{5mm}}
\item BSS runaways are expected to have increased helium
abundance\footnote{The abundances of other elements are increased as
well, in particular that of nitrogen which can be enhanced by a factor
4--5 (Van Rensbergen priv.\ comm.)} and large rotational velocity:
when the primary fills its Roche lobe, mass and angular momentum is
transferred to the star that will become the runaway. The mass
transfer stops when the `primary' has become a helium star (i.e., only
the helium core remains). This process enriches the runaway with
helium, and spins it up (e.g., Packet \cite{pac1981}; van den Heuvel
\cite{heu1985}; Blaauw~\cite{bla1993}).
\item A BSS runaway can become a blue straggler, because it is
      rejuvenated during the mass-transfer period through the fresh
      fuel it receives from the primary.
\item The kinematic age of a BSS runaway star (defined as the time
      since the runaway left its parent group) should be smaller than
      the age of the parent group. The primary of the original binary
      system first evolves for several Myr before it explodes and the
      runaway is ejected.
\end{list}

\paragraph{Dynamical ejection scenario (DES):} In this scenario  
runaway stars are formed through gravitational interactions between
stars in dense, compact clusters. Although binary-single star
encounters produce runaways (e.g., Hut \& Bahcall \cite{hb1983}), the
most efficient interaction is the encounter of two hard binary systems
(Hoffer \cite{hof1983}).  Detailed simulations show that these
collisions produce runaways with velocities up to 200~km~s$^{-1}$
(Mikkola \cite{mik1983a}, \cite{mik1983b}; Leonard \& Duncan
\cite{ld1988}, \cite{ld1990}; Leonard \cite{leo1991}). The outcome of
a binary-binary collision can be (i) two binaries, (ii) one single
star and a hierarchical triple system, (iii) two single stars and one
binary, and (iv) four single stars (Leonard \cite{leo1989}). In most
cases the collision will result in the ejection of two single stars
and one hard binary with an eccentric orbit (Hoffer \cite{hof1983};
Mikkola \cite{mik1983a}). Since the resulting binary is the most
massive end product of the collision it is unlikely to gain a lot of
speed; it might even remain within the parent cluster. This process
naturally leads to a low (0--33\%) runaway binary fraction which is in
qualitative agreement with the observations (e.g, Gies \& Bolton
\cite{gb1986}).  For the DES to be efficient the initial binary
fraction in clusters needs to be large. Recent observations show that
the binary fraction for massive stars in young clusters is $>$50\%
(Abt \cite{abt1983}; Kroupa, Petr \& McCaughrean \cite{kpm1999};
Preibisch et al.\ \cite{pre1999} [approaching~100\%]).

DES runaways have the following characteristics: 
\newcounter{it62}
\begin{list}{\arabic{it62}:}{
     \usecounter{it62}
     \setlength{\topsep}{1mm}
     \setlength{\labelwidth}{3mm}
     \setlength{\labelsep}{2mm}
     \setlength{\itemindent}{0mm}
     \setlength{\parsep}{0mm}
     \setlength{\itemsep}{0mm}
     \setlength{\leftmargin}{5mm}}
\item DES runaways are formed most efficiently in a high-density
      environment, e.g., in young open clusters. They may also
      originate in OB associations. These birth sites of massive stars
      are unbound stellar groups and therefore expand (e.g., Blaauw
      \cite{bla1952a,bla1978}; Elmegreen \cite{elm1983}; 
      Kroupa \cite{kro2000a}), so DES runaways
      must have been ejected very soon after their formation. The
      kinematic age and the age of the parent association are thus
      nearly equal.
\item DES runaways are not expected to show signs of binary evolution
      such as large rotational velocities and increased helium
      abundance. However, Leonard (\cite{leo1995}) suggested that some
      binary-binary encounters produce runaways consisting of two
      stars that merged during the interaction. These would have
      enhanced helium abundances and large rotational velocities (Benz
      \& Hills \cite{bh1987}, but see Lombardi, Rasio \& Shapiro
      \cite{lrs1995}).
\item DES runaways are expected to be mostly single stars.
\end{list}

\TabOne
\bigskip
\noindent
Which of the two formation processes is responsible for runaway stars
has been debated vigorously. Both mechanisms create stars with large
peculiar velocities which enable them to travel far from their parent
group: a velocity of 100~km~s$^{-1}$ corresponds to $\sim$100~pc in
only 1~Myr. The relative importance of the two scenarios can be
established by (i) studying the statistical properties of the ensemble
of runaway stars, or by (ii) investigating individual runaways in
detail. The former approach is based on differences in the general
runaway characteristics predicted by each scenario.  This requires a
large, complete database of runaway stars, which is, to date,
unavailable (e.g., Moffat et al.\ \cite{mof1998}). Here we therefore
follow the latter, individual approach by retracing the orbits of
runaway stars back in time. The objects encountered by a runaway along
its path (e.g., an open cluster, an association, other runaways, or a
neutron star), and the times at which these encounters occurred, provide
information about its formation. Evidence for the BSS as the formation
mechanism for single runaway stars is to find a runaway and a neutron
star (pulsar) which occupied the same region of space at the same time
in the past.  Evidence for the DES is to find a common site of origin
for the individual components of the encounter, e.g., a pair of
runaways and a binary, in a dense star cluster.

The individual approach requires highly accurate positions
($\alpha$,$\delta$,$\pi$) and velocities
($\mu_{\alpha\ast}$,$\mu_\delta$,$\mathv_\mathrm{rad}$).  Here
$\alpha$ denotes right ascension, $\delta$ declination, $\pi$
parallax, $\mu_{\alpha\ast} = \mu_\alpha \cos \delta$ proper motion in
right ascension, $\mu_\delta$ proper motion in declination, and
$\mathv_\mathrm{rad}$ the radial velocity.  The milli-arcsecond (mas)
accuracy of Hipparcos astrometry (ESA \cite{esa1997}) allows specific
investigations of the runaway stars within $\sim$700~pc.  Positions
and proper motions of similar accuracy are now available as well for
some pulsars through timing measurements and VLBI observations (e.g.,
Taylor, Manchester \& Lyne \cite{tml1993}; Campbell
\cite{cam1995}). The Hipparcos data also significantly improved and
extended the membership lists of the nearby OB associations (de Zeeuw
et al.\ \cite{zee1999}), and of some nearby young open clusters.  The
resulting improved distances and space velocities of these stellar
aggregates make it possible to connect the runaways and pulsars to
their parent group, and, in some cases, to identify the specific
formation scenario (Hoogerwerf, de Bruijne \& de Zeeuw \cite{hbz2000};
de Zeeuw, Hoogerwerf \& de Bruijne \cite{zhb2000}).  Pre-Hipparcos
data (e.g., Blaauw \& Morgan \cite{bm1954}; Paper I; Blaauw
\cite{bla1993}; van Rensbergen, Vanbeveren \& de Loore
\cite{rvl1996}) allowed identification of the parent groups for some
runaways, but generally lacked the accuracy to study the orbits of the
runaways in detail (but see Blaauw \& Morgan \cite{bm1954} and Gies \&
Bolton \cite{gb1986}).

We define a sample of nearby runaways and pulsars with good astrometry
in \S\ref{sec:samples}, and then analyse two cases in depth:
$\zeta$~Oph and PSR~J1932$+$1059 in \S\ref{sec:SNinUS} and AE~Aur,
$\mu$~Col and $\iota$~Ori in \S\ref{sec:orion}. We apply the method
developed in these sections to the entire sample of runaways and
pulsars in \S\ref{sec:class-runaways}, \S\ref{sec:new-runaways}, and
\S\ref{sec:noparent}.  We discuss helium abundances, rotational
velocities and the blue straggler nature of runaways in
\S\ref{sec:helium} and \S\ref{sec:blue}, and summarize our conclusions
in \S\ref{sec:end}.\looseness=-2

\section{Nearby runaway stars and pulsars}\label{sec:samples}
The parent group is known for about a dozen `classical' runaway stars
(Paper I; Blaauw \cite{bla1993}). The Hipparcos Catalogue
contains these stars, as well as many additional O and B stars which
were known in 1982 to have large radial velocities, including 153 of
the 162 runaway candidates in Hipparcos Proposal 141\footnote{From a
list of O and B stars with $v_{\rm rad} > 30$ km s$^{-1}$ provided by
the late Jan van Paradijs.} (de Zeeuw et al.\ \cite{zee1999}). Many of
these objects are located beyond $\sim$700~pc, where the Hipparcos
parallax measurement is of modest quality. For this reason we
restricted ourselves to a sample of nearby runaway stars, and added to
this the (few) nearby pulsars with measured proper motions.

\FigOne
\subsection{Selection of the sample}\label{sec:selection}
We started with all 1118 O to B5 stars in the Hipparcos Catalogue
which have radial velocities listed in the Hipparcos Input Catalogue
(Turon et al.\ \cite{tur1992}). Next we only considered those stars
which have significant parallaxes ($\pi - 2 \sigma_\pi > 0$~mas) and
proper motions ($\sigma_\mu / \mu \le 0.1$), and space velocities
larger than 30~km~s$^{-1}$ with respect to the standard of rest of the
runaway. For the last requirement we corrected the runaway velocity
for Solar motion and Galactic rotation (Dehnen \& Binney
\cite{db1997}). The somewhat arbitrary choice of the velocity limit of
30~km~s$^{-1}$ minimizes the contamination of the sample by normal O
and B stars (\S\ref{sec:intro}). These criteria yield 54 runaway
candidates (five of which are classical runaways)\footnote{The sample
criteria exclude the known runaways, e.g., $\alpha$~Cam, 68~Cyg
(Blaauw \cite{bla1993}), HIP~35707 (Comer\'on, Torra \& G\'omez
\cite{ctg1998}), and HIP~60350 (Maitzen et al.\ \cite{mai1998}) which
all have distances larger than 1~kpc.}.  This new sample does not
contain the nearby runaways $\zeta$~Oph and $\xi$~Per. The former is
not selected because its space velocity is smaller than 30~km~s$^{-1}$
(although its velocity relative to its parent group Sco OB2 is larger,
cf.\ \S\ref{sec:SNinUS}) and the latter is not selected because
$\sigma_\mu / \mu > 0.1$.  However, since the runaway nature of these
two stars is well established (e.g., Paper I) we
included them in our sample, bringing the total to 56. The Hipparcos
numbers and space velocities of these 56 stars are listed in
Table~\ref{tab:01}.  Panel {\it a} of Figure~\ref{fig:01}
shows the histogram of the derived space velocities.

We selected a sample of nearby pulsars from the Taylor, Manchester \&
Lyne (\cite{tml1993}) catalogue, as updated on {\tt
http://pulsar.princeton.edu/}. It contains 94 pulsars with known
proper motions and distances. Only seven of these meet our distance
($D \lta 1$~kpc) and proper motion ($\sigma_\mu / \mu < 0.1$)
constraints (see panel {\it b} of Figure~\ref{fig:01}).  Most
pulsar distances are derived from the dispersion measure. These
distances are unreliable, especially for nearby objects, since they
depend on the local properties of the ISM. 
For one nearby pulsar, PSR J0953$+$0755, high precision VLBA
measurements became available recently (Brisken et al.\
\cite{bri2000}). We added this pulsar to our sample.
The eight pulsars are listed in Table~\ref{tab:01}, together with
Geminga, a nearby neutron star which is not a pulsar, for which an
accurate proper motion is known (Caraveo et al.\
\cite{car1996}).\looseness=-2

\TabTwo

Our sample of nearby runaway stars and compact objects is severely
incomplete. The Hipparcos Catalogue is complete to $V = 7.3$--9~mag,
with the limit depending on Galactic latitude and spectral type (2163
of the 3622 O to B5 stars have $V > 7.3$~mag).  The data available for
the O and B stars is inhomogeneous and incomplete, e.g., less than a
third of the O to B5 stars in the Catalog has a measured radial
velocity.  We have excluded those with large $\mathv_\mathrm{rad}$ but
insignificant proper motions, as their retraced orbits are uncertain.
The beamed nature of the radio emission from pulsars hides many from
observation, and not all of those that do radiate in our direction
have been found. Of these, only a few have an accurately measured
proper motion and a reliable distance.

\subsection{Nearby OB associations and open clusters}\label{sec:obclus}

We adopt the positions and mean space motions of the OB associations
within 700~pc of the Sun as derived by de Zeeuw et al.\
(\cite{zee1999}) from Hipparcos measurements.  For the open clusters
we compiled a list from the {\tt WEBDA} catalogue ({\tt
http://obswww.unige.ch/webda/}), and consider only those which are
young ($\tau < 50$~Myr) and with distances less than 700~pc as likely
parent groups. The age requirement is comparable to the age of the
oldest runaways we consider here (B5V). Typical pulsar ages are less
than 50~Myr (e.g., Blaauw \& Ramachandran \cite{br1998}). This
selection yields nineteen open clusters (see
Table~\ref{tab:02}), of which five are already covered in the
study of the nearby associations by de Zeeuw et al.\ (\cite{zee1999}).
To obtain the space motion of these clusters we use the {\tt WEBDA}
member stars listed in the Hipparcos Catalogue to obtain reliable
astrometry, and those in the Hipparcos Input Catalogue to obtain the
radial velocity. In this way we are able to construct a more or less
reliable space motion for seven of the fourteen remaining open
clusters (those labeled ``Y'' or ``?'' in Table~\ref{tab:02}
which summarizes the results).

\subsection{Orbits}\label{sec:orbits}
Traditionally, the orbits of runaway stars have been traced back in
time using straight lines through space. This is sufficiently accurate
for identification of the parent group for times up to a few Myr and
distances less than a few hundred pc. To make sure we include the
effect of the Galactic potential, we use a fourth-order Runge--Kutta
numerical integration method, with a fixed time-step of $10\,000$ yr,
to calculate the orbit. The Galactic potential we use consists of (i)
a logarithmic potential for the halo, (ii) a Miyamoto--Nagai potential
for the disk, and (iii) a Plummer potential for the bulge of the
Galaxy.  The potential predicts Oort constants $A =
13.5$~km~s$^{-1}$~kpc$^{-1}$ and $B = -12.4$~km~s$^{-1}$~kpc$^{-1}$
and a circular velocity $\mathv_{\rm circ} = 219.8$~km~s$^{-1}$ at
$R_0 = 8.5$~kpc. These values agree with those which Feast \&
Whitelock (\cite{fw1997}) obtained using Hipparcos data: $A =
14.82\pm0.84$~km~s$^{-1}$~kpc$^{-1}$, $B =
-12.37\pm0.64$~km~s$^{-1}$~kpc$^{-1}$, $\mathv_{\rm circ} =
231.2\pm16.2$~km~s$^{-1}$ at $R_0 = 8.5$~kpc.  Since the volume
covered in the orbit integration is typically a few hundred pc, and
the time of the integration is typically less than 10~Myr,
perturbations of the orbits caused by small-scale structure in the
disk are negligible. 

Before integrating the orbit, we correct the observed velocity
$\vec{\mathv}_\ast$ for (i) the Solar motion with respect to the Local
Standard of Rest, $\vec{\mathv}_\mathrm{lsr}$ (Dehnen \& Binney
\cite{db1997}), and (ii) the Galactic rotational velocity of the Local
Standard of Rest, $\vec{\mathv}_\mathrm{gr}$ (Binney \& Tremaine
\cite{bt1987}, p.\ 14). The stellar velocity
$\vec{\mathv}_\mathrm{gal}$ relative to the Galactic reference frame
is then given by
\begin{equation}
\vec{\mathv}_\mathrm{gal} = \vec{\mathv}_\ast + \vec{\mathv}_\mathrm{lsr} +
                            \vec{\mathv}_\mathrm{gr}.
\end{equation}
To retrace the orbit, we reverse the velocity and integrate forward in
time. We calculate the distance of a star as $1/\pi$, where $\pi$ is
the trigonometric parallax. Since we use the individual parallax, we
cannot correct this distance for possible biases (e.g., Smith \&
Eichhorn \cite{se1996}).

\subsection{Identification of parent groups}
We calculate the past orbit of each of the 56 runaway stars listed in
Table~\ref{tab:01} for 10~Myr. We do this $10\,000$ times for each
star, in order to sample the error ellipsoid of the measured
parameters, defined by the covariance matrix of the Hipparcos
astrometry and the error in the radial velocity measurement.
Retracing the orbit of a pulsar is more difficult, because the radial
velocity is unknown. We therefore cover a range of radial velocities
of $\mathv_\mathrm{rad} = 0 \pm 500$~km~s$^{-1}$ in the orbit
integrations for the pulsars. Figure~\ref{fig:02} shows the
positions of the runaways and pulsars on the sky, together with their
orbits, retraced back for only 2~Myr so as not to confuse the diagram.
Three orbits are shown for each pulsar: for $\mathv_\mathrm{rad} =
0$~km~s$^{-1}$ (filled square), $\mathv_\mathrm{rad} =
200$~km~s$^{-1}$ (open square), and $\mathv_\mathrm{rad} =
-200$~km~s$^{-1}$ (open star).

We also retrace the orbits of the set of nearby OB associations and
open clusters defined in \S\ref{sec:obclus}.  These groups have
typical linear dimensions of 10--30~pc. We consider a group to be a
possible site of origin for a runaway or pulsar if the minimum
separation between the runaway/pulsar and the group was less than
10~pc at some time in the past 10~Myr. With this definition, we find a
parent group for 21 of the 56 runaways.  These stars are indicated by
the filled circles in Figure~\ref{fig:02}, and include the seven
classical runaways in the sample. We discuss them in detail in
\S\S\ref{sec:SNinUS}--\ref{sec:new-runaways} below. Six of the nine
neutron stars possibly traversed one of the nearby stellar groups;
these are PSR~J0826$+$2637, PSR~J0835$-$4510, PSR~J1115$+$5030,
PSR~J1239$+$2453, PSR~J1932$+$1059, and Geminga (objects 1, 2, 4, 6,
8, and 9 in Figure~\ref{fig:02}). We discuss them in
\S\ref{sec:SNinUS} and \S\ref{sec:new-runaways}, and identify the
parent group for four of them. Table~\ref{tab:03} summarizes the
data for the 22 runaways, four pulsars, and Geminga. The pulsars and
runaways for which we cannot identify a parent group are discussed
further in \S\ref{sec:noparent}.
\FigTwo 
\clearpage 

\TabThree
\noindent

\section{A binary supernova in Upper Scorpius}\label{sec:SNinUS}

\subsection{$\zeta$ Oph and PSR~J1932$+$1059}\label{sec:zoph}
$\zeta$~Oph is a single O9.5Vnn star, and was first identified as a
runaway originating in the Sco~OB2 association by Blaauw
(\cite{bla1952b}).  Based on its proper motion, which points away from
the association, its radial velocity, and the large space velocity
($\sim$30~km~s$^{-1}$), Blaauw suggested that $\zeta$~Oph might have
formed in the center of the association $\sim$3~Myr ago. Later
investigations (e.g., Paper I; Blaauw \cite{bla1993}; van 
Rensbergen et al.\ \cite{rvl1996}) showed that $\zeta$~Oph either
became a runaway $\sim$1~Myr ago in the Upper Scorpius subgroup of
Sco~OB2, or 2--3~Myr ago in the Upper Centaurus Lupus subgroup (cf.\
de Zeeuw et al.\ \cite{zee1999}). 

If $\zeta$~Oph is a BSS runaway, as suggested by its high helium
abundance ($\epsilon=0.16$, corresponding to a mass fraction $X =
0.577$ of H) and large rotational velocity (348~km~s$^{-1}$), and if
the binary dissociated after the supernova explosion, we might be able
to identify the associated neutron star. None of the pulsars in
Figure~\ref{fig:02} was ever inside the Upper Centaurus Lupus
subgroup, but two could have originated from the Upper Scorpius
subgroup: PSR~J1239$+$2453 and PSR~J1932$+$1059\footnote{Recently, 
Walter (\cite{wal2000}) suggested that RX~J185635$-$3754 as another 
candidate neutron star that could have encountered $\zeta$~Oph in the 
past. We show in Appendix~B that this is unlikely.}.

We first consider PSR~J1239$+$2453. Its estimated distance is
$\sim$560~pc. It passed within about 20~pc of the Upper Scorpius
region $\sim$1~Myr ago if and only if its (unknown) radial velocity is
large and positive ($\sim$650~km~s$^{-1}$). With a tangential velocity
of $\sim$300~km~s$^{-1}$ (the proper motion is 114 mas yr$^{-1}$), the
space velocity would have to be over 700 ~km~s$^{-1}$, which is
uncomfortably large. Furthermore, while 1~Myr is consistent with the
kinematic age for $\zeta$~Oph, it is in conflict with the characteristic age
($P/(2\dot{P}) = 23$~Myr) of the pulsar. The latter is an uncertain age
indicator, but the difference between the two times is so large that
we consider it unlikely that PSR~J1239$+$2453 was associated with
$\zeta$~Oph. The pulsar is currently at a Galactic latitude of
$86^\circ$, i.e, at $z\sim$560~pc above the Galactic plane. Typical
$z$-oscillation periods of pulsars are of order 100~Myr (e.g., Blaauw
\& Ramachandran \cite{br1998}), so that maximum height is reached
after $T_{1/4}\sim25$~Myr. Taking the characteristic age at face value
suggests the pulsar is near its maximum height above the plane, had a
$z$-velocity of about 30 km~s$^{-1}$, and was not formed in the Upper
Scorpius association (age $\sim$5~Myr), but was born $\sim$25~Myr ago
in the Galactic plane outside the Solar neighbourhood.

The path of the other pulsar, PSR~J1932$+$1059 (earlier designation
PSR~B1929$+$10), also passed the Upper Scorpius association some
1--2~Myr ago. The characteristic age of this pulsar is only
$\sim$3~Myr, consistent with the kinematic age of $\zeta$~Oph within
the uncertainties. The present $z$-velocity of the pulsar
($\sim$40~km~s$^{-1}$ away from the Galactic plane) predicts a maximum
distance away from the plane of 680~pc and $T_{1/4} \sim 28$~Myr. The
pulsar is presently located only $\sim$10~pc below the plane. Since it
presumably formed close to the plane, this means that PSR~J1932$+$1059
either formed recently or well over 50~Myr ago. Considering that both
the characteristic age and the typical pulsar ages (up to
$\sim$50~Myr) (Blaauw \& Ramachandran \cite{br1998}) are significantly
smaller than $\sim$50~Myr, we conclude that the pulsar formed
recently.  Upper Scorpius is the only site of star formation along the
past trajectory of the pulsar. We thus consider PSR~J1932$+$1059 a
good candidate for the remnant of the supernova which caused the
runaway nature of $\zeta$~Oph.

\subsection{Data}
Table~\ref{tab:03} summarizes the data for $\zeta$~Oph and PSR
J1932$+$1059. The radial velocity of the pulsar is unknown.  The
pulsar proper motion listed by Taylor et al.\ (\cite{tml1993}) was
calculated from timing measurements (Downs \& Reichley \cite{dr1983}).
More accurate proper motions can be obtained from VLBI observations;
Campbell (\cite{cam1995}) measured a provisional proper motion and
parallax of PSR~J1932$+$1059 of $(\mu_{\alpha\ast},\mu_\delta) =
(96.7$$\pm$$1.7,41.3$$\pm$$3.5)$~mas~yr$^{-1}$ and $\pi =
5$$\pm$$1.5$~mas, respectively, including a full covariance matrix.
These measurements are in good agreement with those of Taylor et al.\
(\cite{tml1993}; see Table~\ref{tab:03} and Figure~\ref{fig:05}).

\subsection{Simulations}\label{sec:us_sim}
Our hypothesis is that $\zeta$~Oph and PSR~J1932$+$1059 are the
remains of a binary system in Upper Scorpius which became unbound when
one of the components exploded as a supernova. Support for this
hypothesis would be to find both objects at the same position at the
same time in the past. Our approach is to calculate their past orbits
and simultaneously determine the separation between the two objects,
$D_\mathrm{min}(\tau)$, as a function of time, $\tau$.  We define
$D_\mathrm{min}(\tau)$ as $|\vec{x}_{\zeta~\mathrm{Oph}} -
\vec{x}_\mathrm{pulsar}|$, where $\vec{x}_j$ is the position of object
$j$. We consider the time $\tau_0$ at which $D_\mathrm{min}(\tau)$
reaches a minimum to be the kinematic age. To take the errors in the
observables into account we calculate a large set of orbits, sampling
the parameter space defined by the errors. We use the Taylor et al.\
(\cite{tml1993}) proper motion for the pulsar. The errors in the
positions of the runaway and the pulsar are negligible, and those in
the proper motions of the two objects and in the parallax of the
runaway are modest ($\le 10$\%). However, the radial-velocity error of
$\zeta$~Oph is considerable (5~km~s$^{-1}$). The distance to the
pulsar has a significant error, and its radial velocity is unknown.
Accordingly, we first determine the region in the
$(\pi_\mathrm{pulsar}, \mathv_\mathrm{rad,pulsar})$ parameter space
for which the pulsar approaches the runaway when we retrace both
orbits. Sampling a grid in $(\pi_\mathrm{pulsar},
\mathv_\mathrm{rad,pulsar})$ while keeping the other parameters fixed,
we find that, for $2 \la \pi_\mathrm{pulsar} \la 6$~mas and $100 \la
\mathv_\mathrm{rad,pulsar} \la 300$~km~s$^{-1}$, the motions of the
pulsar and the runaway are such that their separation decreases as one
goes back in time.

\FigThree
Adopting $\pi_{\rm pulsar} = 4$$\pm$$2$~mas and
$\mathv_\mathrm{rad,pulsar} = 200$$\pm$$50$ km~s$^{-1}$, we calculate
three million orbits for the pulsar and the runaway.  Considering that
the pulsar proper-motion errors might be underestimated (Campbell et
al.\ \cite{cam1996}; Hartman \cite{har1997}), we increased them by a
factor of two.  For each run we create a set of positions and
velocities for the runaway and the pulsar consistent with the
(modified 3$\sigma$) errors on the observables. We also calculated the
orbit of the Upper Scorpius association back in time, using the mean
position and velocity derived by de Zeeuw et al.\ (\cite{zee1999},
their table~2).  $30\,822$ of these simulations resulted in a minimum
separation between the pulsar and the runaway of less than 10~pc. In
$4\,214$ simulations the pulsar and the runaway had a minimum
separation less than 10~pc and were both situated within 10~pc of the
center of the association (the smallest minimum separation found was
0.35~pc). Thus, only a small fraction (0.14\%) of the simulations is
consistent with the hypothesis that the pulsar and the runaway were
once, $\sim$1~Myr ago, close together within the Upper Scorpius
association. We now show that given the measurement uncertainties,
this low fraction is perfectly consistent with the two objects being
in one location in the past. 

Figure~\ref{fig:03} shows the distribution of the minimum
separations, $D_\mathrm{min}(\tau_0)$, and the kinematic ages,
$\tau_0$, of the $4\,214$ simulations mentioned above. The lack of
simulations which yield very small minimum absolute separations is due
to the three-dimensional nature of the problem. Consider the following
case: two objects are located at exactly the same position in space,
e.g., the binary containing the pulsar progenitor and the
runaway. However, the position measurement of each of these objects
has an associated typical error. The distribution of the absolute
separation between the objects, obtained from repeated measurements of
the positions of both objects, can be calculated analytically for a
Gaussian distribution of errors (see Appendix A). The solid line in
Figure~\ref{fig:03} shows the result for an adopted distance
measurement error of 2.5~pc, which agrees very well with our
simulations. The peculiar statistical properties of the sample of
successful simulations make it difficult to give a simple argument to
derive the value of 2.5~pc from the uncertainties in the kinematic
properties of the runaway and the pulsar.  We suspect that the
disagreement between the solid line and histogram for separations
$>$6~pc is most likely due to a slight mismatch between the model and
the actual situation. Even so, Figure~\ref{fig:03} shows that due to
measurement errors, very few simulations will produce a small observed
minimum separation, even when the intrinsic separation is
zero.\looseness=-2

\FigFour

Figure~\ref{fig:04} shows the astrometric parameters of the pulsar
and the runaway at the start of the orbit integration, i.e., the
`present' observables, for the simulations which result in a minimum
separation less than 10~pc occurring within Upper Scorpius. The
parameters of $\zeta$~Oph show no correlations except between the
parallax and the radial velocity. This is expected due to the
degeneracy of these two quantities (a change in stellar distance,
depending on whether it increases or decreases the separation between
the star and the association, can be compensated for by a larger or
smaller radial velocity, respectively).

The parameters of the pulsar behave very differently.  In addition to
the $\pi$ vs.\ $\mathv_\mathrm{rad}$ correlation, we find that the
parallax is also correlated with both of the proper motion
components. As a result, the proper motion components are correlated
with each other. This means that only a subset of the full parameter
space defined by the six-dimensional error ellipsoid of the pulsar
fulfills the requirement that the pulsar and the runaway meet.
Furthermore, if they met, then we know the radial velocity of the
pulsar for each assumed value of its distance. A reliable distance
determination would thus yield an astrometric radial velocity. The
current best distance estimate of the pulsar derived from VLBI
measurements, $\pi = 5\pm1.5$~mas (Campbell \cite{cam1995}), predicts
a radial velocity of 100--200~km~s$^{-1}$. This radial velocity is
comparable to the tangential velocity: $\sim$100~km~s$^{-1}$ (for $\pi
= 5$~mas).

The pulsar proper motions of the $4\,214$ successful simulations are
shown in Figure~\ref{fig:05}, together with the proper-motion
measurements of Lyne, Anderson \& Salter (\cite{las1982}) (dot-dash
line), Taylor et al.\ (\cite{tml1993}) (solid line), and Campbell
(\cite{cam1995}) (dashed line). The measurements show a reasonable
spread, reflecting the difficulty in obtaining pulsar proper motions,
but are consistent, within 3$\sigma$, with the proper motions
predicted by the simulations.

\subsection{Interpretation}
The observed astrometric and spectroscopic parameters of $\zeta$~Oph 
and PSR~J1932$+$1059 are consistent with the assumption that these
objects were very close together $\sim$1~Myr ago
(Figure~\ref{fig:03}).  At that time both were within the boundary
of Upper Scorpius (Figure~\ref{fig:06}), which has a nuclear age of
$\sim$5~Myr (de Geus, de Zeeuw \& Lub \cite{gzl1989}).

Several characteristics of $\zeta$~Oph and Upper Scorpius support the
interpretation that the runaway and the pulsar must have been produced
by the BSS. (i) The HI distribution in the direction of Upper Scorpius
shows an expanding shell-like structure.  De Geus (\cite{geu1992})
argued that the energy output of the stellar winds of the massive
stars in Upper Scorpius is two orders of magnitudes too low to account
for the kinematics of the HI shell, and proposed that a supernova
explosion created the Upper Scorpius HI shell.
(ii) Based on the present-day mass function, de Geus showed that the
initial population of Upper Scorpius most likely contained one star or
binary more than the present population. The estimated mass of this
additional object is $\sim$40~$M_\odot$. Stars of this mass have
main-sequence lifetimes of $\sim$4~Myr, and end their lives in a
supernova explosion.  Since the association has an age of $\sim$5~Myr,
the supernova explosion might have taken place $\sim$1~Myr ago.
(iii) The characteristics of $\zeta$~Oph are indicative of close
binary evolution. The helium abundance is large, and the star has a
large rotational velocity (see \S\ref{sec:intro}).
These facts make it very likely that the same supernova event in Upper
Scorpius created PSR~J1932$+$1059 and endowed $\zeta$~Oph with its
large velocity.

\FigFive

It could be that $\zeta$~Oph and PSR~J1932$+$1059 are not
related. This would imply that the pulsar originated in Upper Scorpius
$\sim$1~Myr ago and that $\zeta$~Oph obtained its large velocity
either in a separate BSS event in Upper Scorpius, or in Upper
Centaurus Lupus, $\sim$3~Myr ago, as suggested by van Rensbergen et
al.\ (\cite{rvl1996}). In the latter case, it would also have to be
formed by the BSS, because of its high helium abundance, large
rotational velocity, and the $\sim$10~Myr difference between the age
of Upper Centaurus Lupus and the kinematic age of $\zeta$~Oph
(\S\ref{sec:intro}).  Given the small probability of finding a runaway
star and a pulsar with orbits that cross, and with {\it both} objects
at the point of intersection at the {\it same} time, we conclude
that $\zeta$~Oph and PSR~J1932$+$1059 were once part of the
same close binary in Upper Scorpius, providing the first direct
evidence for the generation of a single runaway star by the BSS.
%
\FigSix
%

\subsection{Pulsar kick velocity}
If $\zeta$~Oph and PSR~J1932$+$1059 were once part of a binary, then
we can derive a number of properties of this system. For example, the
true age of the pulsar must be the kinematic age of 1~Myr (as compared
to the characteristic age estimate of 3~Myr). It follows that, if no
glitches occurred, the pulsar had a period of 0.18 seconds at birth,
as compared to the current period of 0.22 seconds.

The velocity distribution of the pulsar population is much broader (a
few$\times$100~km~s$^{-1}$) than that of the pulsar progenitors (a
few$\times$10~km~s$^{-1}$). The mechanism responsible for this
additional velocity (the `kick velocity' $\vec{\mathv}_\mathrm{kick}$)
is not well understood (e.g., Lai \cite{lai1999}). The kick velocity
is most likely due to asymmetries in the core of a star just before,
or during, the supernova explosion.

The simulations described in \S\ref{sec:us_sim} provide the velocities
of the runaway star, the pulsar, and the association at the present
time. This makes it possible to determine $\vec\mathv_\mathrm{kick}$
of the neutron star. For the $4\,214$ successful runs we find that the
average velocities with respect to Upper Scorpius are:
$\vec{\mathv}_{\zeta~\mathrm{Oph}} = (-6.4$$\pm$$4.2, 33.8$$\pm$$1.4,
5.8$$\pm$$2.0)$~km~s$^{-1}$ and $\vec{\mathv}_\mathrm{pulsar} =
(48.6$$\pm$$21.7, 222.9$$\pm$$36.1, -70.7$$\pm$$8.4)$\break
km~s$^{-1}$ in Galactic Cartesian coordinates $(U,V,W)$\footnote{$U$
points in the direction of the Galactic centre, $V$ points in the
direction of Galactic rotation, and $W$ points towards the North
Galactic pole.}. 
%

To derive $\vec\mathv_\mathrm{kick}$ we consider a binary with
components of mass $M_1$ and $M_2$ in a circular orbit, in which the
first component (star$_1$) explodes as a supernova and creates a
neutron star. At the time of the explosion, star$_1$ is the least
massive component of the binary, due to the prior mass transfer phase,
and is most likely a helium star. The rapidly expanding supernova
shell, with mass $\Delta M = M_1 - M_{\rm n}$, where $M_{\rm n} =
1.4~M_\odot$ is the typical mass a neutron star, will quickly leave
the binary system. The shell has a net velocity equal to the orbital
velocity of star$_1$ at the moment of the explosion ($v_1$). A net
amount of momentum ($\Delta M \times v_1$) is thus extracted from the
system and the binary reacts by moving in the opposite direction with
a velocity $v = -(\Delta M\times v_1) / (M_2 + M_n)$, the so-called
`recoil velocity'. The binary will remain bound after the explosion
because less than half of the total mass of the system is expelled
($M_1 < M_2$; cf.\ Paper I).  However, if the neutron
star receives a kick in the supernova explosion the binary might
dissociate, depending on the direction and magnitude of the kick
velocity.  We simulate this by using a simple orbit integrator for two
bodies. We determine the semi-major axis and orbital velocities
assuming the binary has a circular orbit and masses $M_1 = 5~M_\odot$
and $M_2 = 15~M_\odot$ ($\zeta$~Oph). We then change the mass of
star$_1$ to $M_n$ and add a kick-velocity to its orbital velocity. We
start the integration at this point and try to reproduce the observed
velocity of $\zeta$~Oph, the pulsar, and the angle between the two
velocity vectors (35$^\circ$).  It turns out that a kick velocity of
order 350~km~s$^{-1}$ is needed in a direction almost opposite to the 
orbital velocity of star$_1$'s prior to the explosion.  This value is
in good agreement with the average pulsar kick velocity found by
Hartman (\cite{har1997}) and Hansen \& Phinney (\cite{hp1997}).  The
current velocity of the pulsar, $\sim$240~km~s$^{-1}$, is more
than 100~km~s$^{-1}$ smaller than the kick it acquired. Our simulations
show that this deceleration is due to the gravitational pull of
$\zeta$~Oph on the pulsar.\looseness=-2

The mass of $\zeta$~Oph used in the above estimate is consistent with
the calibration of Schmidt--Kaler (\cite{sk1982}). The more recent mass 
calibration of Vanbeveren, Van Rensbergen \& De Loore (\cite{vvd1998}) 
suggests 21 $M_\odot$ (Table \ref{tab:03}). This would increase the 
inferred kick velocity to $\sim$400 km~s$^{-1}$.

\section{A dynamical ejection in Orion}\label{sec:orion}

\subsection{AE~Aurigae \& $\mu$~Columbae}
Blaauw \& Morgan (\cite{bm1954}) drew attention to the isolated stars
AE~Aur (O9.5V) and $\mu$~Col (O9.5V/B0V), which move away from the
Orion star-forming region (e.g., McCaughrean \& Burkert \cite{mb2000})
in almost opposite directions with comparable space velocities of
$\sim$100~km~s$^{-1}$ (Figure~\ref{fig:11}, stars 5 and 6 in
Table~\ref{tab:03}). Blaauw \& Morgan suggested that {\it `...\ the
stars were formed in the same physical process 2.6 million years ago
and that this took place in the neighborhood of the Orion Nebula.'}
The past orbits of AE~Aur and $\mu$~Col intersect on the sky near the
location of the massive highly-eccentric double-lined spectroscopic
binary $\iota$~Ori (O9III$+$B1III, see Stickland et al.\
\cite{sti1987}). This led Gies \& Bolton (\cite{gb1986}) to suggest
that the two runaways resulted from a dynamical interaction also
involving $\iota$~Ori: {\it`...\ $\iota$~Ori is the surviving binary
of a binary-binary collision that ejected both AE~Aur and $\mu$~Col.'}

\FigSeven

\subsection{Data}\label{sec:data}
Table~\ref{tab:03} lists the data for AE~Aur, $\mu$~Col, and
$\iota$~Ori. We adopt Stickland's et al.\ (\cite{sti1987}) radial
velocity for $\iota$~Ori ($\vec{\mathv}_\gamma$). For the
radial-velocity errors for AE~Aur and $\mu$~Col we use the largest
errors quoted in either the Catalogue de Vitesses Radiales Moyennes
Stellaires (Barbier--Brossat \cite{bb1989}), the Hipparcos Input
Catalogue (Turon et al.\ \cite{tur1992}), the Wilson--Evans--Batten
Catalogue (Duflot, Figon \& Meyssonnier \cite{dfm1995}), or in the
SIMBAD database.

\subsection{Simulations}\label{sec:story}
To investigate the hypothesis that the three stellar systems, AE~Aur,
$\mu$~Col, and $\iota$~Ori, were involved in a binary-binary
encounter, we retrace their orbits back in time to find the minimum
separation between them. As in \S\ref{sec:SNinUS}, we explore the
parameter space determined by the errors of, and correlations between,
the observables.

Even with the unprecedented accuracy in trigonometric parallaxes
obtained by the Hipparcos satellite, the errors on the individual
distances are rather large: $D_\mathrm{AE~Aur} =
446_{-111}^{+220}$~pc, $D_{\mu\ \mathrm{Col}} =
397_{-\phantom{1}71}^{+110}$~pc, $D_{\iota\ \mathrm{Ori}} =
406_{-\phantom{1}96}^{+185}$~pc. We therefore first determine which
distances are most likely to agree with our hypothesis, and then study
the effect of the measurement errors on the other observables. For
each pair of stars, Figure~\ref{fig:07} shows contours of minimum
separation between the respective orbits as a function of their
present distances.  The distances of the stars for which the orbits
have a small minimum separation are strongly correlated, i.e., if the
distance of star $i$ increases that of star $j$ also needs to increase
to obtain a small minimum separation. We therefore choose to show the
contours of constant minimum separation with respect to this
correlation. The vertical axes thus show offsets from the straight
line in the distance vs.\ distance plane defined by the equation in
the top right of each panel.

We start each simulation with a set of positions and velocities which
are in agreement with the observed parameters and their covariance
matrices ($<$3$\sigma$). Furthermore, we require the distances of the
stars to fall within the 10~pc minimum-separation contours of
Figure~\ref{fig:07}. We then calculate the orbits of AE~Aur,
$\mu$~Col, and $\iota$~Ori. We define the separation between the three
stellar systems, $D_\mathrm{min}(\tau)$, as the maximum deviation of
the objects from their average position, i.e., $D_\mathrm{min}(\tau) =
\mathrm{max}|\vec{x}_{j} - \bar{\vec{x}}|$ for $j = $ AE~Aur,
$\mu$~Col, and $\iota$~Ori, where $\bar{\vec{x}} = \frac{1}{3}
(\vec{x}_{\mathrm{AE\ Aur}} + \vec{x}_{\mu\ \mathrm{Col}} +
\vec{x}_{\iota\ \mathrm{Ori}})$ is the mean position and $\vec{x}_j$
the position of star $j$. The time $\tau_0$ at which
$D_\mathrm{min}(\tau)$ reaches a minimum is considered to be the time
of the encounter, i.e., the kinematic age.\looseness=-2

%
\FigEight
%
We computed 2.5 million orbits, of which 114 yielded
$D_\mathrm{min}(\tau_0) < 1$~pc with $\tau_0 =$ 2--3~Myr. One of the
simulations resulted in a minimum separation of 0.019~pc which is
equal to $4\,000$~AU (Figure~\ref{fig:08}). The small number of
simulations with small minimum separations is due to (i) the large
number of parameters involved (i.e., 18) and (ii) the
three-dimensional nature of the problem (cf.\ \S\ref{sec:us_sim}).

We have numerically determined the distribution of the minimum
separations $D_\mathrm{min}$ of three points drawn from a
three-dimensional Gaussian error distribution (the analytic results of
Appendix A are valid only for two Gaussians).  We randomly draw three
points from three spherical three-dimensional Gaussians (with standard
deviation $\sigma$) and determine $D_\mathrm{min}$. The Gaussians have
the same mean positions. The resulting distribution for $\sigma =
4$~pc resembles the real one remarkably well (Figure~\ref{fig:08}).
A distance uncertainty of four pc is consistent with the
$\sim$2~km~s$^{-1}$ uncertainties in the velocities of the runaways
and $\iota$~Ori: 2~km~s$^{-1}$ over $\sim$2~Myr results in a
displacement of $\sim$4~pc.  Thus, the data and their errors are
consistent with the hypothesis that $\sim$2.5~Myr ago AE Aur,
$\mu$~Col, and $\iota$~Ori were in the same small region of space.

\FigNine

\subsection{Interpretation}
The nominal observed properties of the runaway stars AE~Aur and
$\mu$~Col and the binary $\iota$~Ori are consistent with a common
origin $\sim$2.5~Myr ago. The most likely mechanism that created the
large velocities of the runaways and the high eccentricity of the
$\iota$~Ori binary is a binary-binary encounter, as suggested by Gies
\& Bolton (\cite{gb1986}). The normal rotational velocities of both
runaways (25~km~s$^{-1}$ and 111~km~s$^{-1}$) and the normal helium
abundance of AE~Aur (Table~\ref{tab:03}, see Blaauw \cite{bla1993},
figure~6) also suggest that these runaways were formed by the
dynamical ejection scenario. The helium abundance of $\mu$~Col is
unknown.
\subsection{Parent cluster}\label{sec:parent}
To find the cluster, or region of space, where the encounter between
AE~Aur, $\mu$~Col, and $\iota$~Ori took place we assume that the
center of mass velocity of the three objects is identical to the mean
velocity $\vec{\mathv}_\mathrm{clus}$ of the parent cluster.  Then
\begin{equation}
\vec{\mathv}_\mathrm{clus} = \frac{\sum_j M_j \vec{\mathv}_j}{\sum_j M_j}, 
\label{eq:2}
\end{equation}
for $j =$ AE Aur, $\mu$ Col, and $\iota$ Ori. For each star we
estimate the mass by interpolating the mass vs.\ spectral-type
calibration of Schmidt--Kaler (\cite{sk1982}, table~23). We obtain
$15.9~M_\odot$ for AE~Aur and $\mu$~Col, and $22.9~M_\odot$ and
$14.9~M_\odot$ for the primary and secondary of $\iota$~Ori,
respectively ($37.8~M_\odot$ for the binary system).  We use the
cluster velocity ($\vec{\mathv}_\mathrm{clus}$) and the mean position
of the three stellar systems at the moment of the encounter to
integrate the orbit of the ensemble of stars $\tau_0$ Myr into the
future. The position and velocity at the end of this integration
should coincide with the present-day properties of the parent
cluster. We extend the Monte Carlo simulations described in
\S\ref{sec:story} to include the integration of the orbit of the
`cluster' forward in time. Figures~\ref{fig:09} and \ref{fig:10}
summarize the results. Panel {\it b} of Figure~\ref{fig:10} shows
that the distances of the three stars and the predicted cluster
distance are tightly correlated (see also Figure~\ref{fig:07}); all
distances increase when the cluster distance increases. A consequence
of this tight correlation is that as soon as the distance to one of
the objects is known, all other distances are fixed.

\subsubsection{Biases and measurement errors}\label{sec:biases}
Two effects influence the mean cluster properties as predicted by the
Monte Carlo simulations. First, it is easier to hit a target from
close by than from far away, i.e., a larger range of velocities
(within the errors) is consistent with the encounter hypothesis when
the distance between the star and the encounter point is small (the
`aiming effect').
%
%
We simulate this effect in the following way. We assume a range of
cluster distances, 350--500~pc. For each distance we use
Figure~\ref{fig:09} (the gray dots in the first row) to determine
the other phase-space coordinates of the parent (position on the sky,
proper motion, and radial velocity). With these `observables' we
calculate the three-dimensional velocity of the cluster, corrected for
Solar motion, and determine its position at a time $\tau_0$ (see first
row in Figure~\ref{fig:09}) in the past. We neglect the variation
of the Galactic potential, ignore Galactic rotation, and use the
linear velocity, to speed up the calculations.  This past position of
the cluster combined with the present three-dimensional positions of
AE Aur, $\mu$ Col, and $\iota$ Ori (based on the present positions on
the sky and the distances from Figure~\ref{fig:10} panel {\it b})
gives the velocities of the three stellar systems today, using
$\tau_0$ as the time difference. These `observed' properties are then
used as input for the Monte Carlo simulations described above to
investigate the influence of the aiming effect on the predicted
cluster distance. The circles in Figure~\ref{fig:10} panel {\it c}
display the bias in the cluster distance.

Secondly, the trigonometric distance of $\mu$ Col, $D_{\mu\
\mathrm{Col}} = 397_{-\phantom{1}71}^{+110}$~pc, is smaller
(2$\sigma$) than the observed photometric distance, $\sim$750~pc
(e.g., Gies \cite{gie1987}). The photometric distance is reliable
since $\mu$~Col is located in a region free of interstellar
absorption. This difference between the trigonometric distance and the
`real' distance results in an additional bias towards smaller
distances for the stars and the cluster. In our Monte Carlo simulation
we draw the parallaxes, like all other observables, from a Gaussian
centred on the observed value and with a width equal to the observed
error. For the Hipparcos distance of $\mu$~Col this means that less
than $\sim$10\% of the random realizations will be consistent with the
photometric distance\footnote{The Hipparcos parallax of $\mu$~Col
(2.52$\pm$0.55~mas) deviates more than 1.5$\sigma$ from the
photometric parallax ($\sim$1.3~mas). The random realisations of
$\pi_{\mu~{\rm Col}}$ which are consistent with $\pi_{\rm phot}$ thus
need to fall outside the $-1.5$$\sigma$ confidence level which occurs
in fewer than 10\% of the cases.}.  And because the distances of the
three stellar systems and the cluster are correlated (see
Figure~\ref{fig:10} panel {\it b}), the other stars also need to be
at smaller distances for the encounter to take place.  This effect
will result in a mean cluster distance (the mean of the Monte Carlo
simulations) which is underestimated.  We simulated this effect in a
similar manner as the aiming effect. The results on the mean cluster
distance in the Monte Carlo simulations, aiming effect and the
parallax of $\mu$~Col, are shown as the triangles in
Figure~\ref{fig:10} panel {\it c}.

\FigTen
%
%
\subsubsection{Cluster properties and identification}
Taking the biases on the cluster distance into account, we reconstruct
the present-day properties of the parent cluster of the stars AE~Aur,
$\mu$~Col, and $\iota$~Ori. The mean cluster distance from our
Monte Carlo simulations is 339~pc (right most panel in the second row
of Figure~\ref{fig:09}). The cluster distance corrected for biases is
425--450~pc. Using this distance we determine the other
properties of the cluster (first row of Figure~\ref{fig:09}), and
summarize them in Table~\ref{tab:04}. Figure~\ref{fig:09}
(right most panel of the first row) indicates that the encounter
happened 2.5~Myr ago; this obviously is a lower limit to the age of
the cluster. Figure~\ref{fig:11} shows the region of the sky where the
parent cluster should be located: the Orion Nebula. The black contours
in the bottom panel show the distribution of $(\ell,b)$ of the parent
cluster obtained from the Monte Carlo simulations.

\TabFour
 
Of all clusters in this active star-forming region the Trapezium
cluster (NGC 1976) is the most likely parent cluster for the following
reasons.  
\newcounter{it63}
\begin{list}{\arabic{it63}:}{
     \usecounter{it63}
     \setlength{\topsep}{1mm}
     \setlength{\labelwidth}{3mm}
     \setlength{\labelsep}{2mm}
     \setlength{\itemindent}{0mm}
     \setlength{\parsep}{0mm}
     \setlength{\itemsep}{0mm}
     \setlength{\leftmargin}{5mm}}
\item The cluster is young. Palla \& Stahler (\cite{ps1999}) find a
      mean age of $\sim$2~Myr based on theoretical pre-main-sequence
      tracks, and established that the first stars formed not more
      than 5 Myr ago.  Thus, the Trapezium is old enough to have
      produced the runaways.
\item The Trapezium is one of the most massive, dense clusters in the
      Solar neighbourhood. Estimates for the stellar density are
      $>20\,000$~stars~pc$^{-3}$ for the inner 0.1--0.3 pc (e.g.,
      McCaughrean \& Stauffer \cite{ms1994}; Hillenbrand \& Hartmann
      \cite{hh1998}). These high stellar densities favor dynamical
      interactions within the cluster core.
\item The Trapezium shows a strong mass segregation (Zinnecker,
      McCaughrean \& Wilking \cite{zmw1993}; Hillenbrand \& Hartmann
      \cite{hh1998}). Five of the six stars more massive than
      $10 M_\odot$ are in the centre. This concentration of massive
      stars increases the probability for dynamical interactions
      between these stars.
\item The binary fraction in the Trapezium cluster is at least as high
      as that of the Solar-type field stars, i.e., $\sim$60\% (Prosser
      et al.\ \cite{pro1994}; Petr et al.\ \cite{pet1998}; Simon, Cose
      \& Beck \cite{scb1999}; Weigelt et al.\ \cite{wei1999}). This
      means that enough binary systems are available for binary-binary
      or binary-single-star interactions to become efficient in
      expelling stars from the cluster.
\end{list}

The mean astrometric properties and the radial velocity of the
Trapezium agree perfectly with those predicted by our Monte Carlo
simulation. The distance to the Trapezium is estimated to be
450--500~pc (Walker \cite{wal1969}; Warren \& Hesser \cite{wh1977a},
\cite{wh1977b}, \cite{wh1978}; Genzel \& Stutzki \cite{gs1989}); we
predict 425--450~pc. The observed radial velocity of the Trapezium is
23--25~km~s$^{-1}$ (Johnson \cite{joh1965}; Warren \& Hesser
\cite{wh1977a}, \cite{wh1977b}; Abt, Wang \& Cardona \cite{awc1991};
Morrell \& Levato \cite{ml1991}); we predict $\sim$28~km~s$^{-1}$.
The absolute proper motion of the Trapezium is ill-determined, but is
known to be small (e.g., de Zeeuw et al.\ \cite{zee1999}). We
collected all stars, within a 0\fdg4 by 0\fdg4 region centred on the
Trapezium, based on the Tycho~2 Catalogue (H{\o}g \cite{hog2000}), and
plot the proper motions in Figure~\ref{fig:10} {\it a}. The proper
motions agree with the predicted cluster proper motion.

Table~\ref{tab:04} shows that the predicted position on the sky of the
parent cluster does not fully agree with the position of the Trapezium
(see Figure~\ref{fig:11}). Here it is important to remember that we
did not allow for any errors on the stellar masses used in
eq.~(\ref{eq:2}). We investigate the effect of mass errors by changing
the masses and running a new set of Monte Carlo simulations. We find
that (i) the results are insensitive to the mass of $\iota$~Ori: a
change as large as $\pm$5~$M_\odot$ produces no noticeable change in
the cluster properties, and (ii) the sky position of the parent
cluster and its proper motion depend on the mass ratio of AE~Aur and
$\mu$~Col. Changing the mass of $\mu$~Col by $-1~M_\odot$ or the mass
of AE~Aur by $+1~M_\odot$ shifts the predicted sky position of the
parent cluster to that of the Trapezium cluster
(Figure~\ref{fig:11}). A mass change in the other direction,
$+1~M_\odot$ for $\mu$~Col and $-1~M_\odot$ for AE~Aur, creates a
similar shift in the opposite direction. There are indications from
spectral-type determinations that $\mu$~Col is indeed slightly less
massive than AE~Aur. Most spectral-type determinations of $\mu$~Col
give O9.5V; however, Blaauw \& Morgan (\cite{bm1954}) and Paper I
quote B0V and Houk (\cite{hou1982}) quotes B1IV/V. 

\clearpage
\FigEleven

\clearpage

We note that the calibration of Vanbeveren et al.\
(\cite{vvd1998}) gives a mass of 38.6 $M_\odot$ for $\iota$~Ori,
similar to that found with the Schmidt--Kaler calibration, but
increases the masses of AE Aur and $\mu$~Col to 21.1 $M_\odot$. This
does not change our results, as it is the ratio of the runaway masses
that determines the predicted current position of the parent cluster.

In summary, the position, distance, proper motion, and radial velocity
of the Trapezium cluster fall within the range predicted by our Monte
Carlo simulations. Furthermore, the youth, extreme stellar density,
mass segregation, and the high binary fraction make it the best
candidate for the parent cluster of the runaways AE~Aur and $\mu$~Col
and the binary $\iota$~Ori. Finally, it is the only likely candidate
in this region of the sky.

\section{53 Ari, $\xi$ Per, $\zeta$~Pup, and 
         $\lambda$ Cep}\label{sec:class-runaways}
The previous sections gave a specific example of each formation
mechanism for runaway stars, and described our orbit retracing methods
in detail.  We now consider the three other classical runaways, as
well as $\zeta$~Pup, and we discuss the likely formation
mechanisms. The results are summarized in Table~\ref{tab:05}.

\subsection{53~Arietis \& Orion~OB1 \hfill {\rm (star 2)}}
Blaauw (\cite{bla1956}) classified 53~Arietis (HIP~14514) as a runaway
star based on its proper motion which is directed away from the Orion
association. He deduced that 53~Ari left the Orion association
$\sim$4.8~Myr ago and predicted the radial velocity of the star (at
that time unknown) to be $\sim$18~km~s$^{-1}$. The discrepancy in
kinematic ages of AE~Aur and $\mu$~Col (\S\ref{sec:orion}) and that of
53~Ari indicates that it is not related to the same event that created
AE~Aur and $\mu$~Col (\S\ref{sec:orion}) but is another runaway from
the Orion star-forming region.  Table~\ref{tab:03} lists the
observables of 53~Ari\footnote{The Hipparcos Input Catalogue (Turon et
al.\ \cite{tur1992}) lists the radial velocity from the Catalogue de
Vitesses Radiales Moyennes Stellaires (Barbier--Brossat \cite{bb1989})
which is incorrect ($-8.5$~km~s$^{-1}$). The radial velocity in the
Troisieme Catalogue Bibliographique de Vitesses Radiales Stellaires
(Barbier--Brossat, Petit \& Figon \cite{bpf1994}) is also
incorrect. They list a radial velocity corrected for Solar motion of
15.3~km~s$^{-1}$ adopted from Sterken (\cite{ste1988}).  The
uncorrected radial velocity is 24.2~km~s$^{-1}$.}.

\FigTwelve

The Ori~OB1 association has four subgroups: {\sl a}, {\sl b}, {\sl c},
and {\sl d} (Blaauw \cite{bla1964}; Brown, de Geus \& de Zeeuw
\cite{bgz1994}). We do not consider subgroup {\sl d} (the Trapezium)
as a possible parent group of 53~Ari, since this subgroup is younger
than the runaway (\S\ref{sec:orion}).  The ages of the other
subgroups are: 8--12~Myr for subgroup {\sl a}, 2--5~Myr for subgroup
{\sl b}, and $\sim$4~Myr for subgroup {\sl c} (Warren \& Hesser
\cite{wh1977a}, \cite{wh1977b}; Brown et al.\ \cite{bgz1994}).

\paragraph{Simulations:}
We performed a set of simulations as in \S\ref{sec:samples}, retracing
orbits for each subgroup ({\sl a}, {\sl b}, {\sl c}).  The kinematic
age of 53~Ari from subgroup {\sl a} is $\sim$4.3~Myr
(Figure~\ref{fig:12}).  This means that the subgroup was
$\sim$6~Myr old when 53~Ari became a runaway star. This very likely
rules out the DES as the formation mechanism (see \S\ref{sec:intro}).
However, there is little direct evidence in favor of the BSS.  The
helium abundance of 53~Ari is unknown and its observed rotational
velocity is small ($\mathv_\mathrm{rot} \sin i = 10$~km~s$^{-1}$), but
this could be caused by a near pole-on orientation. We did not find a
neutron star associated with 53~Ari, but our sampling of the nearby
compact objects is severely limited
(\S\ref{sec:selection}).\looseness=-2

If subgroup {\sl b} is the parent association the kinematic age for
53~Ari is $\sim$4.8~Myr. This is comparable to the canonical age of
the subgroup, and excludes the BSS as a production mechanism for
53~Ari (see \S\ref{sec:intro}). If Ori~OB1 {\sl b} is the parent group
of 53~Ari then the kinematic age is $\sim$4.8~Myr and the formation
mechanism is most likely the DES.  However, the most recent age
determination (Brown et al.\ \cite{bgz1994}) gives $1.7\pm1.1$~Myr. If
Ori~OB1 {\sl b} is indeed this young then the subgroup is younger than
53~Ari and cannot be the parent group.

For subgroup {\sl c} we find that the minimum separation between the
subgroup centre and the runaway was never smaller than 15~pc, while
the simulations for the other two subgroups {\sl a} and {\sl b} yield
minimum separations as small as 1~pc.  The space motion of Ori~OB1 is
mostly directed radially away from the Sun, and the proper motion
component is relatively small. The Hipparcos data did not allow de
Zeeuw et al.\ (\cite{zee1999}) to discriminate between the different
subgroups in their selection procedure; they only give one proper
motion and radial velocity for the whole Orion complex. It is possible
that subgroup {\sl c} has a motion that differs slightly from that of
the other two subgroups, so that it cannot be ruled out as a candidate
parent group.  The age of subgroup {\sl c}, $\sim$5~Myr, is similar to
the kinematic age of 53~Ari. By the argument given above this suggests
that if Ori~OB1 {\sl c} is the parent association of 53~Ari, then the
formation mechanism is most likely the DES.

In order to decide which of the Ori~OB1 subgroups is the parent group
of 53~Ari, we need to know the distances and velocities of the
subgroups and the runaway star with a better accuracy than is now
available.  Figure~\ref{fig:12} could then be used to pin down the
parent group, and the mechanism which is responsible for the runaway
nature of 53~Ari. Since subgroup {\sl a} is the only one for which the
BSS is indicated, finding a pulsar originating from subgroup {\sl a}
at the same time as 53~Ari would also clinch the issue.

\subsection{$\xi$~Persei \& Perseus~OB2 \hfill {\rm (star 3)}}
The O7.5III star $\xi$~Persei (HIP~18614) lies within the boundary of
the Per~OB2 association on the sky. This positional coincidence, and
the low density of early-type stars near Per~OB2, led Blaauw
(\cite{bla1944}) to propose $\xi$~Per as a member of the association.
At that time it was thought that the large radial-velocity difference
between $\xi$~Per and Per~OB2 ($\sim$40~km~s$^{-1}$) was due to
uncertainties in the measurement of $\mathv_\mathrm{rad}$ for
$\xi$~Per.  However, when the radial velocity of $\xi$~Per was
confirmed, its membership of the Per~OB2 association became doubtful
(Blaauw \cite{bla1952a}). In Paper I, Blaauw classified
$\xi$~Per as a runaway star from the Per~OB2 association, which
naturally explains the discrepancy of the radial velocity of the star
and the association. $\xi$~Per was the first star to be classified as
runaway based on its $\mathv_\mathrm{rad}$ alone. Most other runaways
were recognized because their proper motions were directed away from
an association. The parent group, Per~OB2, has an age of $\sim$7~Myr
(e.g., Seyfert, Hardie \& Grenchik \cite{shg1960}; de Zeeuw \& Brand
\cite{zb1985}).

\paragraph{Data:}
We adopt $\mathv_\mathrm{rad}=58.8$~km~s$^{-1}$ for $\xi$~Per
(Bohannan \& Garmany \cite{bg1978}; 
Garmany, Conti \& Massey \cite{gcm1980}; 
Stone \cite{sto1982}; 
Gies \& Bolton \cite{gb1986}). 
This value differs by 10~km~s$^{-1}$ from those quoted in the
Hipparcos Input Catalogue (67.1~km~s$^{-1}$, Turon et al.\
\cite{tur1992}) and the WEB catalogue (70.1~km~s$^{-1}$, Duflot et
al.\ \cite{dfm1995}), which derive from the value listed in the
General Catalogue of Radial Velocities (70.1~km~s$^{-1}$, Wilson
\cite{wil1953}). We take the radial-velocity error to be
5~km~s$^{-1}$; this is equal to the amplitude of the velocity
variations induced by the non-radial pulsations of $\xi$~Per (de Jong
et al.\ \cite{jon1999}). The rotational velocity and helium abundance
are $\mathv_\mathrm{rot} \sin i = 200$~km~s$^{-1}$ and $\epsilon =
0.18$, respectively (see also Table~\ref{tab:03}).

%
\FigThirteen
\paragraph{Simulations:}
Our orbit calculations (\S\ref{sec:samples}) show that the kinematic
age of $\xi$~Per is $\sim$1~Myr (Figure~\ref{fig:13}). At that time
the star was located $\sim$5~pc from the center of Per~OB2, well
inside the association. Figure~\ref{fig:13} also shows that the
present distance of the runaway is 360~pc, assuming 318~pc as the
distance of Per~OB2 (de Zeeuw et al.\ \cite{zee1999}). This distance
for $\xi$~Per is consistent with the Hipparcos parallax at the
2$\sigma$ level.

We infer that the BSS is responsible for the runaway nature of
$\xi$~Per based on (i) the 6~Myr age of Per~OB2 at the time that
$\xi$~Per was ejected, (ii) the high helium abundance of $\xi$~Per,
(iii) its blue straggler nature (\S\ref{sec:blue}), and (iv) the large
rotational velocity (see \S\ref{sec:intro}). Further evidence of a
supernova explosion in the Per~OB2 association is provided by a shell
structure containing HI, dust, OH, CH, and other molecules (Sancisi
\cite{san1970}; Sancisi et al.\ \cite{san1974}). This feature has been
interpreted as a supernova shell which is physically connected to the
Per~OB2 association. We have not found a pulsar counterpart.

$\xi$~Per presently illuminates the California Nebula (NGC~1499),
resulting in an HII emission region. The distance of this nebula is
hard to determine (350--525~pc; Bohnenstengel \& Wendker
\cite{bw1976}; Sargent \cite{sar1979}; Klochkova \& Kopylov
\cite{kk1985}; Shull \& van Steenberg \cite{ss1985}), but must be
similar to that of $\xi$~Per, i.e., $\sim$360~pc.

\subsection{$\zeta$~Puppis, Vela~R2, Vela~OB2 \& Trumpler 10 \break \null \hfill {\rm (star 10)}}
The O4I star $\zeta$~Puppis (HIP~39429) is the brightest and nearest
single O star to the Sun. Its location outside any known association
and its large space velocity ($\sim$60~km~s$^{-1}$) led Upton
(\cite{upt1971}) to propose $\zeta$~Pup as a runaway star. In spite of
the many groups of young, massive stars in the direction of
$\zeta$~Pup, Upton could not make a unique identification of the
parent association. He proposed Vel~OB2 as the most likely
candidate. In a recent study of the stars and ISM in the direction of
Vela, Sahu (\cite{sah1992}) proposed the R association Vel~R2 as a
possible parent for $\zeta$~Pup, based on the retraced path of the
runaway on the sky. For our simulations we considered Vel~OB2, Vel~R2,
the open cluster NGC~2391, and the cluster/association
Tr~10 (de Zeeuw et al.\ \cite{zee1999}) which also lies in the same
direction as Vel~R2.

\paragraph{Data:}
Table~\ref{tab:03} summarizes the data for $\zeta$~Pup. The position
and velocity for the associations Vel~OB2 and Tr~10 are adopted from
de Zeeuw et al.\ (\cite{zee1999}). Only two of the Vel~R2 members are
contained in the Hipparcos Catalogue.  Their distances,
$411_{-143}^{+473}$ (HIP~43792) and $294_{-61}^{+107}$ (HIP~43955),
have large uncertainties or do not agree well with the canonical
distance of Vel~R2, $\sim$870~pc (Herbst \cite{her1975}). We are
therefore unable to obtain meaningful phase-space coordinates of
Vel~R2.

%
\FigFourteen
%
\paragraph{Simulations:}
Although we are unable to run our simulations for Vel~R2, we conclude
that it is not a likely candidate parent association for
$\zeta$~Pup. The distance difference between Vel~R2 ($\sim$870~pc) and
$\zeta$~Pup $(\sim$400~pc, its canonical distance) is too large. The
relative radial velocity between the star and association is
$\sim$40~km~s$^{-1}$ ($\mathv_\mathrm{rad, \zeta~Pup} =
-23.9$~km~s$^{-1}$ and $\mathv_\mathrm{rad, Vel~R2} \sim
20$~km~s$^{-1}$)\footnote{We assume that Vel~R2 has a motion similar
to all other groups in the direction of the constellation Vela, i.e.,
$\sim$20~km~s$^{-1}$, almost independent of distance.}. The
differences in distance, $\sim$400~pc, and velocity,
$\sim$40~km~s$^{-1}$, between $\zeta$~Pup and Vel~R2 yield a kinematic
age of $\sim$10~Myr. This is older than the expected life time of
$\zeta$~Pup (van Rensbergen et al.\ \cite{rvl1996}). Hence Vel~R2 is
not likely to be the parent group.

Our simulations also show that Vel~OB2 is not the parent
association. The minimum separation between the association and the
runaway star is never smaller than 40~pc for reasonable association
distances. Since the association radius is, at maximum, 30~pc, we
conclude that $\zeta$~Pup has never been inside the boundaries of
Vel~OB2. We similarly rule out NGC~2391 as parent group. 

The simulations for the Trumpler~10 group result in minimum
separations of $\ge$10~pc. The inferred kinematic age is $\sim$2~Myr
(Figure~\ref{fig:14}). Ten parsec is comparable to the radius of
Tr~10, so we cannot unambiguously identify or exclude it as the parent
association. Furthermore, if $\zeta$~Pup was in or near Tr~10, then
its current distance must be 250--350~pc (Figure~\ref{fig:14}),
which is smaller than the canonical distance of 400~pc.

The Vela region contains many young stellar clusters, and suffers from
a fair amount of extinction (although $\zeta$~Pup itself is almost
unreddened). It is therefore reasonable to assume that we have not yet
identified the parent group of $\zeta$~Pup. Similar conclusions were 
obtained by Vanbeveren et al.\ (\cite{vvd1998}), and Vanbeveren, De Loore 
\& Van Rensbergen (\cite{vdv1998}). 
%

\FigFiveteen
\subsection{$\lambda$~Cephei, Cepheus~OB2 \& Cepheus~OB3 \hfill {\rm (star 22)}}
$\lambda$~Cep (HIP~109556) was first classified as a runaway star by
Blaauw (Paper I). He noted that its position on the sky, the
direction of the proper motion, and the radial velocity were
consistent with an origin in the association Cep~OB2 (see, e.g.,
figure~22 in de Zeeuw et al.\ \cite{zee1999}).  This luminous
supergiant (O6I(n)f) is located below Cep~OB2 (in Galactic
coordinates) and is traveling away from the Galactic plane and
Cep~OB2. Several other stellar groups are located near Cep~OB2, and of
these, Cep~OB3 may also be a possible parent of $\lambda$~Cep. We
consider both associations.

\paragraph{Data:}
The data for $\lambda$~Cep are given in Table~\ref{tab:03}. The
phase-space coordinates of Cep~OB2 are adopted from de Zeeuw et al.\
(\cite{zee1999}). Unfortunately, the Hipparcos data did not allow
these authors to obtain meaningful results for Cep~OB3 which is at a
distance of $\sim$730~pc (Crawford \& Barnes \cite{cb1970}).  To
estimate the phase-space coordinates of Cep~OB3 we used the mean
position, proper motion, and radial velocity for the Cep~OB3 members
of Blaauw, Hiltner \& Johnson (\cite{bhj1959}): $(\ell,b) =
(110\fdg50,2\fdg91)$; $(\mu_{\ell\ast},\mu_b) =
(-2.72$$\pm$$0.28,-1.78$$\pm$$0.30)$~mas~yr$^{-1}$;
$\mathv_\mathrm{rad} = -17.4$$\pm$$3.0$~km~s$^{-1}$.

\paragraph{Simulations:}
The orbit calculations show that Cep~OB2 cannot be the parent group of
$\lambda$~Cep. The simulations do result in small minimum separations
between the two, but only for $D_{\lambda~\mathrm{Cep}} \sim 250$~pc
and a kinematic age of $\sim$9~Myr. These values do not agree with the
observed photometric (860~pc, Gies \cite{gie1987}) and trigonometric
distances ($505_{-95}^{+153}$~pc, Hipparcos) of $\lambda$~Cep, nor
with the age of the association (5--7~Myr, de Zeeuw et al.\
\cite{zee1999}). Moreover, $\lambda$~Cep is a massive O supergiant, and its 
lifetime cannot be more than a few million years. 

When we run the simulations using Cep~OB3 as the parent group we also
obtain minimum separations $<$10~pc.  Figure~\ref{fig:15} shows
that the expected distance of the runaway is now $\sim$450~pc and that
the kinematic age is $\sim$4.5~Myr. This is a little on the large side
for the nominal lifetime of a 40 $M_\odot$ star, but might not be
impossible. Cep~OB3 consists of two subgroups with ages of 5.5
(subgroup {\sl b}) and 7.5~Myr (subgroup {\sl a}) (e.g., Jordi,
Trullols \& Galad\'{\i}--Enr\'{\i}quez \cite{jtg1996}).  Considering
the high helium abundance and large rotational velocity of
$\lambda$~Cep, subgroup {\sl a} is a likelier parent of the runaway
than subgroup {\sl b}, since the age difference between the subgroup
and the runaway is 3~Myr for {\sl a}. For subgroup {\sl b} this is only 1~Myr,
leaving little time for binary evolution.  We conclude that
$\lambda$~Cep is likely to have become a runaway star as the result of
a supernova explosion in a binary system in subgroup {\sl a} of
Cep~OB3 $\sim$4.5~Myr ago.

\section{Identification of new parent groups}\label{sec:new-runaways}

The orbit retracing technique allows us to identify the (likely)
parent group for thirteen `new' single runaways, one new pair, two
more pulsars, and Geminga from the samples defined in
\S\ref{sec:samples}.  Little is known about most of these objects, so
our discussion is relatively brief. The results are summarized in
Table~\ref{tab:05}.

\subsection{HIP~3881 \& Lacerta~OB1 \hfill {\rm (star 1)}}
The orbit of the spectroscopic binary $\nu$~Andromedae (B5V+F8V)
intersected the Lac~OB1 association 8--10~Myr ago. If this is the
parent group, then the kinematic age is comparable to the (uncertain)
age of the association ($\sim$10~Myr).  This, together with a normal
rotational velocity for the primary (80~km~s$^{-1}$) and the binary
nature (two main-sequence stars) of this runaway, suggests that it was
formed by dynamical ejection.

\subsection{HIP~22061, HIP~29678, Geminga \& the $\lambda$~Ori region \null \hfill {\rm (stars 4 and 7 and neutron star 9)}} 
The two stars form an analog of the runaway pair AE~Aur and
$\mu$~Col (\S\ref{sec:orion}). They also move in opposite directions
and with similar space motions: 86.5~km~s$^{-1}$ for HIP~22061 (B2.5V)
and 63.0~km~s$^{-1}$ for HIP~29678\footnote{HIP~29678 is a visual
double star. Gies \& Bolton (\cite{gb1986}) did not detect any radial velocity
variations.} (B1V) (velocities are with respect to their own standard
of rest).  Retracing the orbits of HIP~22061 and HIP~29678, we find
that the two stars were close together $\sim$1.1~Myr ago, suggesting
that these two runaways were also formed by the DES. We did not find a
possible third party (i.e., another runaway in the neighbourhood or a
massive binary).

Applying the principle of conservation of linear momentum at the time
of the encounter, as we did in \S\ref{sec:orion}, we can predict the
properties of the parent cluster. We can only use these two stars and
not three as in \S\ref{sec:parent}. We find that the parent cluster
should be located around $(\ell,b) \sim (196\fdg5,-12\fdg0)$. This
coincides with the $\lambda$~Orionis star-forming region (e.g., Gomez
\& Lada \cite{gl1998}; Dolan \& Mathieu \cite{dm1999}), which contains
at least three young stellar clusters (the $\lambda$~Ori cluster and
the clusters associated with the dark clouds B30 and B35) and is
surrounded by the $\lambda$~Orionis ring. Several authors have
suggested that this expanding ring of molecular clouds is the result
of a supernova explosion $\sim$0.35~Myr ago (e.g., Cunha \& Smith
\cite{cs1996}). The predicted cluster position does not coincide with
one of the three star clusters. Furthermore, the predicted radial
velocity of the cluster, $\sim$10~km~s$^{-1}$, differs significantly
from that of the $\lambda$~Ori clusters, $\sim$24~km~s$^{-1}$.
However, these differences might be erased if a third body (either a
single star or a binary) was involved (cf.\
\S\ref{sec:orion})\footnote{Dolan \& Mathieu (\cite{dm1999}) report
that long term monitoring of $\lambda$~Ori's radial velocity indicates
that this star is a binary with a systematic velocity
($\mathv_\mathrm{rad} \sim 18$~km~s$^{-1}$) which differs by
6--8~km~s$^{-1}$ from that of the clusters in the $\lambda$~Ori
region. However, the simulations show that $\lambda$~Ori cannot be
related to the event that created the runaways HIP~22061 and
HIP~29678.}.

The conclusion that the DES is the acting mechanism for these runaways
is supported by (i) the youth of the clusters in the $\lambda$~Ori
star-forming region, 2--6~Myr (Dolan \& Mathieu \cite{dm1999}), (ii)
the density of these clusters (Dolan \& Mathieu), and (iii) the small
rotational velocity of HIP~29678, $<$25~km~s$^{-1}$ (Morse, Mathieu,
Levine \cite{mml1991}).

It is worth mentioning that HIP~22061 and HIP~29678 are not the only
objects running away from the $\lambda$~Orionis region. Frisch
(\cite{fri1993}) and Smith, Cunha \& Plez (\cite{scp1994}) suggested
that the neutron star Geminga also originated from this star-forming
region (Figure~\ref{fig:02}; but see Bignami \& Cara\-veo
\cite{bc1996}).  Moreover, the age of Geminga ($\sim$$350\,000$~yr)
agrees well with the time of the supernova explosion which created the
$\lambda$~Orionis ring.

\subsection{HIP~38455 \& Collinder 135 \hfill {\rm (star 8)}}
The ellipsoidal variable star HIP~38455 (B2V) moves away from the open
cluster Collinder 135 with a velocity of almost 40~km~s$^{-1}$, mostly
in the radial direction. The orbits of the
runaway and cluster intersected each other about 3~Myr ago; this is
significantly smaller than the age of the open cluster:
$\sim$35~Myr. This difference, and the large rotational velocity
($\mathv_\mathrm{rot} \sin i = 212$~km~s$^{-1}$), suggest that the
runaway originated via the BSS. 
Ellipsoidal variable stars show brightness variation because of their
non-spherical shapes. The deformation of the star is thought to be due
to tidal interactions with a companion star (Beech \cite{bee1985}).
The fact that HIP~38455 is double (cf.\ H\"afner \& Drechsel
\cite{hd1986}) suggests that the kick velocity of the compact object
created in the supernova explosion was small and that the binary
remained bound. This system might in the future become a high-mass
X-ray binary (\S\ref{sec:intro}).

\subsection{HIP~38518 \& Vela~OB2 \hfill {\rm (star 9)}} 
This B0.5Iab supergiant is located behind the Vel~OB2 association and
is currently moving away from it. Retracing the orbit shows that
$\sim$6~Myr ago the star was located inside the Vel~OB2
association. This kinematic age is smaller than the association age
(10--15~Myr). Adding to this the large rotational velocity
($\mathv_\mathrm{rot} \sin i = 220$~km~s$^{-1}$) {\it and} the blue
straggler character of HIP~38518 (see \S\ref{sec:blue} for a further
discussion), we infer that the BSS is the preferred scenario.

\subsection{HIP~42038 \& Upper Centaurus Lupus or IC 2391 \break \null \hfill {\rm (star 11)}}
The path of the B3V runaway star HIP~42038 traversed the open cluster
IC~2391 $\sim$6~Myr ago and the Upper Centaurus Lupus association
$\sim$8~Myr ago. Little is known about HIP~42038, making it difficult
to determine its origin. The only available criterion is that the
kinematic age is smaller than both the age of IC~2391 (45~Myr) and the
age of Upper Centaurus Lupus (13~Myr) favoring the BSS.
 
\subsection{HIP~46950 \& IC 2602 \hfill {\rm (star 12)}}
Somewhere between 2~Myr and 10~Myr ago the B1.5IV star occupied the
same space as the open cluster IC~2602. The only bit of information
available for the identification of the runaway's origin is the
difference between the kinematic age and the age of the open cluster
($\sim$25~Myr). This would suggest that the BSS is the origin.

\subsection{HIP~48943 \& Lower Centaurus Crux \hfill {\rm (star 13)}} 
HIP~48943 is a runaway from the Lower Centaurus Crux association. The
orbit of the B5Ve star intersected the association center $\sim$4~Myr
ago. The runaway has a rotational velocity of 230~km~s$^{-1}$ and its
kinematic age is significantly smaller than the age of Lower Centaurus
Crux, $\sim$10~Myr. The BSS is thus the most likely formation
mechanism.

\subsection{HIP~49934 \& IC 2391 or IC 2602 \hfill {\rm (star 14)}}
The retraced orbit of the emission-line star HIP~49934 (B2IVnpe)
intersects two open clusters: IC~2391 $\sim$3~Myr ago and IC~2602
$\sim$6~Myr ago. The large difference between the nuclear ages of both
clusters (45~Myr and 25~Myr, respectively) and the kinematic age
indicates that the BSS is the most likely origin of HIP~49934. This
assumption is supported by the large rotational velocity
($\mathv_\mathrm{rot} \sin i = 280$~km~s$^{-1}$).

\subsection{HIP~57669 \& IC 2602 \hfill {\rm (star 15)}}
About 3~Myr ago the emission-line star HIP~57669 left the open cluster
IC~2602. This B3Ve star has a large rotational velocity
($\mathv_\mathrm{rot} \sin i = 251$~km~s$^{-1}$) and its kinematic age
differs significantly from the age of IC~2602. Both these points
suggest the BSS as the origin of HIP~57669.

\subsection{HIP~69491 \& Upper Centaurus Lupus or Cepheus~OB6 \hfill {\rm (star 16)}}
This candidate runaway is an eclipsing binary (B5V). Its path
traverses the Upper Centaurus Lupus association and the Cep~OB6
association. The respective kinematic ages are 3 and 10~Myr. Both these
kinematic ages are smaller than the association ages of 13~Myr for Upper
Centaurus Lupus and $\sim$50~Myr for Cep~OB6. This discrepancy in ages
combined with the binary nature of the candidate runaway excludes both
the standard BSS and the DES and might suggest that neither Upper
Centaurus Lupus nor Cep~OB6 is the parent of HIP~69491. The BSS is
highly unlikely because the eclipses imply two objects of similar size
and not, for example, a main-sequence star and a compact object. The
DES is excluded because both associations were no longer compact/dense
enough for dynamical encounters to be efficient at the time of
ejection.

What other mechanisms do exist to create a fast moving
($\mathv_\mathrm{space} = 77$~km~s$^{-1}$) binary system? One
possibility is a supernova explosion in a triple system consisting of
a hard binary and a third star with a larger semi-major axis (i.e., a
stable triple system). This would result in either (i) a hard binary
moving at moderate speed ($<$30~km~s$^{-1}$) or (ii) a fast runaway
and a normal field star. In the latter case one of the stars in the
binary explodes and creates a fast runaway. The third star, being
weakly bound to the system would hardly be affected by the
explosion. In the former case the single star explodes causing the
binary to start moving at the orbital speed it had within the triple
system. This velocity should be small since the binary is much more
massive than the third star. However, neither case would create a
runaway binary-system like HIP~69491. Whereas it is likely that the
star originated in either Upper Centaurus Lupus or Cep~OB6, the
mechanism that formed this runaway remains unknown.

\subsection{HIP~76013 \& Lower Centaurus Crux \hfill {\rm (star 17)}}
$\kappa^1$~Apodis is a B1npe emission-line star, and is the brightest
component of a visual double system. This star has a radial velocity
of 62~km~s$^{-1}$ and is moving away from the Galactic plane. Its
orbit intersects the Lower Centaurus Crux subgroup of Sco~OB2 2--3~Myr
ago. Because of the large difference in kinematic age and association
age ($\sim$10~Myr) the BSS is the most likely explanation for the
runaway nature of $\kappa^1$~Aps.

\subsection{HIP~82868 \& IC 2602 \hfill {\rm (star 19)}}
Little is known about the B3Vnpe star HIP~82868 whose orbit intersects
that of the IC~2602 open cluster some 6~Myr ago.  A firm
identification of the formation mechanism is difficult, since we only
know that the kinematic age differs from the age of IC~2602
(25~Myr). This suggests that HIP~82686 is a BSS runaway.

\subsection{HIP~91599 \& Perseus~OB2 or Perseus~OB3 \break \null \hfill {\rm (star 20)}} 
HIP~91599 is a known runaway star (B0.5V; Vitrichenko, Gershberg \&
Metik \cite{vgm1965}); however, its parent association/cluster has
never been identified. The simulations show that HIP~91599 originates
from Per~OB2 or Per~OB3. The two associations have very different ages
($\sim$7~Myr for Per~OB2 and $\sim$50~Myr for Per~OB3). The kinematic
age of HIP~91599 is $\sim$8~Myr and $\sim$6~Myr for Per~OB2 and
Per~OB3, respectively. Since we lack information on the rotational
velocity and the helium abundance, we are unable to conclude whether
HIP~91599 is a DES runaway from Per~OB2 or a BSS runaway from Per~OB3.

\subsection{HIP~102274 \& Cepheus~OB2 \hfill {\rm (star 21)}} 
The B5 star HIP~102274 was located at the center of the Cep~OB2
association between two and three Myr ago, when the association was
3--4 Myr old. This kinematic age coincides with the time of the
supernova explosion proposed by Kun, Bal\'azs \& T\'oth
(\cite{kbt1987}) to explain the characteristics of the Cepheus bubble,
a ring-like structure of infrared emission.  Taken together, this is
strong circumstantial evidence for HIP~102274 being a BSS runaway.

\subsection{PSR~J0826$+$2637, PSR~J1115$+$5030 \& Perseus~OB3 \hfill {\rm (pulsars 1 and 4)}}
The orbits of the pulsars J0826$+$2637 and J1115$+$5030 intersect that
of the Per~OB3 association. This group contains the $\alpha$~Persei
open cluster, and is often referred to as the Cassiopeia--Taurus
association (de Zeeuw et al.\ \cite{zee1999}).  The simulations show
that if Per~OB3 is the parent of PSR~J0826$+$2637, its kinematic age
is $\sim$1~Myr and its radial velocity is $\sim$100~km~s$^{-1}$. For
PSR~J1115$+$5030 we predict a kinematic age of $\sim$1.5~Myr and a
radial velocity of $\sim$150~km~s$^{-1}$. The characteristic ages
($P/(2\dot{P})$) of the pulsars are 4.9~Myr and 10~Myr for
PSR~J0826$+$2637 and PSR~J1115$+$5030, respectively. The unknown
radial velocity of these pulsars makes it difficult to prove beyond
doubt that these pulsars were born in the Per~OB3 association.
Although Figure~\ref{fig:02} shows that the orbits projected on the
sky do not differ much for different radial velocities, and that they
cross Per~OB3, they may also cross the paths of other, more distant,
associations or clusters not shown in the Figure.

If the two pulsars orginated in Per~OB3, then the initial periods
would be 0.47 s for J0826$+$2637 and 1.53 s for J1115$+$5030, assuming
no glitches occurred. The latter value is large, which might indicate
that this pulsar traveled longer, from another site of origin.

It is not unlikely to find many pulsars associated with Per~OB3 since
its age, $\sim$50~Myr, is comparable to the main-sequence life-time of
an 8~$M_\odot$ star. These are the least massive stars to explode as a
supernova. Since the moment at which a star explodes,
$\tau_\mathrm{SN}$, depends on its mass, ($\tau_\mathrm{SN} \sim
M^{-\alpha}$, where $\alpha > 0$ and $M > 8~M_\odot$) and the number
of stars of mass $M$, $N(M)$, also depends on the mass ($N(M) \sim
M^{-\beta}$, where $\beta > 0$), the number of supernovae increases
with time ($N_\mathrm{SM} \sim \tau^{\beta/\alpha}$, for $M >
8~M_\odot$). The number of supernovae, and thus the number of pulsars,
will thus increase with time until the stars of $8~M_\odot$ have
exploded as supernovae. Afterwards the pulsar production rate will
drop to almost zero.

\subsection{The Vela pulsar \hfill {\rm (pulsar 2 )}}

PSR~J0835$-$4510 is only $10\,000$~yr old, and therefore has not
traveled far from its birth place ($\sim$$9'$ on the sky), the Vela
star-forming region at $\sim$450~pc. It lies within the boundaries of
the $\sim$10 Myr old Vel~OB2 association (de Zeeuw et al.\
\cite{zee1999}), which is the likely parent group.

\FigSixteen

\section{Runaways and pulsars without parents?}\label{sec:noparent}
The sample of runaways and pulsars we have analysed here is severely
incomplete (see \S\ref{sec:selection}).  Furthermore, we have only a
limited knowledge of where massive stars form in the Solar
neighbourhood. The case of $\zeta$~Pup described in
\S\ref{sec:class-runaways} shows that some local parent groups may not
yet have been found. Beyond $\sim$500~pc, only the large complexes
such as Cygnus~OB1/OB2 or Sco~OB1 are well documented. The accuracy of
the three-dimensional positions and velocities of these distant
star-forming regions is poor, making determination of reliable orbits
very difficult.  Runaway stars of spectral types B0 and later are
especially difficult to link to a parent group. These stars have
main-sequence lifetimes of up to several tens of Myr and can therefore
travel far (several kpc) from their places of origin. A similar
argument holds for the pulsars. We suspect that most of the runaway
stars and pulsars in our sample for which we were not able to identify
a parent group originated outside the Solar neigbourhood.

An example is the B2.5V star 72~Col, HIP~28756 (van Albada
\cite{alb1961}; the asterisk at $[\ell,b] \sim [238^\circ,-23^\circ]$
in Figure~\ref{fig:02}). It has a peculiar velocity of
$\sim$200~km~s$^{-1}$ and its parent association is Sco~OB1, at a
distance of $\sim$2~kpc (Humphreys \cite{hum1978}). Van Albada derived
a kinematic age of $\sim$14~Myr for 72~Col, based on a simple model of
Galactic rotation (Kwee, Muller \& Westerhout \cite{kmw1954}).  This
star does not appear in \S\S\ref{sec:class-runaways}
and~\ref{sec:new-runaways} because its path did not carry it through
one of the nearby associations. The Solar neighbourhood thus not only
contains runaways for which the parent associations are also nearby,
but it also contains runaways which originated far from the Sun.

Another star that immediately catches the eye in Figure~\ref{fig:02}
is HIP~94899 (the asterisk at $[\ell,b] \sim [341^\circ,-26^\circ]$).
This double star of spectral type B3Vn has a radial velocity of
151~km~s$^{-1}$, and its path seems to cross the Per~OB3
association. However, our simulations show that the runaway never
comes within 40~pc of the association, implying that this system must
have another, unknown, parent.

We have seen in \S\ref{sec:SNinUS} that PSR~J1239$+$2453 most likely
originates outside the Solar neighbourhood.  We did not find a parent
group for the remaining pulsars because (i) an unreasonably large
radial velocity ($>$500~km~s$^{-1}$) is necessary for the paths of the
pulsar and parent group to intersect (PSR~J1135$+$1551), just as we
found for PSR~J1239$+$2453, or (ii) the past orbit simply does not
intersect any of the nearby young stellar groups (PSR~J0953$+$0755 and
PSR~J1456$-$6845).

\section{Helium abundance versus rotational velocity}\label{sec:helium}
Blaauw (\cite{bla1993}) pointed out that most of the reliable runaways
known at the time have high helium abundances ($\epsilon \ga 0.13$)
and large rotational velocities ($\mathv_\mathrm{rot} \sin i \ga
150$~km~s$^{-1}$), whereas most of the non-runaways have normal helium
abundances and rotational velocities. Only one classical runaway star,
AE~Aur, does not follow this trend, and this was already suspected to
be the result of the DES rather than the BSS. Blaauw suggested that
the $(\epsilon,\mathv_\mathrm{rot} \sin i)$-diagram of the small
sample of known runaway stars supported the conjecture that most
runaways are produced by the BSS, since these characteristics are
natural consequences of close binary evolution (\S\ref{sec:intro}; van
den Heuvel \cite{heu1985}).

As shown in Table~\ref{tab:03}, only five of the 23 runaways listed
there have a measurement of $\epsilon$, and six do not have a measured
rotational velocity. In order to pursue Blaauw's suggestion, we
therefore constructed a sample of O stars with known rotational
velocities (Penny \cite{pen1996}) and helium abundances (Kudritzki \&
Hummer \cite{kh1990}; Herrero et al.\ \cite{her1992}). We also
determined, based on Hipparcos astrometry and Hipparcos Input
Catalogue radial velocities, the space velocities of these stars with
respect to their local standard of rest. The
$(\epsilon,\mathv_\mathrm{rot} \sin i)$ diagram in the left panel of
Figure~\ref{fig:16} shows that these O stars can roughly be divided
into three groups:
(i) those with small rotational velocities, $\mathv_\mathrm{rot} \sin
i \la 80$~km~s$^{-1}$, and normal helium abundances, $\epsilon \sim
0.09$,
(ii) those with moderate rotational velocities, $80 \la
\mathv_\mathrm{rot} \sin i \la 150$~km~s$^{-1}$, and normal to high
helium abundances, $0.09 \la \epsilon \la 0.2$, and
(iii) those with large rotational velocities, $\mathv_\mathrm{rot}
\sin i \ga 150$~km~s$^{-1}$, and high helium abundances, $\epsilon \ga
0.14$.
The symbols in the left panel of Figure~\ref{fig:16} are chosen
according to the magnitude of the space velocity. The stars
represented by filled circles have space velocities
$\mathv_\mathrm{space} \ge 30$~km~s$^{-1}$, and the open circles have
$\mathv_\mathrm{space} < 30$~km~s$^{-1}$; the other symbols indicate
stars for which no Hipparcos data are available (asterisks) or for
stars with insignificant Hipparcos data (starred).

\FigSeventeen
The left panel of Figure~\ref{fig:16} does not show an obvious
separation between runaways (filled circles) and non-runaways (open
circles). However, based on the available data for each star
(position, radial velocity, distance modulus, cluster membership) it
is possible to decide whether the star is a runaway. The result is
shown in the right panel of Figure~\ref{fig:16}; it turns out that
many of the stars with insignificant Hipparcos data are runaways
(including some classical runaways), and that some of the stars
indicated as high-velocity stars in the left panel are members of a
cluster which has a peculiar motion. The resulting
$(\epsilon,\mathv_\mathrm{rot} \sin i)$ diagram now shows a clear
separation between the runaway stars and the normal O stars. Except
for AE~Aur (\S\ref{sec:orion}) all runaways have high helium
abundances, $\epsilon \ga 0.12$, and large rotational velocities,
$\mathv_\mathrm{rot} \sin i \ga 150$~km~s$^{-1}$. Only one star in
this area of the $(\epsilon,\mathv_\mathrm{rot} \sin i)$ diagram is
not indicated as a runaway: the double star HIP~113306 (OV7n). This
star is a member of Cep~OB3 ($\epsilon = 0.17$, $\mathv_\mathrm{rot}
\sin i \sim 359$~km~s$^{-1}$).

We thus confirm Blaauw's conclusion that massive runaways
predominantly have high helium abundances and large rotational
velocities, suggesting that they are formed mainly by the
binary-supernova scenario. However, this conclusion is based on a
limited sample which is by no means statistically complete. A
systematic survey of the radial velocities, rotational velocities and
chemical abundances of the early-type stars in the Solar neighbourhood
is highly desirable.

\section{Blue stragglers}\label{sec:blue}
One of the characteristics of runaway stars produced according to the
binary-supernova scenario is that these stars are expected to be blue
stragglers (\S\ref{sec:intro}). The mass transfer in close binary
systems from the primary to the secondary prior to the supernova
explosion deposits a large amount of hydrogen onto the future
runaway. This new supply of fuel makes the runaway appear younger than
the association/cluster in which it was born, i.e., it appears
rejuvenated.

Figure~\ref{fig:17} shows the colour vs.\ absolute magnitude
diagrams of the parent clusters discussed in this paper. The
association members (dots) have been de-reddened following the
Q-method; only the early-type members (A0 and earlier) are shown. The
solid lines denote the Schaller et al.\ (\cite{sch1992}) isochrones
for Solar metallicity and a standard mass loss rate for the ages of
the associations. The runaway stars are denoted by starred symbols.  A
runaway can appear in more than one panel if two or more possible
parents have been identified. The $B-V$ colour and absolute magnitude
have been determined using the spectral type of the runaways
(Table~\ref{tab:03}; Schmidt--Kaler \cite{sk1982}). Three stars in
Figure~\ref{fig:17} clearly are blue stragglers: HIP 38518,
$\xi$~Per, and $\lambda$~Cep; and three others could be blue
stragglers depending on the correct identification of the parent:
$\zeta$~Pup, HIP~49934, and HIP~91599. The latter three stars have
uncertain parent identifications (see \S\ref{sec:class-runaways} and
\S\ref{sec:new-runaways}). The blue straggler nature of the former
three stars confirms their identification as BSS runaway (see
Table~\ref{tab:05}). The star $\zeta$~Oph has also been claimed to
be a blue straggler (e.g., Blaauw \cite{bla1993}). However, the Upper
Scorpius panel shows that $\zeta$~Oph is the bluest star of the group,
but it lies on the main sequence, as also found by de Geus et al.\
(\cite{gzl1989}) on the basis of $uvby\beta$ photometry.

By contrast to the BSS runaways, those produced by the DES are
expected to follow the main sequence of the parent group. These
runaways most likely did not experience a period of binary evolution
in which mass transfer was important. The runaway stars which we
identified securely as DES runaways (AE~Aur, $\mu$~Col, HIP~22061, and
HIP~29678) indeed fall on the main sequence of their parents (the
Trapezium and the $\lambda$~Ori cluster).

\section{Concluding remarks}\label{sec:end}
We have used the Hipparcos astrometry for a sample of nearby candidate
OB runaway stars to locate their parent groups, and to identify their
formation mechanisms. We retraced the orbits of these candidate
runaways, and determined where and when they passed through a possible
parent group. We find that both mechanisms proposed for the production
of runaway stars, the binary-supernova scenario (BSS) and the
dynamical ejection scenario (DES), operate. Table~\ref{tab:05}
summarizes the results.  Even though the number of runaways discussed
in this paper is small, and the weight of evidence varies, we find
that roughly $2/3$ of the runaways is produced by the BSS and $1/3$ by
the DES. This agrees with the results of binary population synthesis
calculations by Portegies Zwart (\cite{por2000}). At present it is not
feasible to extend this study to other runaways. This is mainly due to
large uncertainties in the velocities and distances of the runaways,
and the limited knowledge of star-forming regions with distances
$>$500~pc.

Tracing the runaway orbits back in time provides, for the first time,
direct evidence that both scenarios produce single runaway stars
(Hoogerwerf et al.\ \cite{hbz2000}). The orbit calculations
demonstrate that the runaway $\zeta$~Oph and the progenitor of
PSR~J1932$+$1059 once formed a binary system in the Upper Scorpius
association, and that the neutron star acquired a kick velocity of
$\sim$350~km~s$^{-1}$ in the supernova explosion.  The runaways AE~Aur
and $\mu$~Col, and the binary $\iota$~Ori were involved in a dynamical
interaction (a binary-binary collision) $\sim$2.5~Myr ago, which took
place in the Trapezium cluster.

\TabFive

The current investigation is biased towards finding BSS runaways. This
is mainly due to the fact that the accuracy of the available data, and
our knowledge of the location and motions of star-forming regions,
restrict the study to $\sim$700~pc. The small volume implies that we
are only able to identify runaway stars with small kinematic ages of
0--10~Myr (i.e., runaways which recently left their parent
association). Runaways which were created at an earlier time have most
likely traveled outside our sample limits. Since associations and open
clusters can create BSS runaways during $\sim$50~Myr (approximately
the lifetime of a $8~M_\odot$ star) and DES runaways only in the
inital stages when the group still has a high density, we expect to find
more BSS than DES runaways because there are relatively many more old
parents than young parents in the Solar neighbourhood. This bias is
somewhat weakened by the fact that most dynamical interactions produce
two runaway stars while the binary-supernova mechanism produces only
one.

The creation of runaway stars modifies the mass function of the parent
group at the high-mass end, where the total number of stars is
small. For example, the encounter in Orion described in
\S\ref{sec:orion} removed four stars with a total mass of order 70
$M_\odot$ from the Trapezium cluster, while only six stars more
massive than 10~$M_\odot$ remain.  Derivation of the initial mass
function of young stellar groups from the present-day mass function
without accounting for the associated runaway stars leads to
erroneous results.

Our Hipparcos-based study has identified 56 runaway stars within 700
pc from the Sun, and tripled the subset of these for which a parent
group is known (from 6 to 21).  As mentioned in \S\ref{sec:samples},
less than a third of the O--B5 stars in the Hipparcos Catalog have a
measured radial velocity. Obtaining these is likely to result in
another factor of three increase in the size of the sample, so that
statistical studies become possible.

The next major step in our understanding of the origin of runaway
stars will come when large datasets of micro-arcsecond ($\mu$as)
astrometry and accurate radial velocities (1--2~km~s$^{-1}$) become
available.  Distances accurate to a few parsec will allow for a final
confirmation or rejection of the genetic link between runaways and
their parents (e.g., Figure~\ref{fig:12}).  These data will become
available over the next two decades with the launches of several
astrometric satellites (FAME, SIM, GAIA). These aim to obtain $\mu$as
astrometry for a large number of stars, from $10\,000$ stars with SIM
to 1 billion stars with GAIA. Besides astrometry, accurate radial
velocities are also required; unfortunately, there is no dedicated
effort to obtain these for a large number of O and B stars. 

The BSS and DES can produce runaway stars with spectral types beyond
B5 (e.g., Kroupa \cite{kro2000b}; Portegies Zwart \cite{por2000}).
These will be harder to find, as the velocity distribution of the
later-type stars in the Galactic disk is broader than for the O--B5
stars, and the fractional production of low-mass runaways is
small. Identifying their parent groups is also harder, because these
stars may have traveled for much longer times. However, $\mu$as
accuracy astrometry complemented with accurate radial velocities will
undoubtedly reveal such objects, and will provide further constraints
on the binary fraction and the binary mass-ratios in open clusters and
associations.

Figure~\ref{fig:01}{\it b} shows that there are 19 additional
pulsars within one kpc for which an accurate proper motion is not
available. A systematic program to measure these might allow the
detection of more examples of pairs such as $\zeta$~Oph and
PSR~J1932$+$1059. It would also improve the characterisation of the
pulsar population as a whole. VLBI techniques hold the promise of
achieving sub-mas astrometry (positions, proper motions, and
parallaxes) in the near future.

%
%
\begin{acknowledgements}
It is a pleasure to thank Bob Campbell for a discussion on VLBI proper
motions of pulsars, Rob den Hollander for writing an early version of
the software used here, Nicolas Cretton for providing the Galactic
potential used in the orbit integrations, and Ed van den Heuvel, Lex
Kaper, Michael Perryman, the referee Walter van Rensbergen, and in
particular Adriaan Blaauw, for stimulating comments and suggestions.
This research was supported by the Netherlands Foundation for Research
in Astronomy (NFRA) with financial aid from the Netherlands
Organization for Scientific Research (NWO).
\end{acknowledgements}
%
%

\appendix

\section{Distribution of differences}\label{ap:a}
\FigApp
%
In this appendix we record the distribution of the difference between
two observables, each of which has a Gaussian error distribution. We
define two observable quantities, $x_1$ and $x_2$, which are both
distributed normally, $G_1(x_1;\mu_1,\sigma_1)$ and
$G_2(x_2;\mu_2,\sigma_2)$, respectively, where $\mu_1$ and $\sigma_1$
are the mean and standard deviation of $G_1$, and $\mu_2$, $\sigma_2$
for $G_2$. We define $\Delta = |\hat{x}_1 - \hat{x}_2|$ as the
absolute difference between the observables, where $\hat{x}_1$ and
$\hat{x}_2$ are the measurements of $x_1$ and $x_2$. We calculate the
distribution of the difference, $F(\Delta)$, in the one-
($F_\mathrm{1D}(\Delta){\rm d}\Delta$), two-
($F_\mathrm{2D}(\Delta){\rm d}\Delta$), and three-dimensional
($F_\mathrm{3D}(\Delta){\rm d}\Delta$) case. In all cases we took
$\mu_1 = 0$ and $\mu_2 = \mu$ and $\sigma_1 = \sigma_2 = \sigma$. The
former simplification is harmless since only the difference $\mu_1 -
\mu_2$ is important. The results are:

\begin{eqnarray}
F_\mathrm{1D}(\Delta) &=& \!\frac{1}{\sqrt{4 \pi \sigma^2}} 
    \left\{ \exp \left [ -\frac{(\Delta+\mu)^2}{4 \sigma^2} \right ]  
           \right.                                \nonumber \\
   &\phantom{=}& \qquad\quad  \left. 
   +  \exp \left [ -\frac{(\Delta-\mu)^2}{4 \sigma^2} \right ] \right \}
\label{eq:ap4}
\end{eqnarray}
\begin{equation}
F_\mathrm{2D}(\Delta) = \frac{\Delta}{2 \sigma^2} 
  \exp \left [ -\frac{1}{2} \frac{\Delta^2 + \mu^2}{2\sigma^2} \right ]
  I_0 \left (\frac{\Delta\mu}{2\sigma^2} \right ),
\label{eq:ap8}
\end{equation}
where $I_0$ is a modified Bessel function of order zero, and
\begin{eqnarray}
F_\mathrm{3D}(\Delta) &=& 
  \frac{\Delta}{2 \sqrt{\pi} \sigma \mu}
  \left \{
    \exp \left [ 
    -\frac{1}{2}\frac{(\Delta-\mu)^2}{2\sigma^2}
    \right ] \right.  \nonumber \\
  &\phantom{=}& \qquad\quad   - \left. 
    \exp \left [ 
    -\frac{1}{2}\frac{(\Delta+\mu)^2}{2\sigma^2}
    \right ]
  \right \}.
\label{eq:ap12}
\end{eqnarray}
Applying L'Hospital's rule we calculate the limit of $F_\mathrm{3D}$
for $\mu \rightarrow 0$:
\begin{equation}
\lim\limits_{\mu \rightarrow 0} F_\mathrm{3D}(\Delta){\rm d}\Delta =
  \frac{\Delta^2}{2 \sqrt{\pi} \sigma^3} 
  \exp \left [
  - \frac{\Delta^2}{4\sigma^2}
  \right ].
\end{equation}
Figure~A1 and eqs (\ref{eq:ap8}) and (\ref{eq:ap12}) show
that for the two- and three-dimensional cases there is a zero
probability of measuring the same value for $x_1$ and $x_2$, i.e.,
$\Delta = 0$.

\section{The neutron star RX~J185635$-$3754}\label{ap:b}
\FigAppB

After this paper was submitted, Walter (\cite{wal2000}) suggested that
perhaps the past trajectory of the isolated neutron star
RX~J185635$-$3754 intersected that of $\zeta$~Oph. He used HST WFPC2
observations over a three-year baseline to obtain a very accurate
proper motion and parallax, and approximated the past orbit by a
straight line. He concluded that the neutron star came very near
$\zeta$~Oph 1.15 Myr ago in Upper Scorpius for an assumed radial
velocity of $-$45 km~s$^{-1}$. 

We have used the simulation machinery described in \S\ref{sec:SNinUS}
to analyse this case in the same way as done there for $\zeta$~Oph and
PSR J1932$+$1059. We have again run three million simulations,
covering the range $-50\pm50$ km~s$^{-1}$ for the radial velocity of
the neutron star. Figure B1 presents the resulting distribution of
minimum separations $D_{\rm min}$ and associated kinematic ages
$\tau_0$. This figure can be compared directly with
Figure~\ref{fig:03}. We have seen in \S\ref{sec:SNinUS} that 30822
simulations put PSR J1932$+$1059 within 10 pc of $\zeta$~Oph, and in
4214 of these the encounter took place in Upper Scorpius. By contrast,
only 748 simulations put RX~J185635$-$3754 within 10 pc of $\zeta$~Oph,
about 1.5 Myr ago. None of these encounters occur within 15~pc of
Upper Scorpius. We conclude that it is unlikely that RX~J185635$-$3754
is the remnant of the supernova that gave $\zeta$~Oph its large space
velocity. We suspect that this neutron star formed long ago somewhere
else in the Galactic plane.

\end{document}

%% file: hoogerwerf_figures.tex
\def\FigurePath{./}

\def\FigOne{
\begin{figure*}[!t]
  \begin{center}
  \includegraphics[angle=0.0, width=12.0cm, 
                   clip=true, keepaspectratio=true]
                   {\FigurePath 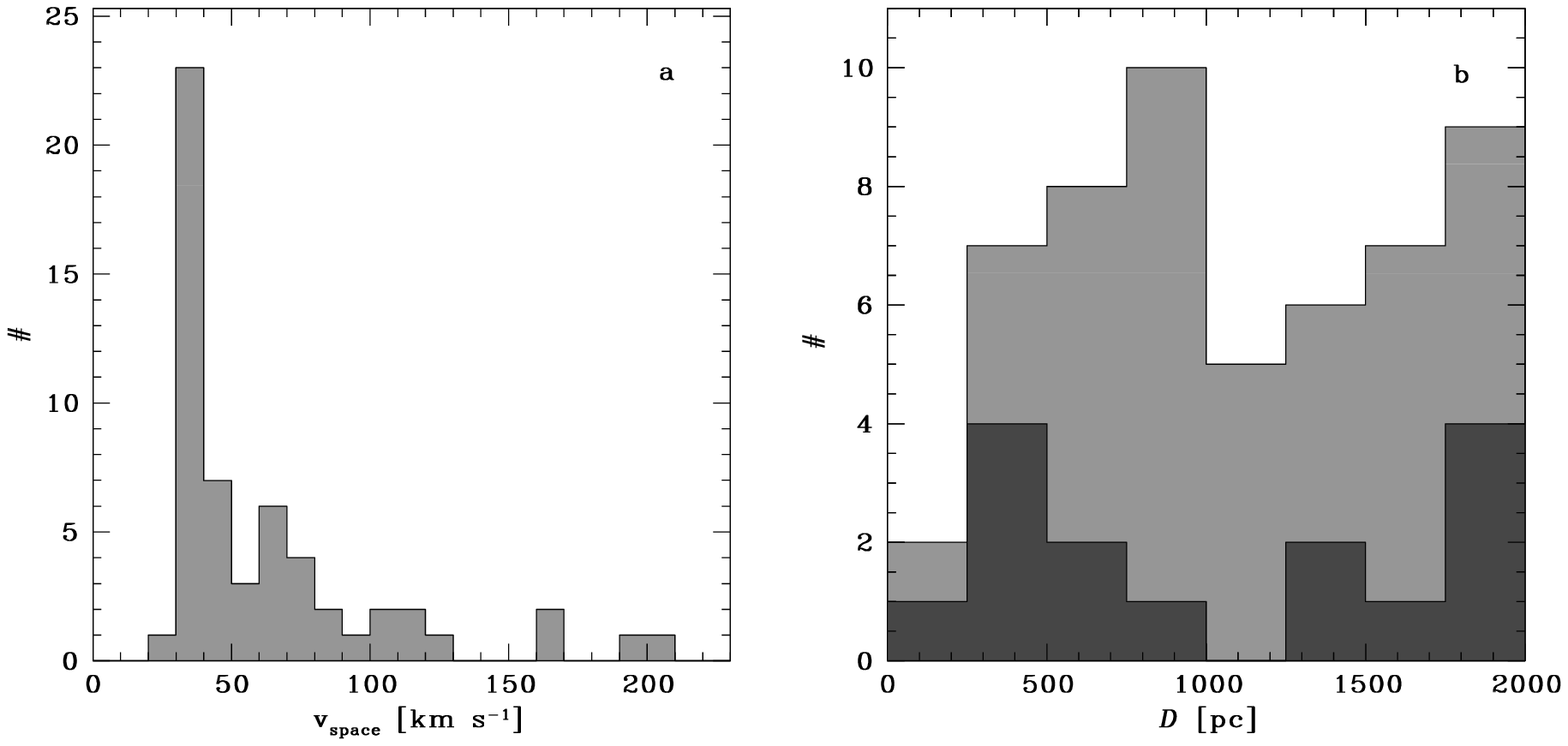}
  \end{center}
  \caption[]{{\it Panel a:} Histogram of the space motions of the
  sample of runaway stars defined in \S\ref{sec:selection}. {\it Panel
  b:} Distribution of pulsars from the Taylor et al.\ (\cite{tml1993})
  catalogue with measured proper motions. The light grey histogram
  shows all pulsars within 2~kpc and the dark grey histogram shows the
  pulsar with accurate proper motions ($\sigma_\mu/\mu < 0.1$). The
  latter, for $D < 1$~kpc, is the pulsar sample defined in
  \S\ref{sec:selection}.}
\label{fig:01}
\end{figure*}}
\def\FigTwo{
\begin{figure*}[!t]
  \begin{center}
  \includegraphics[angle=0.0, width=14cm, 
                   clip=true, keepaspectratio=true]
                   {\FigurePath 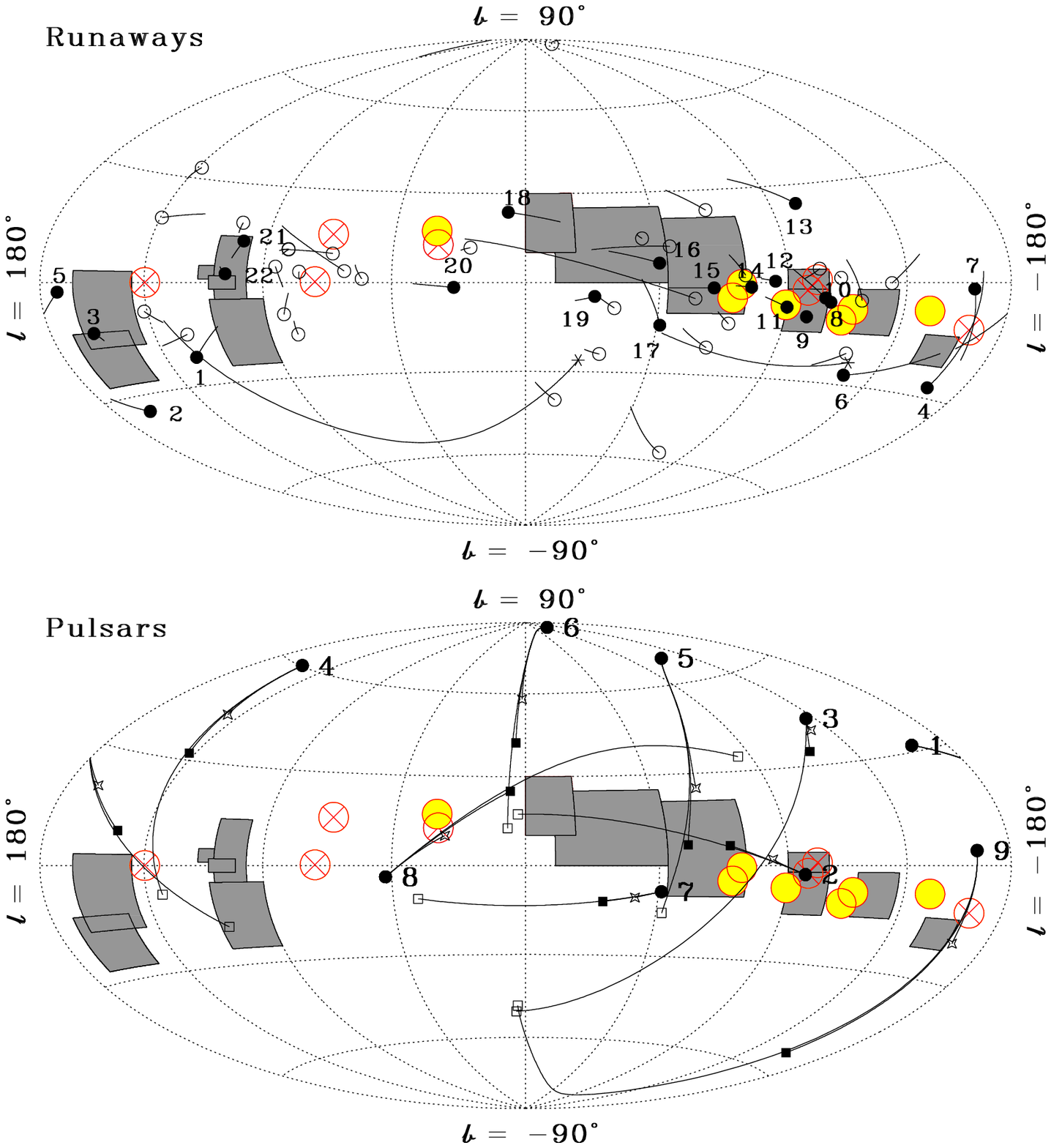}
  \vskip -5mm
  \end{center}
  \caption[]{{\it Top:} Sample of runaway stars defined in
  \S\ref{sec:selection}, in Galactic cordinates. The open circles
  denote the present positions of the runaways, and the arcs show
  their past orbits, calculated for 2~Myr. The filled circles are the
  runaways for which we can identify the parent association. The
  numbers refer to the entries in Table~\ref{tab:03}. The asterisks
  indicate two additional runaways (72~Col, HIP~94899 [left most of
  the two asterisks]) discussed in \S\ref{sec:noparent}. The grey
  fields outline the nearby OB associations (de Zeeuw et al.\
  1999). From left to right and from top to bottom: Per~OB3
  ($\alpha$~Persei), Per~OB2, Cep~OB3, Cep~OB2, Cep~OB6, Lac~OB1,
  Upper Scorpius, Upper Centaurus Lupus, Lower Centaurus Crux, Tr~10,
  Vel~OB2, Col~121, and Ori~OB1. The open clusters are identified by
  the filled light-grey circles for those with reliable positions and
  velocities, and by the open, crossed circles for the remaining
  clusters. The positions and designations of the clusters can be
  found in Table~\ref{tab:02}.  {\it Bottom:} Pulsar sample defined in
  \S\ref{sec:selection}, in Galactic coordinates. The filled circles
  indicate the present positions of the pulsars. The past orbits of
  pulsars, calculated for 2~Myr, are shown for three different assumed
  radial velocities: $0$~km~s$^{-1}$ (filled squares),
  $200$~km~s$^{-1}$ (open squares), $-200$~km~s$^{-1}$ (open stars).
  The pulsars are labeled 1 through 8; 1: J0826$+$2637, 2:
  J0835$-$4510 (Vela pulsar), 3: J0953$+$0755, 4: J1115$+$5030, 5:
  J1136$+$1551, 6: J1239$+$2453, 7: J1456$-$6843, 8:
  J1932$+$1059. Number 9 is the neutron star Geminga. The associations
  and open clusters typically move comparatively little in 2~Myr.}
\label{fig:02}
\end{figure*}}
\def\FigThree{
\begin{figure}[t]
  \includegraphics[angle=0.0, width=8.5cm, 
                   clip=true, keepaspectratio=true]
                   {\FigurePath 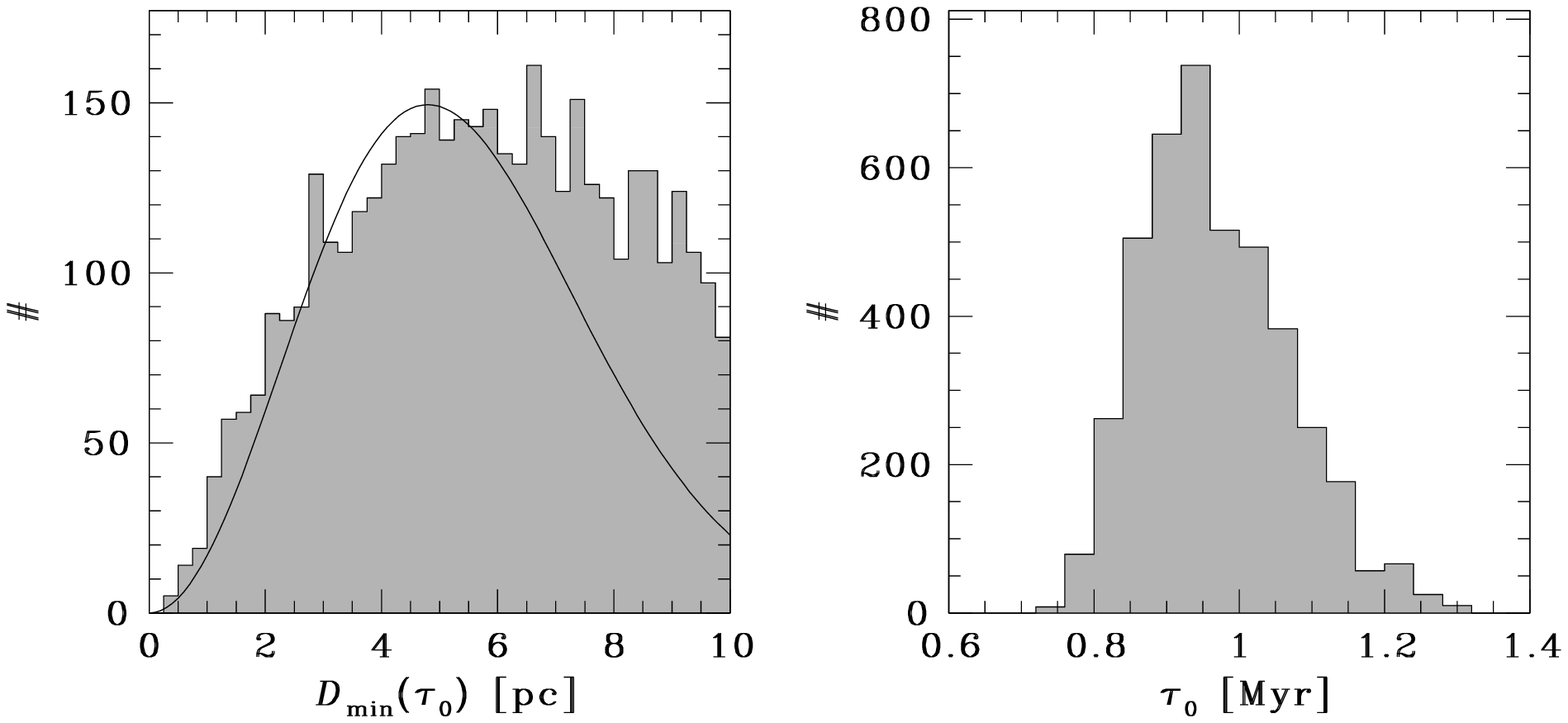}
  \caption[]{{\it Left:} Distribution of minimum separations,
  $D_\mathrm{min}(\tau_0)$, between $\zeta$~Oph and
  PSR~J1932+1059. The solid line denotes the expected distribution of
  $D_\mathrm{min}$, see \S\ref{sec:us_sim}. {\it Right:}
  Distribution of the times $\tau_0$ at which the minimum separation 
  was reached.}
\label{fig:03}
\end{figure}}
\def\FigFour{
\begin{figure*}[t]
  \centerline{ 
  \includegraphics[angle=0.0, width=13.5cm, 
                   clip=true, keepaspectratio=true] 
                   {\FigurePath 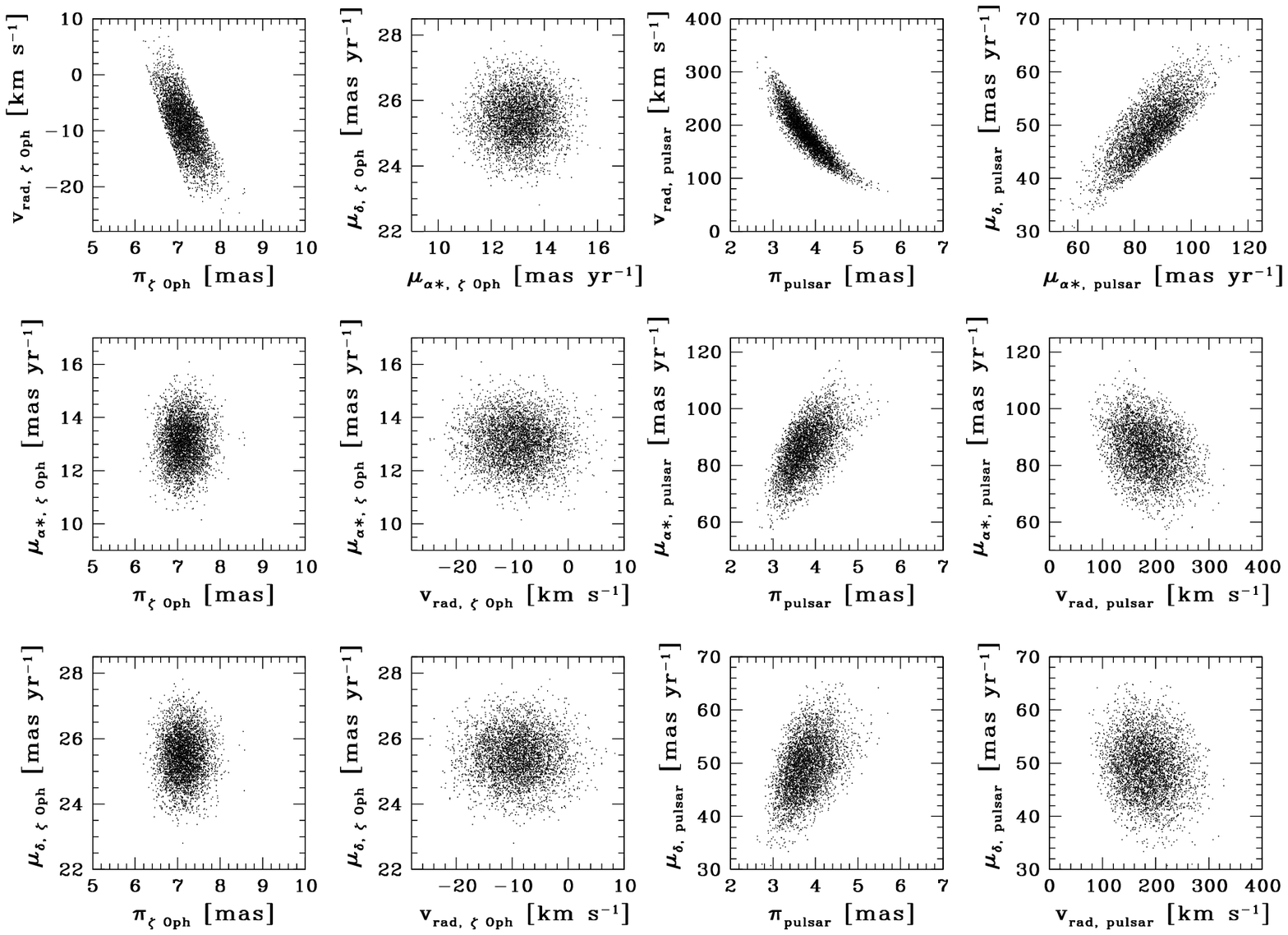}}
  \caption[]{Astrometric parameters of the runaway $\zeta$~Oph and the
  pulsar PSR~J1932$+$1059 at the start of each of the $4\,214$
  simulations for which the minimum separation between the runaway and
  the pulsar was less than 10~pc, and both of them were within 10~pc of
  the center of Upper Scorpius sometime in the past.}
\label{fig:04}
\end{figure*}}
\def\FigFive{
\begin{figure}[t]
  \includegraphics[angle=0.0, width=8.5cm, 
                   clip=true, keepaspectratio=true]
                   {\FigurePath 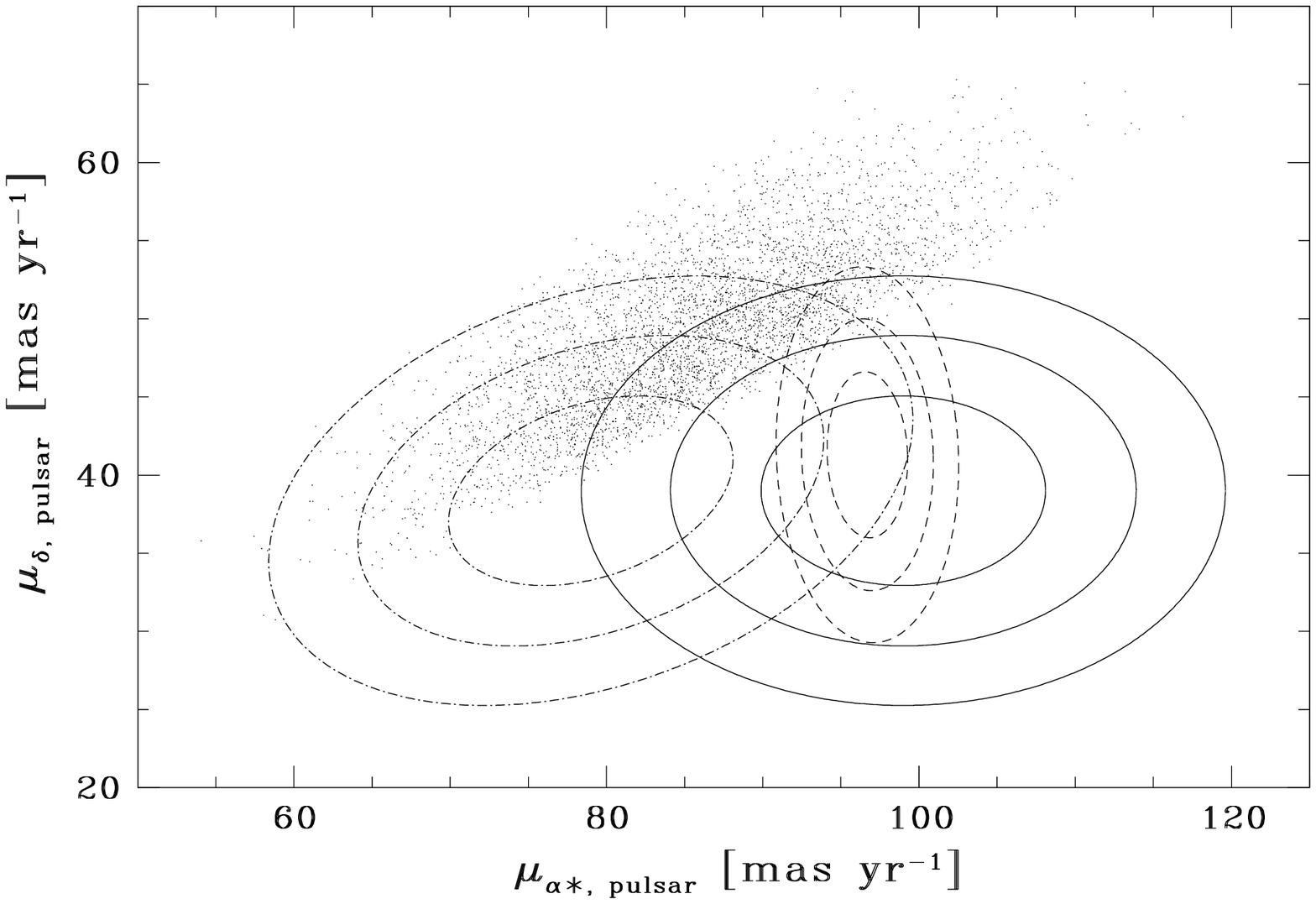}
  \caption[]{Proper motions of the pulsar PSR~J1932$+$1059 at the
  start of the $4\,214$ successful simulations (dots; see
  \S\ref{sec:us_sim}), and the proper-motion measurements of the
  pulsar: dot-dash line denotes Lyne, Anderson \& Salter (1982), solid
  line denotes Taylor, Manchester \& Lyne (1993), and the dashed line
  denotes Campbell (1995). The contours indicate the 1, 2, and
  3$\sigma$ confidence levels.}
\label{fig:05}
\end{figure}}
\def\FigSix{
\begin{figure}[t]
  \includegraphics[angle=0.0, width=8.5cm, 
                   clip=true, keepaspectratio=true]
                   {\FigurePath 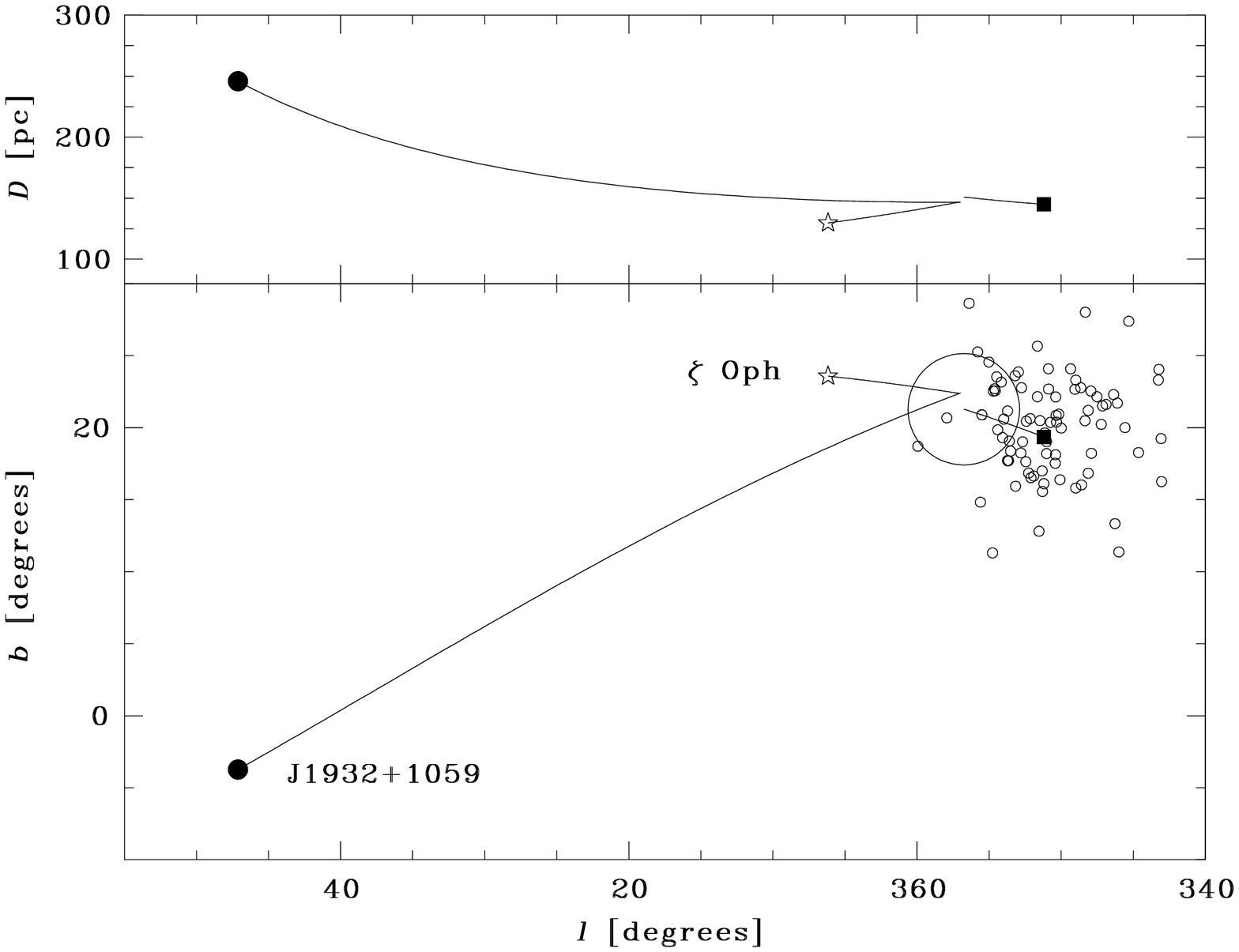}
  \caption[]{The orbits of $\zeta$~Oph, PSR~J1932$+$1059, and Upper
  Scorpius. The present positions are denoted by a star for the
  runaway, a filled circle for the pulsar, and by a filled square for
  the association. The {\it top} panel shows the distance vs.\
  Galactic longitude of the stars. The {\it bottom} panel shows the
  orbits projected on the sky in Galactic coordinates. The small open
  circles in the bottom panel denote the present-day positions of the
  O, B, and A-type members of Upper Scorpius, taken from de Zeeuw et
  al.\ (\cite{zee1999}). The large circle denotes the position of the
  association at the time of the supernova explosion, and has a 10~pc
  radius. This figure assumes a set of space motions consistent with
  the common origin hypothesis.}
\label{fig:06}
\end{figure}}
\def\FigSeven{
\begin{figure}[t]
  \includegraphics[angle=0.0, width=8.5cm, 
                   clip=true, keepaspectratio=true]
                   {\FigurePath 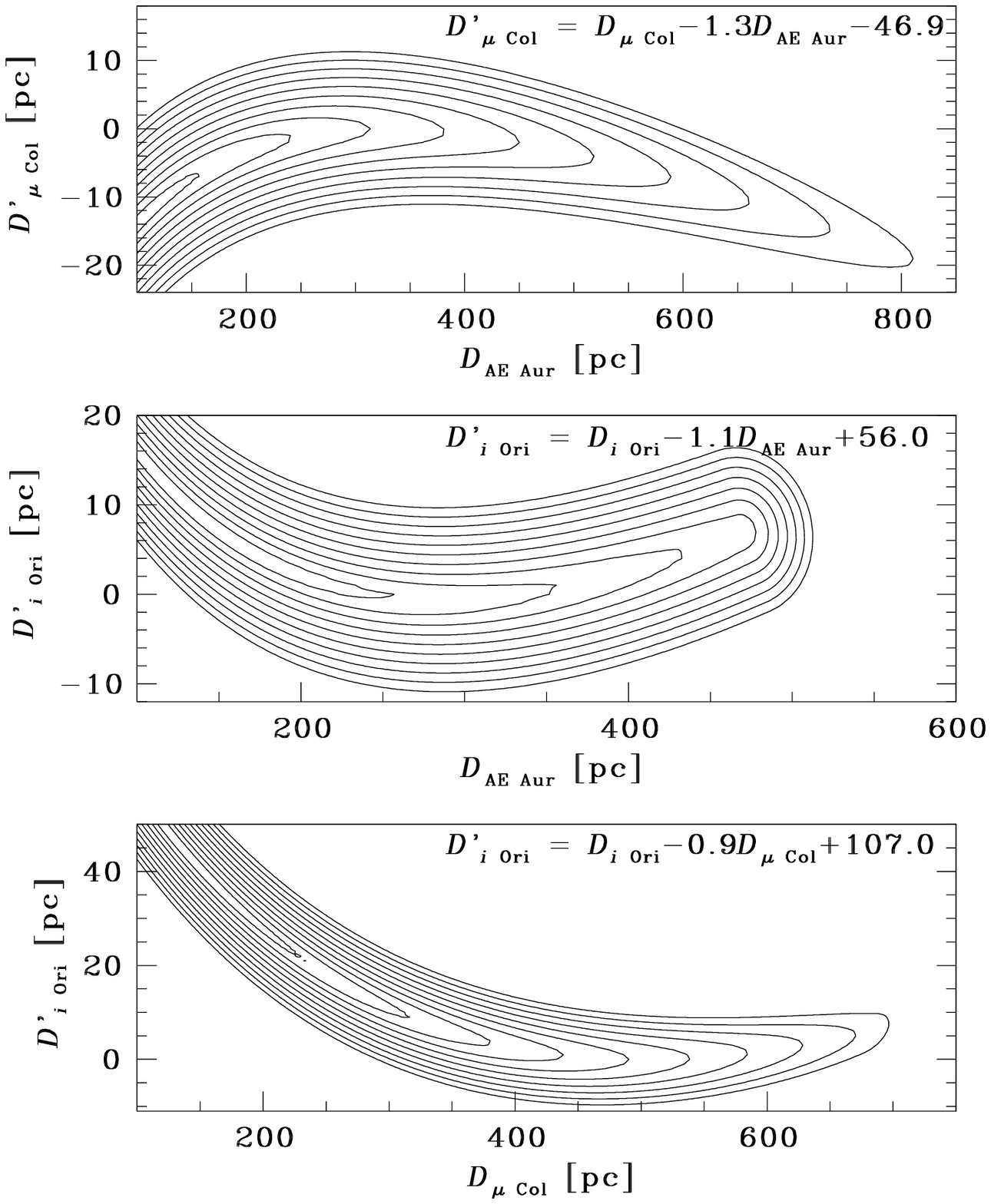}
  \caption[]{Contours of minimum separation for the pairs
  AE~Aur--$\mu$~Col ({\it top}), AE~Aur--$\iota$~Ori ({\it middle}),
  and $\mu$~Col--$\iota$~Ori ({\it bottom}). The contours are spaced
  every 1~pc with the outermost contour being 10~pc. The ordinates
  represent straight lines in the distance--distance plane, defined 
  as indicated in the top right of each panel.}
\label{fig:07}
\end{figure}}
\def\FigEight{
\begin{figure}[t]
  \begin{center}
  \includegraphics[angle=0.0, width=8.5cm, 
                   clip=true, keepaspectratio=true]
                   {\FigurePath 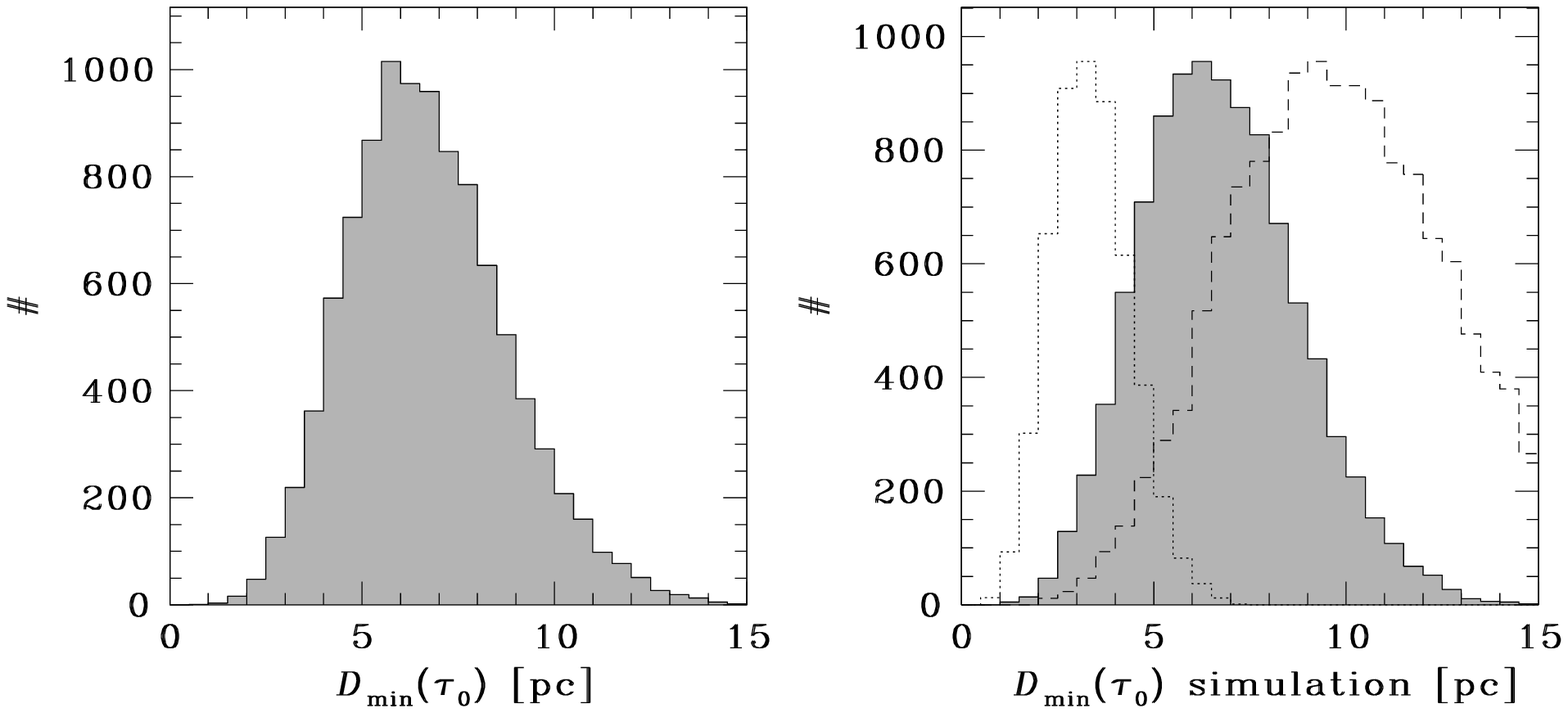}
  \end{center}
  \caption[]{{\it Left:\/} Distribution of minimum separations between
AE~Aur, $\mu$~Col, and $\iota$~Ori, $D_\mathrm{min}(\tau_0)$, of
$10\,000$ Monte Carlo simulations of the stellar encounter. {\it
Right:\/} The $D_\mathrm{min}(\tau_0)$ distribution for three randomly
drawn points from three spherical Gaussians with standard deviations
of $\sigma = 4$~pc (solid line and shaded), $\sigma = 2$~pc (dotted
line), and $\sigma = 6$~pc (dashed line). The $D_\mathrm{min}(\tau_0)$
distribution for $\sigma = 4$~pc is a good representation of the
distribution in the left panel. Four pc is the typical spread
in the end positions of the orbits due to the errors on the present
day velocity ($\sim$2~km~s$^{-1}$). The dotted and dashed histograms
have been normalized such that their shapes can be compared with the
solid histogram.}
\label{fig:08}
\end{figure}}
\def\FigNine{
\begin{figure*}[t]
  \includegraphics[angle=0.0, width=\textwidth, 
                   clip=true, keepaspectratio=true]
                   {\FigurePath 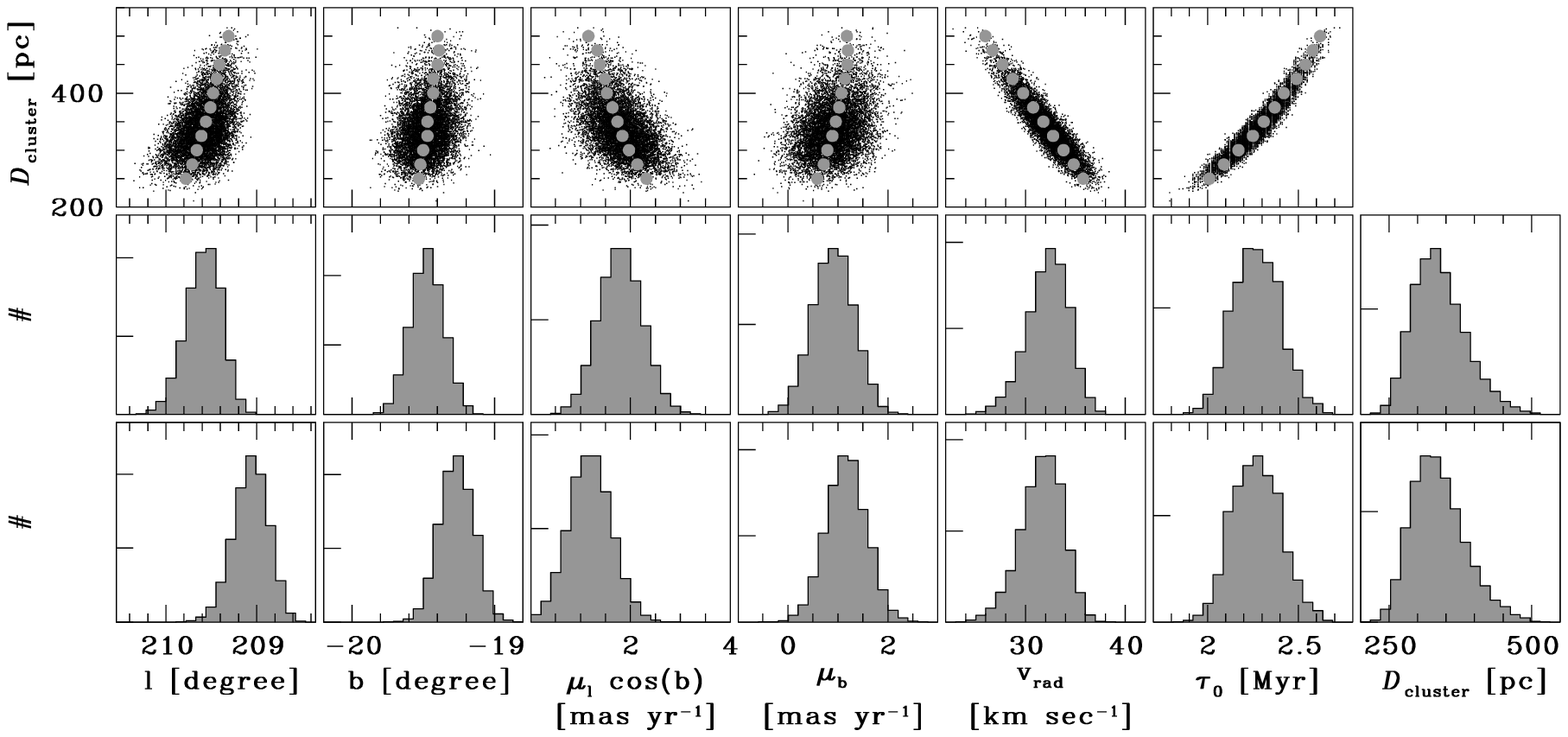}
  \caption[]{Properties of the parent cluster of the runaways AE~Aur and
$\mu$~Col and the binary $\iota$~Ori obtained from our Monte Carlo
simulations. {\it First row:} Cluster properties plotted vs.\ the
cluster distance. The grey dots denote the median values of the
cluster properties for distance-bins of 25 pc. {\it Second row:}
Histograms of the predicted cluster properties. The tick marks on the
vertical axis have a spacing of 1000. {\it Third row:} Histograms of the
predicted cluster properties when the mass of $\mu$~Col is changed by
$-1~M_\odot$. Note that the distance, time, and radial-velocity
histograms do not change significantly. The tick-mark spacing along the
vertical axis is similar to that in the second row. }
\label{fig:09}
\end{figure*}}
\def\FigTen{
\begin{figure*}[t]
  \includegraphics[angle=0.0, width=\textwidth, 
                   clip=true, keepaspectratio=true]
                   {\FigurePath 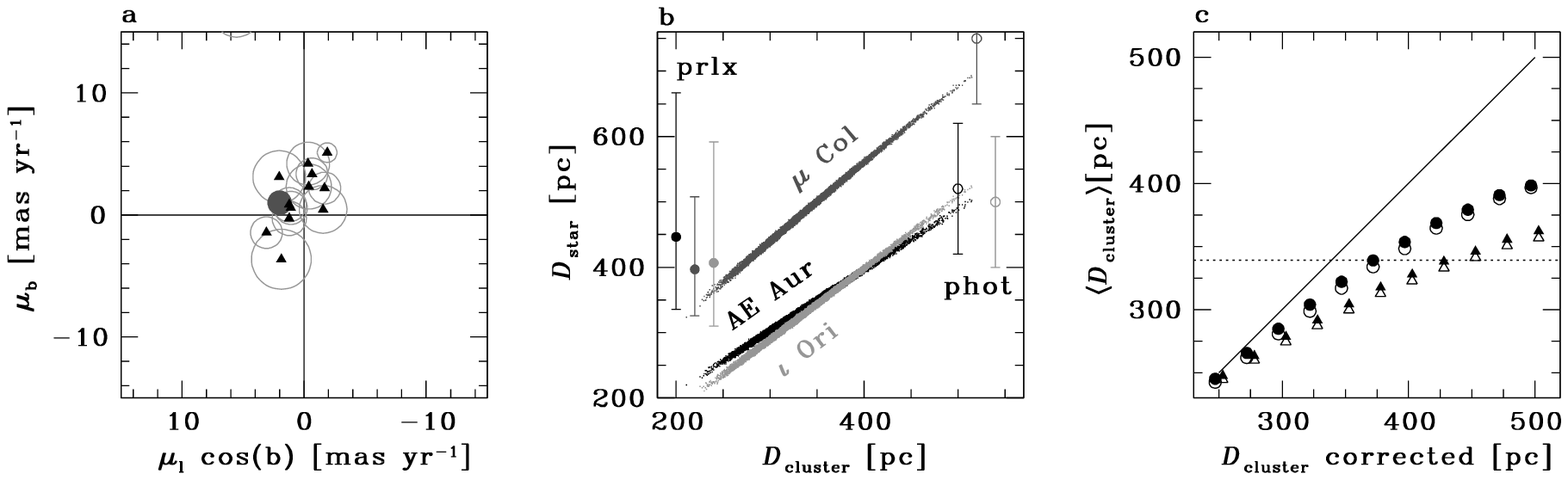}
  \caption[]{Properties of the parent cluster of the runaways AE~Aur
and $\mu$~Col and the binary $\iota$~Ori. Panel {\it a}: Proper
motions and their errors (grey circles) for all stars in the Tycho~2
Catalogue (filled triangles) within an area of 0\fdg4 by 0\fdg4
centred on the Trapezium cluster. The large grey dot denotes the
average of the predicted cluster proper motion for the Monte Carlo
simulations. Panel {\it b}: Distances of the runaway stars as a
function of the cluster distance in the Monte Carlo simulations. The
different grey scales are labeled in the panel. The filled circles and
their error bars denote the observed distances of the stars derived
from the Hipparcos parallaxes (prlx) and the open circles denote the
distances derived from photometry (phot) (Gies 1987). Panel {\it c}:
the biases on the predicted cluster distance as discussed in
\S\ref{sec:biases}. The filled and open symbols denote the mean and
median, respectively, of the cluster distance distributions. The
circles include only the first bias, the aiming effect, and the
triangles include both the aiming effect and the `incorrect'
Hipparcos parallax of $\mu$ Col. For clarity, the circles and
triangles are displaced $-3$ and $3$~pc, respectively. The dotted line
indicates the mean cluster distance based on the Monte Carlo
simulation.\looseness=-2}
\label{fig:10}

\end{figure*}}
\def\FigEleven{
\begin{figure*}[t]
  \begin{center}
  \includegraphics[angle=0.0, width=15.5cm, 
                   clip=true, keepaspectratio=true]
                   {\FigurePath 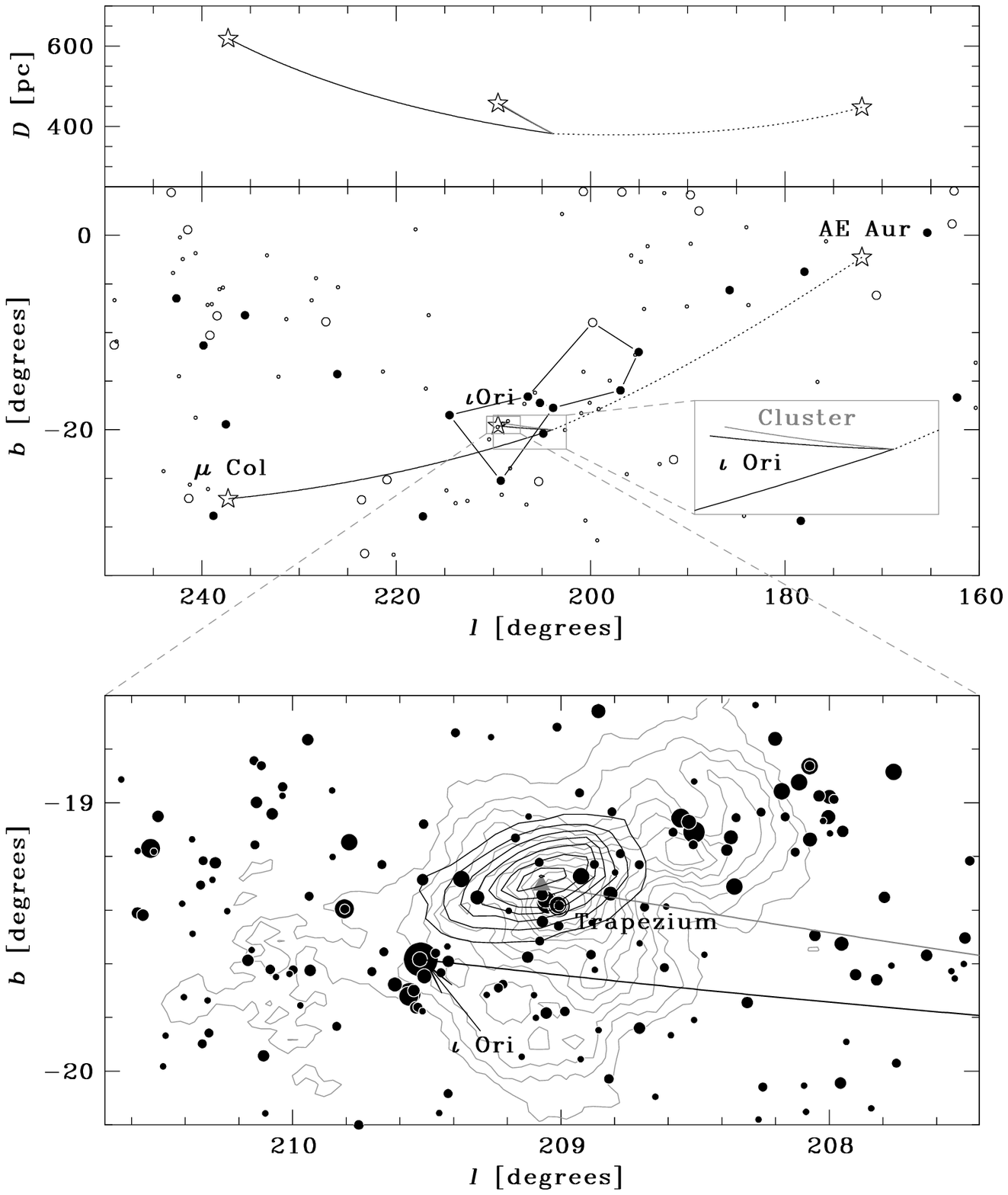}
  \end{center}
\vskip -0.5truecm
  \caption[]{{\it Top \& middle:\/} Orbits, calculated back in time,
of the runaways AE~Aur (dotted line) and $\mu$~Col (solid line) and
the binary $\iota$~Ori for one of the Monte Carlo simulations
described in the text. The {\it top} panel shows the distance vs.\
Galactic longitude of the stars. The {\it middle} panel shows the
orbits projected on the sky in Galactic coordinates. The starred
symbols depict the present position of the three stars. The stars met
$\sim$2.5~Myr ago. Using conservation of linear momentum, the orbit of
the parent cluster (grey solid line, see blow up) is calculated from
the time of the assumed encounter to the present. The large circles
denote all stars in the Hipparcos Catalogue brighter than $V =
3.5$~mag; filled circles denote O and B stars, open circles denote
stars of other spectral type. The small circles denote the O and B
type stars with $3.5~\mathrm{mag} \le V \le 5$~mag (cf.\ figure~1 in
Blaauw \& Morgan 1954). The Orion constellation is indicated for
reference. {\it Bottom:\/} The predicted position of the parent
cluster (black contours) together with all stars in the Tycho
Catalogue (ESA 1997) in the field down to $V = 12.4$ mag. The size of
the symbols scales with magnitude; the brightest star is
$\iota$~Ori. The Trapezium and $\iota$~Ori are indicated. The black
and dark grey lines are the past orbits of $\iota$~Ori and the
Trapezium, respectively (see top panel). The triangle denotes the
predicted present-day position of the parent cluster for this
particular simulation. The grey contours display the IRAS 100 micron
flux map, and mainly outline the Orion Nebula.}
\label{fig:11}
\end{figure*}}
\def\FigTwelve{
\begin{figure}[t]
  \includegraphics[angle=0.0, width=8.5cm,
                   clip=true, keepaspectratio=true]
                   {\FigurePath 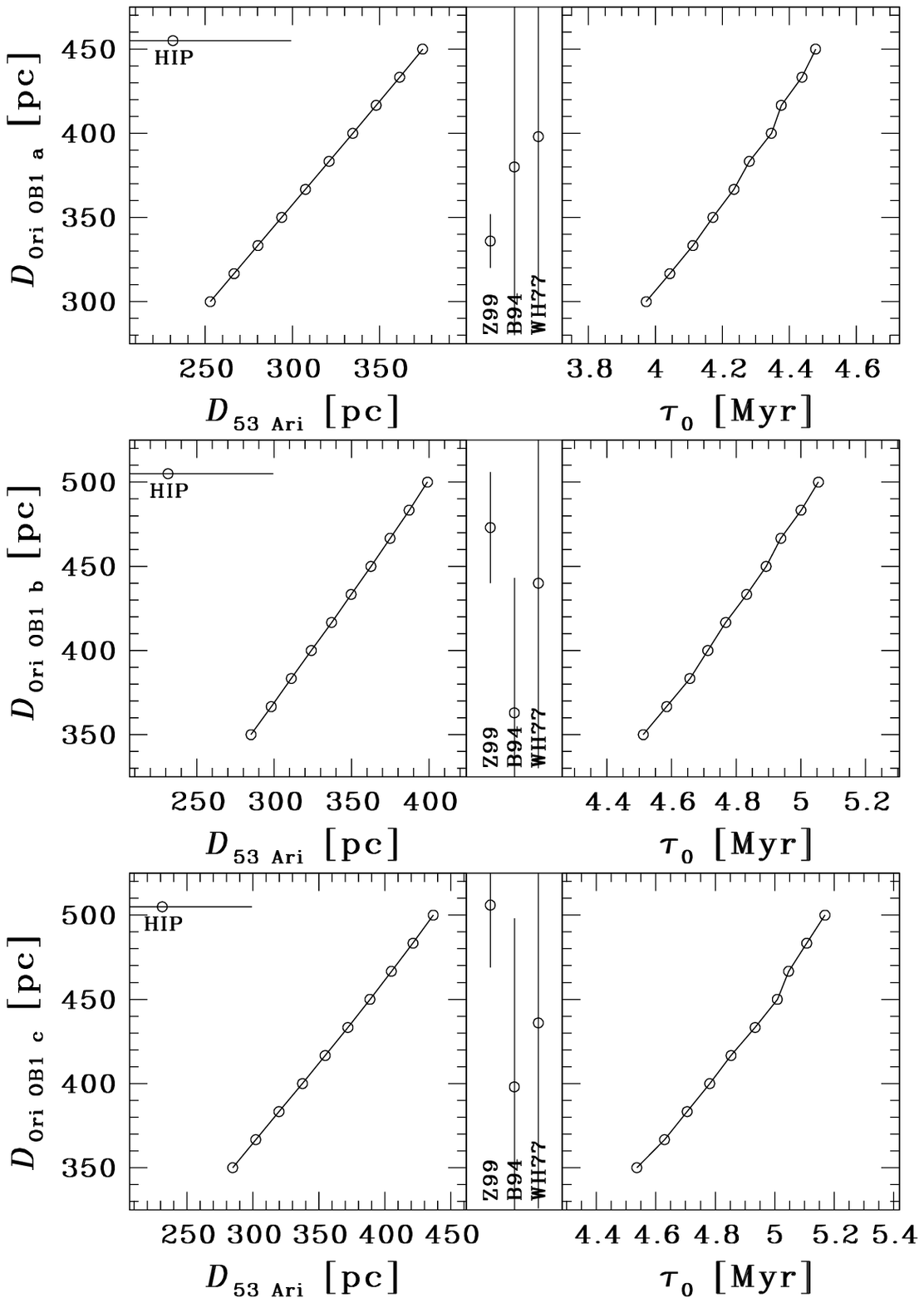}
  \caption[]{Predicted distance ({\it left}) and kinematic age ({\it
  right}) of 53~Arietis as a function of the distance of the parent
  association: Ori~OB1 subgroup {\sl a} ({\it top}), Ori~OB1
  subgroup {\sl b} ({\it middle}), and Ori~OB1 subgroup {\sl c}
  ({\it bottom}). The Hipparcos distance and its 1$\sigma$ error are
  indicated in the left panels. Distance estimated by Warren \& Hesser
  (1977a; 1977b, WH77), Brown et al.\
  (1994, B94), and de Zeeuw et al.\ (1999, Z99)
  are indicated in the {\it middle} panels.}
\label{fig:12}
\end{figure}}
\def\FigThirteen{
\begin{figure}[t]
  \includegraphics[angle=0.0, width=8.5cm, 
                   clip=true, keepaspectratio=true]
                   {\FigurePath 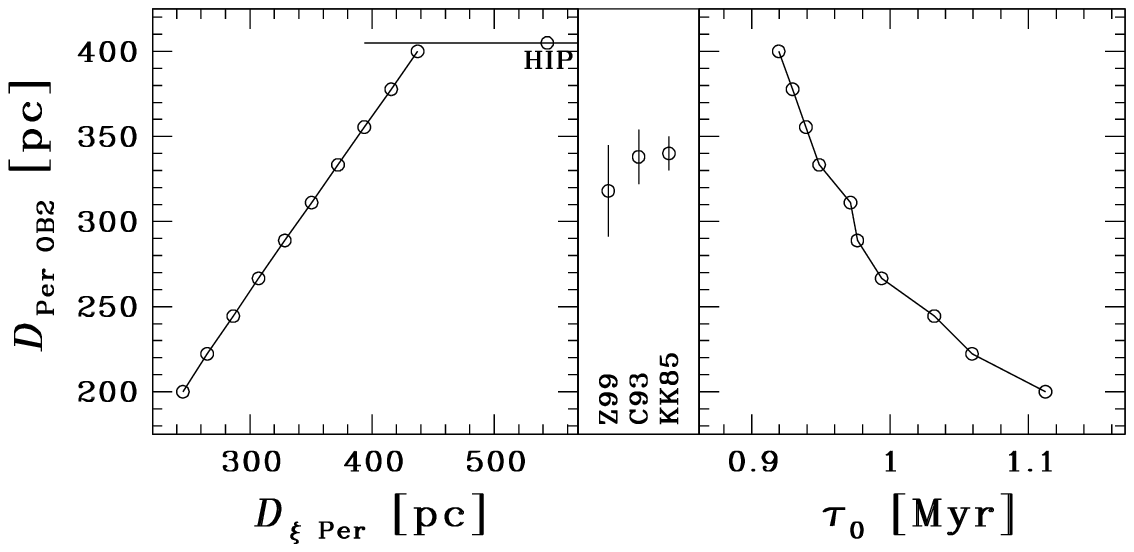}
  \caption[]{Predicted distance ({\it left}) and kinematic age ({\it
  right}) of $\xi$~Per as a function of the distance of Per~OB2. The
  Hipparcos distance and its 1$\sigma$ error are indicated in the left
  panel. Distance estimates for Per~OB2 by Klochkova \& Kopylov (1985, 
  KK85), {\v C}ernis (1993, C93), and de Zeeuw et al.\
  (1999, Z99) are indicated in the {\it middle}
  panel.}
\label{fig:13}
\end{figure}}
\def\FigFourteen{
\begin{figure}[t]
  \includegraphics[angle=0.0, width=8.5cm, 
                   clip=true, keepaspectratio=true]
                   {\FigurePath 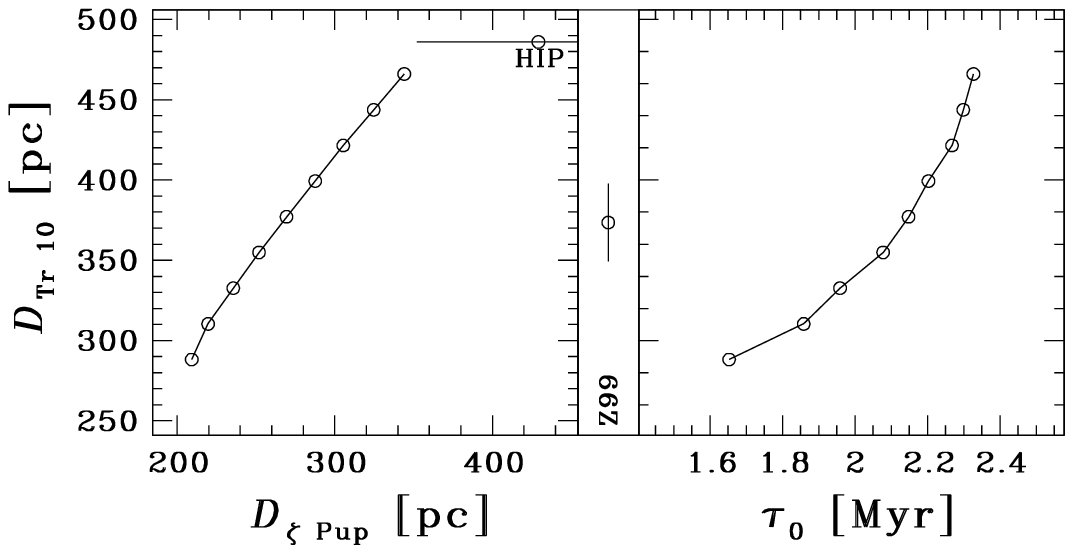}
  \caption[]{Predicted distance ({\it left}) and kinematic age ({\it
  right}) of $\zeta$~Pup as a function of the distance of Tr~10. The
  Hipparcos distance and its 1$\sigma$ error are indicated in the left
  panel. The distance estimate by de Zeeuw et al.\ (1999, Z99) is indicated 
  in the {\it middle} panel.}
\label{fig:14}
\end{figure}}
\def\FigFiveteen{
\begin{figure}[t]
  \includegraphics[angle=0.0, width=8.5cm, 
                   clip=true, keepaspectratio=true]
                   {\FigurePath 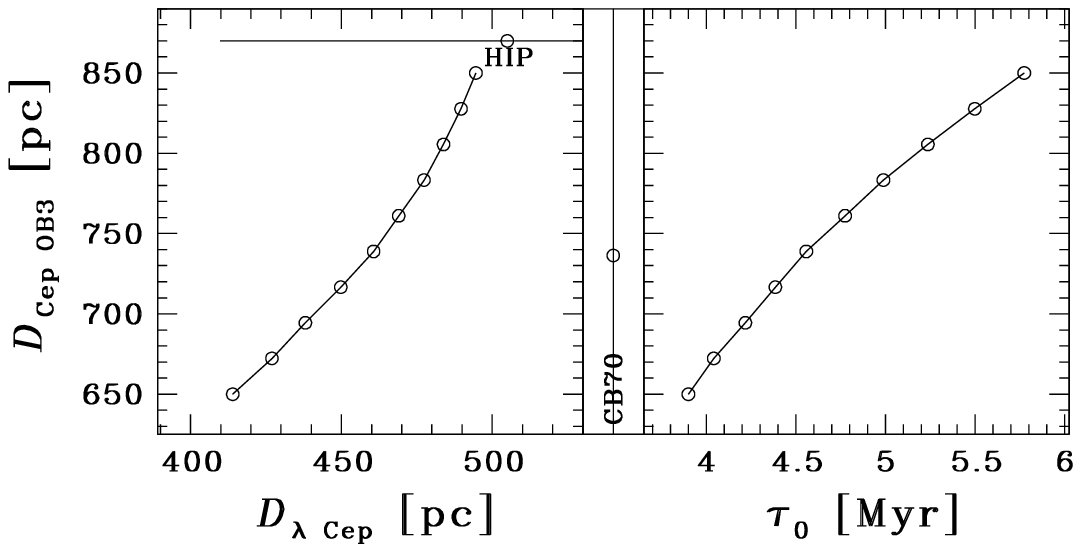}
  \caption[]{Predicted distance ({\it left}) and kinematic age ({\it
  right}) of $\lambda$~Cep as a function of the distance of Cep~OB3. The
  Hipparcos distance and its 1$\sigma$ error are indicated in the left
  panel. The distance estimate by Crawford \& Barnes (1970, CB70) is 
  indicated in the {\it middle} panel.}
\label{fig:15}
\end{figure}}
\def\FigSixteen{
\begin{figure*}[t]
  \begin{center}
  \includegraphics[angle=0.0, width=15cm, 
                   clip=true, keepaspectratio=true]
                   {\FigurePath 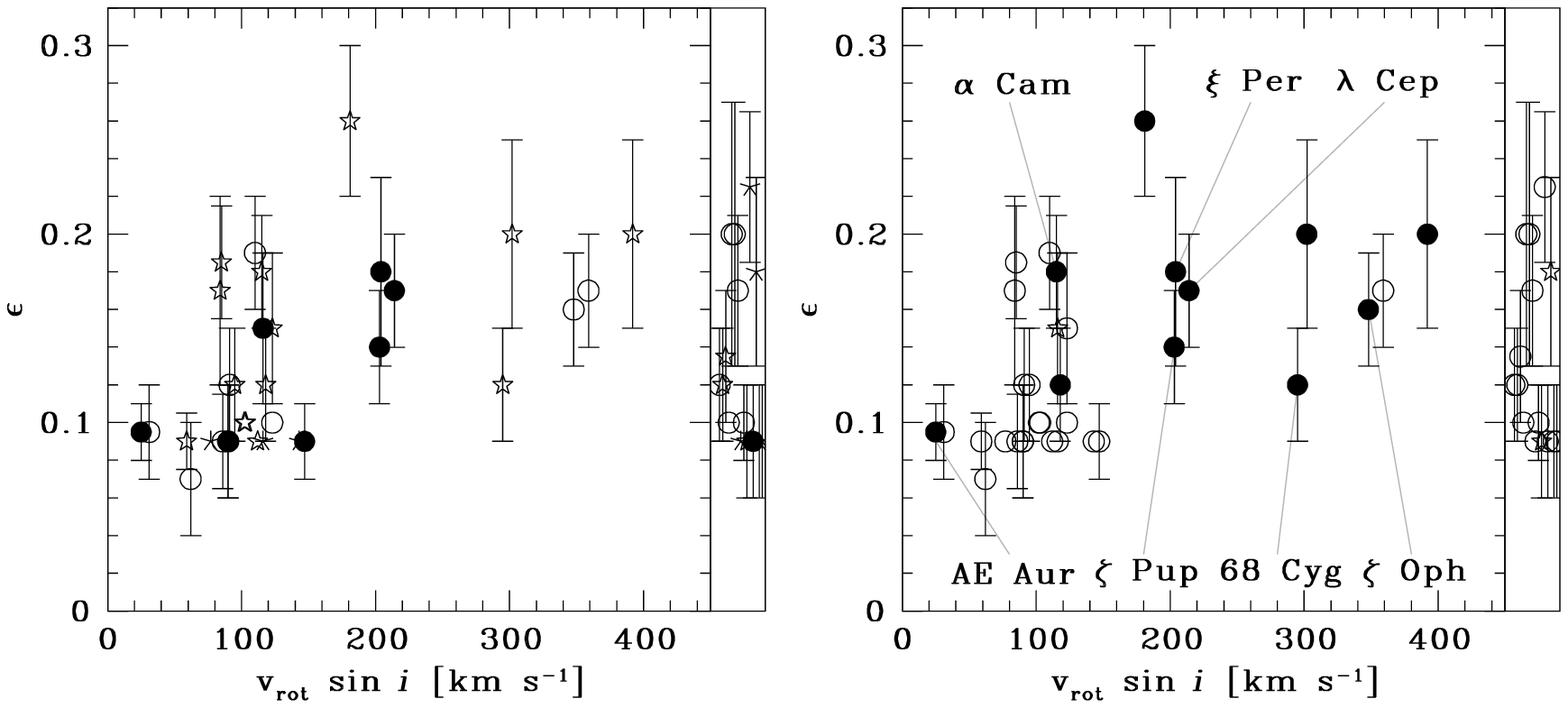}
  \end{center}
  \caption[]{Helium abundance ($\epsilon$: \# relative to Hydrogen)
   versus rotational velocity for O stars. {\it Left:} open circles:
   stars with $|\mathv| \le 30$~km~s$^{-1}$, filled circles: stars with
   $|\mathv| > 30$~km~s$^{-1}$, stars: stars with $\pi - \sigma_\pi <
   0$~mas, and asterisks: stars without Hipparcos data.  {\it Right:} open
   circles: non-runaway stars, filled circles: runaway stars, and stars:
   doubtful runaways. The right of each panel display the helium abundances
   of stars with an unknown rotational velocity. See text for
   details.}
\label{fig:16}
\end{figure*}}
\def\FigSeventeen{
\begin{figure*}[t]
  \begin{center}
  \includegraphics[angle=0.0, width=15.5cm, 
                   clip=true, keepaspectratio=true]
                   {\FigurePath 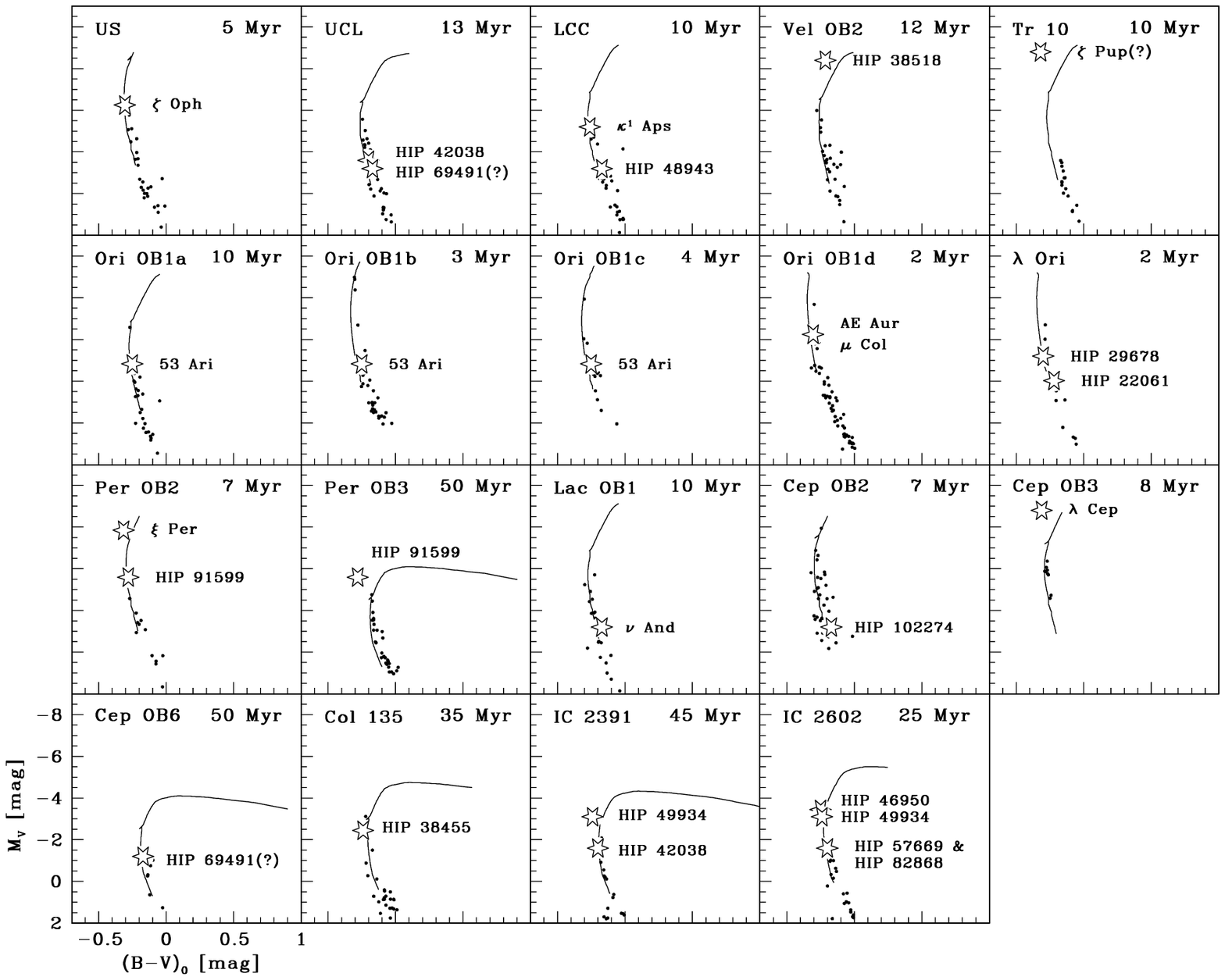}
  \end{center}
  \caption[]{Colour vs.\ absolute magnitude diagrams of the runaways
  (stars) and their parent association/cluster (small dots). The
  association/cluster members have been de-reddened using the
  Q-method. The colours and absolute magnitudes of the runaways have
  been determined using their spectral types (Table~\ref{tab:03};
  Schmidt--Kaler 1983). The isochrones are from Schaller et al.\
  (1992) for Solar metallicity and standard mass loss. The ages of the
  associations are indicated in the top right of each panel (US:
  Upper Scorpius, UCL: Upper Centaurus Lupus, LCC: Lower Centaurus
  Crux).}
\label{fig:17}
\end{figure*}}

\def\FigApp{
\begin{figure*}[t]
  \begin{center}
  \includegraphics[angle=0.0, width=14cm, 
                   clip=true, keepaspectratio=true]
                   {\FigurePath 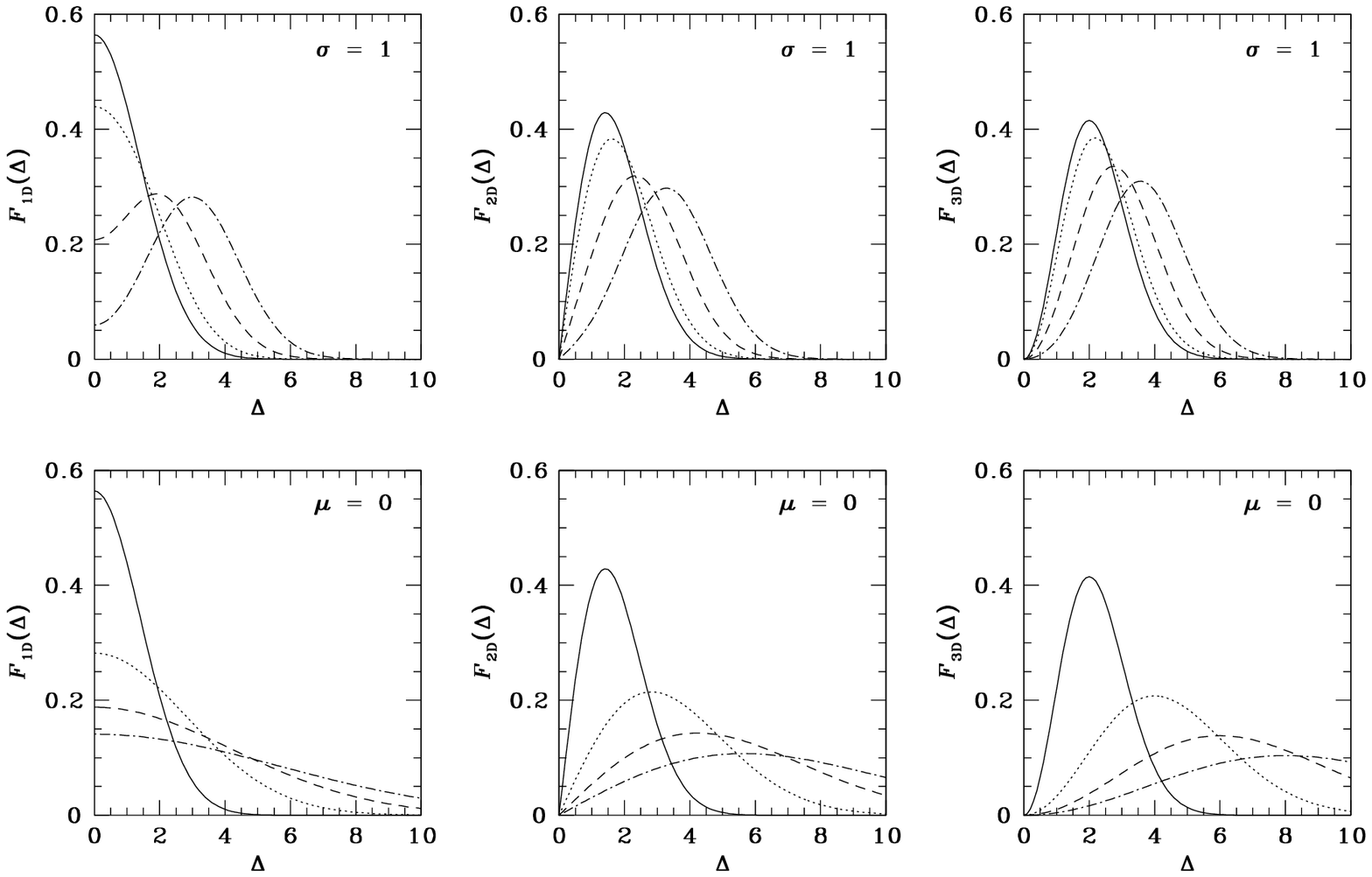}
  \end{center}
 \caption[]{{\it Left to right:} The distribution of absolute
  differences between two observable quantities, $\Delta$, in one
  ($F_\mathrm{1D}$), two ($F_\mathrm{2D}$), and three
  ($F_\mathrm{3D}$) dimensions, taking into account the measurement
  errors. {\it Top:} $\mu = 0$, 1, 2, and 3 are represented by the
  solid, dotted, dashed, and dot-dashed lines, $\sigma$ is indicated
  in the top right of each panel. {\it Bottom:} $\sigma = 1$, 2, 3,
  and 4 are represented by the solid, dotted, dashed, and dot-dashed
  lines, $\mu$ is indicated in the top right of each panel.}
\label{fig:app}
\end{figure*}}

\def\FigAppB{
\begin{figure}[t]
  \includegraphics[angle=0.0, width=8.5cm, 
                   clip=true, keepaspectratio=true]
                   {\FigurePath 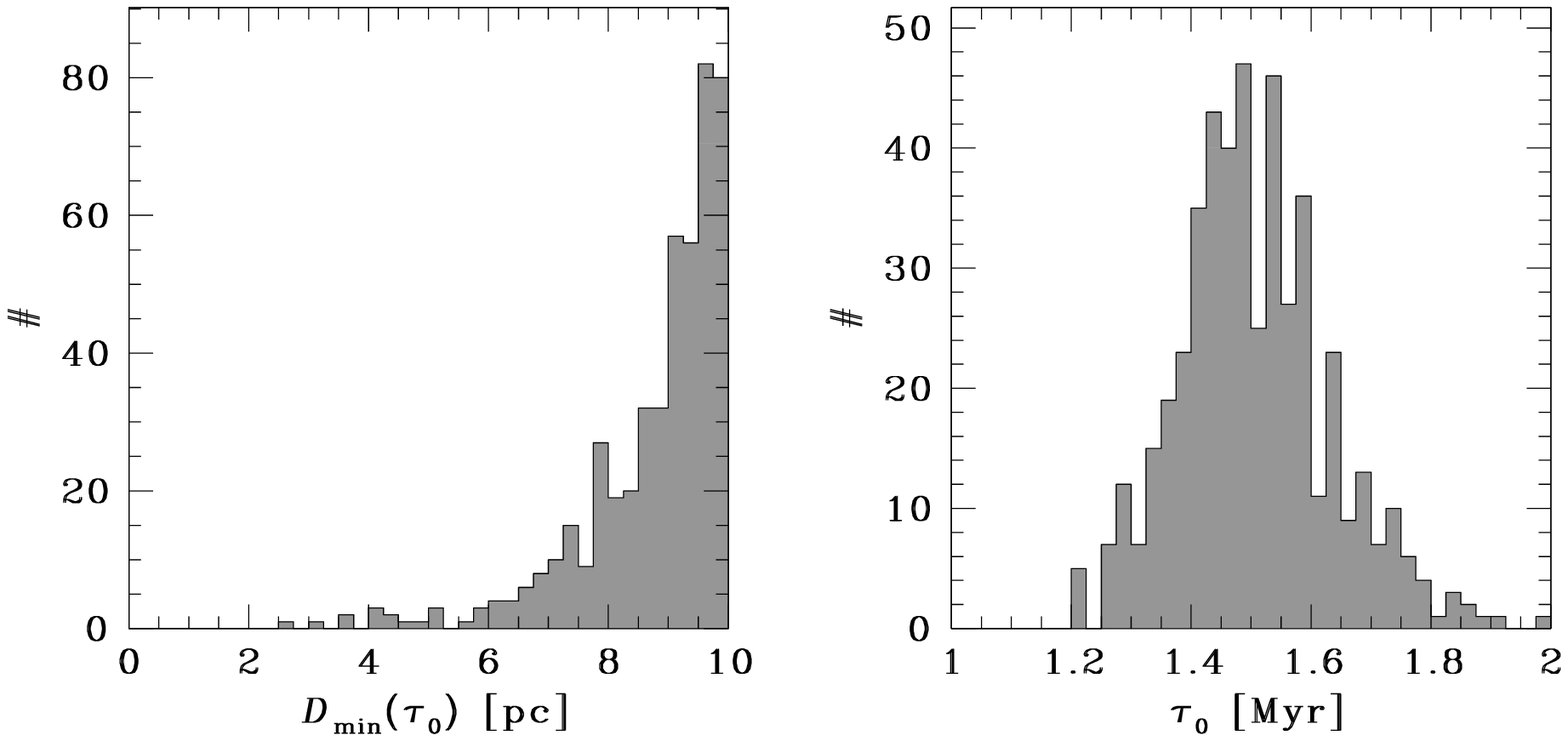}
  \caption[]{{\it Left:} Distribution of minimum separations,
  $D_\mathrm{min}(\tau_0)$, between $\zeta$~Oph and
  RX~J185635$-$3754. The solid line denotes the expected distribution of
  $D_\mathrm{min}$, see \S\ref{sec:us_sim}. {\it Right:}
  Distribution of the times $\tau_0$ at which the minimum separation 
  was reached.}
\label{fig:b1}
\end{figure}}

\def\TabOne{
\begin{table*}[t]
  \caption[]{The nearby runaway stars and pulsars with accurate
  astrometry. HIP indicates the number of the runaway star in the
  Hipparcos Catalogue, $\mathv_{\rm space}$ indicates the space motion
  of the runaway star relative to Galactic rotation (in km~s$^{-1}$),
  and PSR indicates the pulsar identifications.}  \label{tab:01}
  \begin{center}
  \relsize{-1}
\begin{tabular}{rrrrrrrrrrrrr}\hline\hline
\multicolumn{1}{c}{HIP} & 
\multicolumn{1}{c}{$\mathv_{\rm space}$} & 
\multicolumn{1}{c}{HIP} & 
\multicolumn{1}{c}{$\mathv_{\rm space}$} & 
\multicolumn{1}{c}{HIP} & 
\multicolumn{1}{c}{$\mathv_{\rm space}$} & 
\multicolumn{1}{c}{HIP} & 
\multicolumn{1}{c}{$\mathv_{\rm space}$} & 
\multicolumn{1}{c}{HIP} & 
\multicolumn{1}{c}{$\mathv_{\rm space}$} & 
\multicolumn{1}{c}{HIP} & 
\multicolumn{1}{c}{$\mathv_{\rm space}$} & 
\multicolumn{1}{c}{PSR} \cr\hline
  3478&   80.4& 28756&  196.6& 43158&   57.2& 61602&   30.2& 91599&   44.7&101350&   36.4& J0826$+$2637\cr
  3881&   32.1& 29678&   63.0& 45563&  125.9& 62322&   43.9& 92609&   31.0&102274&   46.1& J0835$-$4510\cr
  9549&  107.9& 30143&   55.5& 46928&   45.3& 66524&  112.7& 94899&  162.8&103206&   32.3& J0953$+$0755\cr
 10849&   50.0& 35951&   34.9& 46950&   32.1& 69491&   77.2& 94934&   94.0&105811&   38.3& J1115$+$5030\cr
 14514&   39.4& 36246&   32.1& 48715&   34.5& 70574&  205.3& 95818&   34.7&106620&   45.2& J1136$+$1551\cr
 18614&   64.9& 38455&   41.4& 48943&   35.2& 76013&   69.0& 96115&  165.6&109556&   74.0& J1239$+$2453\cr
 20330&   34.7& 38518&   31.1& 49934&   31.2& 81377&   23.5& 97774&   35.0&      &       & J1456$-$6843\cr
 22061&   86.5& 39429&   62.4& 52161&   34.8& 82171&   62.9& 97845&   70.3&      &       & J1932$+$1059\cr
 24575&  113.3& 40341&   61.7& 57669&   31.1& 82868&   30.3& 99435&   39.4&      &       &      Geminga\cr
 27204&  107.8& 42038&   31.3& 59607&   78.8& 86768&   30.1& 99580&   55.6&      &       &             \cr\hline\hline
  \end{tabular}
  \relsize{+1}
  \end{center}
\end{table*}
}
\def\TabTwo{
\begin{table*}[t]
  \begin{center} \caption[]{The nearby ($D < 700$~pc), young ($\tau <
  50$~Myr), open clusters listed in the {\tt WEBDA} Catalogue. The
  table gives the designation of the open cluster (Name), its position
  on the sky $(\ell,b)$, the number of member stars contained in the
  Hipparcos (HIP) and Hipparcos Input Catalogues (HIC), its distance,
  proper motion, and radial velocity as obtained from the Hipparcos
  data ($D$, $[\mu_{\ell\ast},\mu_b]$, $\mathv_\mathrm{rad}$), and
  whether the cluster is a candidate parent group in this study
  (Cand.). The candidate status is denoted by ``Y'' for the clusters
  with well-determined positions and velocities, by ``N'' for clusters
  for which not all information is available (either astrometry or
  radial velocity), by ``N$^*$'' if the astrometry does not show a
  clear signature of an open cluster (i.e., a clump in the
  proper-motion vs.\ proper-motion diagram), or by ``?'' if the
  measurements are not very reliable (either because of a small number
  of member stars or a large spread in the data).}\label{tab:02}
  \begin{tabular}{lrrrrrrrrr}\hline\hline
                   \multicolumn{1}{c}{Name}&
                 \multicolumn{1}{c}{$\ell$}& 
                    \multicolumn{1}{c}{$b$}& 
                    \multicolumn{1}{c}{HIP}& 
                    \multicolumn{1}{c}{HIC}& 
                    \multicolumn{1}{c}{$D$}&
       \multicolumn{1}{c}{$\mu_{\ell\ast}$}&
                \multicolumn{1}{c}{$\mu_b$}&
  \multicolumn{1}{c}{$\mathv_\mathrm{rad}$}&
                  \multicolumn{1}{c}{Cand.}\cr
                                           &
                 \multicolumn{1}{c}{[deg.]}& 
                 \multicolumn{1}{c}{[deg.]}& 
                   \multicolumn{1}{c}{[\#]}& 
                   \multicolumn{1}{c}{[\#]}& 
                   \multicolumn{1}{c}{[pc]}&
        \multicolumn{1}{c}{[mas~yr$^{-1}$]}&
        \multicolumn{1}{c}{[mas~yr$^{-1}$]}& 
          \multicolumn{1}{c}{[km~s$^{-1}$]}&
                                           \cr\hline
Collinder 359 &  29.75&    12.54& 10&$-$&          &             &            &             &N\rlap{$^*$}\cr
IC 4665       &  30.61&    17.08& 13&  5&$385\pm40$& $-7.2\pm0.3$&$-3.0\pm0.3$&$-13.5\pm3.0$&           Y\cr
Stephenson 1  &  66.85&    15.51&  0&$-$&          &             &            &             &           N\cr
Roslund 5     &  71.40&     0.25& 13&$-$&          &             &            &             &N\rlap{$^*$}\cr
Stock 7       & 134.68&     0.04&  3&$-$&          &             &            &             &N\rlap{$^*$}\cr
$\alpha$~Persei& \multicolumn{9}{c}{Central part of the Per~OB3 association, contained in the list of de Zeeuw et al.}\cr
IC 0348       & \multicolumn{9}{c}{Associated with Per~OB2}\cr
Collinder 69  & 195.05& $-$12.00&  6&$-$&\multicolumn{4}{l}{$\lambda$~Ori cluster}&          N\rlap{$^*$}\cr
NGC 1976      & \multicolumn{9}{c}{Trapezium cluster: associated with Ori~OB1}\cr
NGC 2232      & 214.36& $-$47.65& 10&  3&$365\pm40$&  $0.7\pm0.5$&$-5.2\pm0.5$&$ 14.6\pm3.0$&           ?\cr
Collinder 121 & \multicolumn{9}{c}{Contained in list of nearby associations of de Zeeuw et al.}\cr
Collinder 140 & 245.18&  $-$7.87& 14&  4&$375\pm40$&$ -7.4\pm0.5$&$-5.5\pm0.5$&$ 22.4\pm3.0$&           Y\cr
Collinder 135 & 248.76& $-$11.20& 19&  4&$300\pm30$&$-10.3\pm0.5$&$-6.8\pm0.5$&$ 16.4\pm3.0$&           Y\cr
Pismis 5      & 259.39&     0.86&  0&$-$&          &             &            &             &           N\cr
Pismis 4      & 262.74&  $-$2.37&  4&  0&          &             &            &             &           N\cr
Trumpler 10   & \multicolumn{9}{c}{Contained in list of nearby associations of de Zeeuw et al.}\cr
IC 2391       & 270.36&  $-$6.88& 24& 13&$150\pm30$&$-33.1\pm0.5$&$-6.0\pm0.5$&$ 15.0\pm3.0$&           Y\cr
vdB-Hagen 99  & 286.56&  $-$0.63&  7&  2&$500\pm50$&$-13.1\pm0.5$&$-6.4\pm0.5$&$ 12.0\pm3.0$&           ?\cr
IC 2602       & 289.60&  $-$4.90& 25&  8&$140\pm10$&$-20.4\pm0.5$& $1.2\pm0.5$&$ 24.1\pm3.0$&           Y\cr \hline
  \end{tabular}
  \end{center}
\end{table*}
}
\def\TabThree{
\begin{landscape}
\begin{table}[t]
  \caption[]{Data for the nearby runaway stars and pulsars discussed
  in this paper. The stars which have an $^\ast$ appended to their HIP
  identifier are the classical runaways (i.e., a parent group was
  already known before this study, see Paper~I. Unless indicated
  otherwise, the position ($\alpha$, $\delta$), proper motion
  ($\mu_{\alpha\ast}$, $\mu_\delta$), and parallax ($\pi$) were taken
  from the Hipparcos Catalogue (ESA 1997), the radial velocity
  ($\mathv_\mathrm{rad}$) from the Hipparcos Input Catalogue (Turon et
  al.\ 1992), the space velocity ($\mathv_\mathrm{space}$) with
  respect to the standard of rest of the runaway, the rotational
  velocity ($\mathv_\mathrm{rot} \sin i$) for the runaways from Penny
  (1996) and the period $P$ for the pulsars in seconds, the spectral
  type from Mason et al.\ (1998) or the Hipparcos Catalogue for the
  runaways and the characteristic age ($\tau = P/(2\dot{P})$) for the
  pulsars, and the helium abundance ($\epsilon$) from Herrero et al.\
  (1992), defined as the number of He atoms relative to H.  The mass
  $M_{\rm SK}$ has been derived from the Schmidt-Kaler (1982) calibration,
  using interpolation. The mass $M_{\rm BB}$ is taken from Vanbeveren, 
  Van Rensbergen \& de Loore (1998). The last column (N) indicates the 
  number of the runaway/pulsar in Figure~\ref{fig:02}. The proper motion 
  and radial velocity are not corrected for Solar motion and Galactic
  rotation. The astrometric data ($\alpha$, $\delta$, $\pi$,
  $\mu_{\alpha\ast}$, and $\mu_\delta$) for the pulsars (the last five
  lines) are taken from the Taylor, Manchester \& Lyne (1993)
  catalogue. Abbreviations used: mas = milli-arcsec; $\mu_{\alpha\ast}
  = \mu_\alpha \cos\delta$.}\label{tab:03}
\bigskip 
\relsize{-2} 
\begin{center}
  \begin{tabular}{rrlrrrrrrrrrrrrr}\hline\hline
  \multicolumn{1}{c}{HIP}& \multicolumn{1}{c}{HD}& Name&
  \multicolumn{1}{c}{$\alpha$ ($^{\rm h~m~s}$)}&
  \multicolumn{1}{c}{$\delta$ ($^{\circ}~'~''$)}&
  \multicolumn{1}{c}{$\pi$}& \multicolumn{1}{c}{$\mu_{\alpha\ast}$}&
  \multicolumn{1}{c}{$\mu_\delta$}&
  \multicolumn{1}{c}{$\mathv_\mathrm{rad}$}&
  \multicolumn{1}{c}{\rlap{\hskip -4.5mm $\mathv_\mathrm{space}$}}&
\multicolumn{1}{c}{\rlap{\hskip -4mm $\mathv_\mathrm{rot} \sin i$}}&
                    \multicolumn{1}{c}{SpT}&
                    \multicolumn{1}{c}{$M_{\rm SK}$}&
                    \multicolumn{1}{c}{$M_{\rm BB}$}&
             \multicolumn{1}{c}{$\epsilon$}&
                      \multicolumn{1}{c}{N}\cr
     &     &     &
 \multicolumn{1}{c}{[J1991.25]}& 
 \multicolumn{1}{c}{[J1991.25]}& 
 \multicolumn{1}{c}{[mas]}& 
 \multicolumn{1}{c}{[mas~yr$^{-1}$]}& 
 \multicolumn{1}{c}{[mas~yr$^{-1}$]}& 
 \multicolumn{1}{c}{[km~s$^{-1}$]}& 
 \multicolumn{1}{c}{\rlap{\hskip -6mm [km~s$^{-1}$]}}& 
 \multicolumn{1}{c}{\rlap{\hskip -4mm [km~s$^{-1}$]}}& & 
 \multicolumn{1}{c}{[$M_\odot$]}& 
 \multicolumn{1}{c}{[$M_\odot$]}& 
 \multicolumn{1}{c}{[\#]}&
                         \cr\hline
                3881& 4727&$\nu$~And       &  $0~49~48.83$& $+41~04~44.2$& $4.80\pm0.75$& $22.68\pm0.53$& $-18.05\pm0.48$&            $-23.9\pm\phantom{1}1.2$&  32.1&    80\rlap{$^a$}&              B5V$+$F8V&  6.9\rlap{$^b$}&      &                &  1\cr 
 14514\rlap{$^\ast$}& 19374&53~Ari         &  $3~07~25.69$& $+17~52~47.9$& $4.32\pm0.98$&$-23.54\pm0.93$&   $9.30\pm0.95$&  $21.2\pm\phantom{1}1.2$\rlap{$^c$}&  39.4&    10\rlap{$^d$}&                  B1.5V&            10.4&   8.5&                &  2\cr 
 18614\rlap{$^\ast$}& 24912&$\xi$~Per      &  $3~58~57.90$& $+35~47~27.7$& $1.84\pm0.70$&  $1.92\pm0.74$&   $2.30\pm0.62$&  $58.8\pm\phantom{1}5.0$\rlap{$^e$}&  64.9&              204&                O7.5III&            33.8&  33.5&            0.18&  3\cr 
               22061& 30112&               &  $4~44~42.16$&  $+0~34~05.4$& $2.94\pm0.86$&$-44.89\pm0.77$& $-29.28\pm0.67$&              $6.0\pm\phantom{1}5.0$&  86.5&                 &                  B2.5V&             8.6&   7.5&                &  4\cr 
 24575\rlap{$^\ast$}& 34078&AE~Aur         &  $5~16~18.15$& $+34~18~44.0$& $2.24\pm0.74$& $-4.05\pm0.66$&  $43.22\pm0.44$&             $57.5\pm\phantom{1}1.2$& 113.3&               25&                  O9.5V&            15.9&  21.1&            0.09&  5\cr 
               26241& 37043&$\iota$~Ori    &  $5~35~25.98$& $-05~54~35.6$& $2.46\pm0.77$&  $2.27\pm0.65$&  $-0.62\pm0.47$&  $28.7\pm\phantom{1}1.1$\rlap{$^f$}&   8.0&    71\rlap{$^g$}& O9III+B1III\rlap{$^h$}& 37.8\rlap{$^i$}&  38.6&                &   \cr 
 27204\rlap{$^\ast$}& 38666&$\mu$~Col      &  $5~45~59.89$& $-32~18~23.0$& $2.52\pm0.55$&  $3.01\pm0.52$& $-22.62\pm0.50$&            $109.0\pm\phantom{1}2.5$& 107.8&              111&                  O9.5V&            15.9&  21.1&                &  6\cr 
               29678& 43112&               &  $6~15~08.46$& $+13~51~03.9$& $2.38\pm0.72$& $24.21\pm0.76$&  $10.65\pm0.49$&             $36.0\pm\phantom{1}5.0$&  63.0& $<$25\rlap{$^j$}&                    B1V&            11.5&  12.0&                &  7\cr 
               38455& 64503&               &  $7~52~38.65$& $-38~51~46.2$& $5.09\pm0.52$& $-9.49\pm0.43$&   $4.02\pm0.42$&            $-31.0\pm\phantom{1}5.0$&  41.4&   212\rlap{$^k$}&                    B2V&             9.4&   8.0&                &  8\cr 
               38518& 64760&               &  $7~53~18.16$& $-48~06~10.6$& $1.68\pm0.50$& $-4.90\pm0.53$&   $5.89\pm0.38$&             $41.0\pm\phantom{1}5.0$&  31.1&   220\rlap{$^d$}&                B0.5Iab&            25.0&  35.1&                &  9\cr 
               39429& 66811&$\zeta$~Pup    &  $8~03~35.07$& $-40~00~11.5$& $2.33\pm0.51$&$-30.82\pm0.44$&  $16.77\pm0.41$&            $-23.9\pm\phantom{1}1.2$&  62.4&              203&                    O4I&                &  67.5& 0.14\rlap{$^l$}& 10\cr 
               42038& 73105&               &  $8~34~09.60$& $-53~04~17.5$& $2.87\pm0.47$&$-12.14\pm0.54$&  $10.13\pm0.48$&            $37.0\pm10.0$&  31.3&                 &                               B3V&             7.9&   7.0&                & 11\cr 
               46950& 83058&               &  $9~34~08.80$& $-51~15~19.0$& $3.50\pm0.53$& $-8.50\pm0.49$&   $6.39\pm0.48$&            $35.0\pm10.0$&  32.1&                 &                            B1.5IV& 10.4\rlap{$^m$}&   9.0&                & 12\cr 
               48943& 86612&               &  $9~59~06.32$& $-23~57~02.8$& $5.19\pm0.77$&$-23.22\pm0.70$&   $5.30\pm0.70$&             $39.0\pm\phantom{1}5.0$&  35.2&   230\rlap{$^d$}&                    B5V&             5.8&      &                & 13\cr 
               49934& 88661&               & $10~11~46.47$& $-58~03~38.0$& $2.52\pm0.50$&$-10.71\pm0.49$&   $6.63\pm0.45$&            $31.0\pm10.0$&  31.2&   280\rlap{$^d$}&                           B2IVnpe&  9.4\rlap{$^m$}&   8.0&                & 14\cr 
               57669& 102776&              & $11~49~41.09$& $-63~47~18.6$& $7.10\pm0.69$&$-17.93\pm0.95$&   $4.44\pm0.63$&             $29.0\pm\phantom{1}2.5$&  31.1&   251\rlap{$^n$}&                    B3V&             7.9&   7.0&                & 15\cr 
               69491& 124195&              & $14~13~39.84$& $-54~37~32.2$& $2.96\pm0.63$&$-18.03\pm0.41$& $-11.15\pm0.41$&            $66.0\pm10.0$&  77.2&                 &                               B5V&             5.8&      &                & 16\cr 
               76013& 137387&$\kappa^1$~Aps& $15~31~30.82$& $-73~23~22.4$& $3.20\pm0.59$&  $0.38\pm0.48$& $-18.28\pm0.55$&             $62.0\pm\phantom{1}5.0$&  69.0&                 &                  B1npe&                &      &                & 17\cr 
 81377\rlap{$^\ast$}& 149757&$\zeta$~Oph   & $16~37~09.53$& $-10~34~01.7$& $7.12\pm0.71$& $13.07\pm0.85$&  $25.44\pm0.72$&             $-9.0\pm\phantom{1}5.5$&  23.5&              348&                O9.5Vnn&            15.9&  21.1&            0.16& 18\cr 
               82868& 152478&              & $16~56~08.85$& $-50~40~29.2$& $4.34\pm0.82$&$-10.21\pm0.84$&  $-9.55\pm0.62$&             $19.0\pm\phantom{1}5.0$&  30.3&                 &                 B3Vnpe&             7.9&   7.0&                & 19\cr 
               91599& 172488&              & $18~40~48.06$& $-08~43~07.5$& $3.61\pm1.16$& $-9.64\pm1.13$& $-22.64\pm0.79$&  $34.1\pm\phantom{1}1.2$\rlap{$^o$}&  44.7&                 &                  B0.5V&            12.7&  13.5&                & 20\cr 
              102274& 197911&              & $20~43~21.62$& $+63~12~32.9$& $1.42\pm0.62$&$-13.72\pm0.53$&  $-3.66\pm0.53$&             $-3.8\pm\phantom{1}5.0$&  46.1&                 &                     B5&                &      &                & 21\cr 
109556\rlap{$^\ast$}& 210839&$\lambda$~Cep & $22~11~30.58$& $+59~24~52.3$& $1.98\pm0.46$& $-7.22\pm0.44$& $-11.06\pm0.39$&            $-75.1\pm\phantom{1}1.2$&  74.0&              214&                    O6I&            40.0&  64.6& 0.17\rlap{$^l$}& 22\cr 
&&&&&&&&&&&&&&\cr
\multicolumn{3}{c}{J0826$+$2637}      &  $8~26~51.31$& $+26~37~25.6$&       $2.6\phantom{\pm11.7}$&  $61\pm3$& $-90\pm2$&  & & $\rlap{\hskip -5mm P =} 0.53$& $\tau = 4.92$& 1.4\rlap{$^p$}& & & 1\cr
\multicolumn{3}{c}{J0835$-$4510}      &  $8~35~20.68$& $-45~10~35.8$&       $2.0\phantom{\pm11.7}$& $-48\pm2$&  $35\pm1$&  & &       0.09&          0.01& 1.4\rlap{$^p$}& & & 2\cr
\multicolumn{3}{c}{J1115$+$5030}      & $11~15~38.35$& $+50~30~13.6$&       $1.9\phantom{\pm11.7}$&  $22\pm3$& $-51\pm3$&  & &       1.65&         10.53& 1.4\rlap{$^p$}& & & 4\cr
\multicolumn{3}{c}{J1932$+$1059}      & $19~32~13.87$& $+10~59~31.8$&       $5.9\phantom{\pm11.7}$&  $99\pm6$&  $39\pm4$&  & &       0.22&          3.10& 1.4\rlap{$^p$}& & & 8\cr
\multicolumn{3}{c}{Geminga\rlap{$^q$}}&  $6~33~54.15$& $+17~46~12.9$&                  $6.4\pm1.7$& $138\pm4$&  $97\pm4$&  & &           &              & 1.4\rlap{$^p$}& & & 9\cr\hline\hline
\noalign{\vskip -5mm}
  \end{tabular}
\end{center}
\relsize{+2}
\end{table}
\relsize{-2}
\noindent Notes: 
$a$: $\mathv_\mathrm{rot} \sin i$ from Slettebak, Wagner \& Bertram
(1997).
$b$: Total mass for the binary: B5V ($5.8~M_\odot$) + F8V ($1.1~M_\odot$).
$c$: $\mathv_\mathrm{rad}$ from Duflot et al.\ (1995). The
Hipparcos Input Catalogue radial velocity is incorrect.
$d$: $\mathv_\mathrm{rot} \sin i$ from Bernacca \& Perinotto (1970).
$e$: We took the average $\mathv_\mathrm{rad}$ from Bohannan \& Garmany
(1978), Garmany, Conti \& Massey (1980), Stone
(1982), and Gies \& Bolton (1986).
$f$: $\mathv_\mathrm{rad}$, i.e., center of mass velocity, from Stickland et
al.\ (1987).
$g$: $\mathv_\mathrm{rot} \sin i$  from Gies (1987).
$h$: Spectral type of the secondary of $\iota$~Ori from Stickland et
al.
$i$: Total mass for the binary $\iota$~Ori. The individual masses are
$22.9~M_\odot$ for the primary and $14.9~M_\odot$ for the secondary.
$j$: $\mathv_\mathrm{rot} \sin i$ from Morse, Mathieu \& Levine (1991).
$k$: $\mathv_\mathrm{rot} \sin i$ from Uesugi \& Fukuda (1970).
$l$: $\epsilon$ from Kudritzki \& Hummer (1990).
$m$ Assumed mass for main-sequence star instead of luminosity class IV.
$n$: $\mathv_\mathrm{rot} \sin i$ from Brown \& Verschueren (1997).
$o$: $\mathv_\mathrm{rad}$ from Gies \& Bolton (1986).
$p$: We take the characteristic mass for a neutron star.
$q$: Position from Caraveo et al.\ (\cite{car1998}); parallax and
proper motion from Caraveo et al.\ (\cite{car1996}).

\relsize{+2}
\vfill
\end{landscape}
}
\def\TabFour{
\begin{table}[t]
  \caption[]{Predicted properties of the parent cluster of the stellar
systems AE~Aur, $\mu$~Col, and $\iota$~Ori: the cluster distance
$D_\mathrm{parent}$, the sky position in equatorial $(\alpha,\delta)$
and Galactic $(\ell,b)$ coordinates, the proper motions
$(\mu_{\alpha\ast},\mu_\delta)$ and $(\mu_{\ell\ast},\mu_b)$, and the
radial velocity $\mathv_\mathrm{rad}$. The predicted distances of the
runaways and $\iota$~Ori {\it if} there was an encounter:
$D_\mathrm{AE~Aur} = 430$~pc, $D_{\mu~\mathrm{Col}} = 600$~pc, and
$D_{\iota~\mathrm{Ori}} = 440$~pc.}\label{tab:04}
  \relsize{-2}
\begin{tabular}{lrcrl}\hline\hline
                               &           & &$M_{\mu~\mathrm{Col}} - 1~M_\odot$& \\ \hline
$D_\mathrm{parent}$            &              425--450&&              425--450& pc\\ 
$(\alpha,\delta)$              &   (84\fdg0,$-$5\fdg8)&&   (83\fdg9,$-$5\fdg2)&  \\
$(\mu_{\alpha\ast},\mu_\delta)$&          (1.7,$-$0.8)&&          (1.7,$-$0.2)& mas~yr$^{-1}$\\
$(\ell,b)$                     & (209\fdg4,$-$19\fdg4)&& (208\fdg9,$-$19\fdg2)& \\
$(\mu_{\ell\ast},\mu_b)$       &             (1.3,1.2)&&             (0.9,1.4)& mas~yr$^{-1}$\\
$\mathv_\mathrm{rad}$               &                  28.3&&                  27.6& km~s$^{-1}$\\
\hline \hline
\end{tabular}
  \relsize{+2}
\end{table}
}
\def\TabFive{
\begin{table}[t]
  \caption[]{Parent associations, kinematic ages ($\tau_0$) and
  formation mechanisms of the runaways and neutron stars discussed in this
  paper. For runaway stars which have more than one possible parent we
  give the results for the individual parent
  associations/clusters. The last column indicates the number of the
  runaway/pulsar in Figure~\ref{fig:02}. Abbreviations: US: Upper
  Scorpius; UCL: Upper Centaurus Lupus; LCC: Lower Centaurus Crux;
  BSS: binary supernova scenario; DES: dynamical ejection
  scenario.}\label{tab:05}
\vskip -2mm
  \relsize{-2}
\begin{tabular}{rllrrr}\hline\hline
\multicolumn{1}{c}{HIP}    & 
\multicolumn{1}{c}{Name}   & 
\multicolumn{1}{c}{Parent} & 
\multicolumn{1}{c}{$\tau_0$}&
\multicolumn{1}{c}{Origin} & Fig.~\ref{fig:02}\cr
      &              &   & [Myr]& \cr\hline
  3881&$\nu$~And     & Lacerta OB1 {\sl b} & 9.0\phantom{1}& DES&  1\cr
 14514&53~Ari        & Orion OB1 {\sl a}   & 4.3\phantom{1}& BSS&  2\cr
      &              & Orion OB1 {\sl b}   & 4.8\phantom{1}& DES&  2\cr
      &              & Orion OB1 {\sl c}   & 5.0\phantom{1}& DES&  2\cr
 18614&$\xi$~Per     & Perseus OB2         & 1.0\phantom{1}& BSS&  3\cr
 22061&              & $\lambda$~Ori SFR   & 1.1\phantom{1}& DES&  4\cr 
 24575&AE~Aur        & Trapezium           & 2.5\phantom{1}& DES&  5\cr
 27204&$\mu$~Col     & Trapezium           & 2.5\phantom{1}& DES&  6\cr
 29678&              & $\lambda$~Ori SFR   & 1.1\phantom{1}& DES&  7\cr
 38455&              & Collinder 135       & 3.0\phantom{1}& BSS&  8\cr
 38518&              & Vela~OB2            & 6.0\phantom{1}& BSS&  9\cr
 39429&$\zeta$~Pup   & ?                   &    \phantom{1}&    & 10\cr
 42038&              & UCL                 & 8.0\phantom{1}& BSS& 11\cr
      &              & IC 2391             & 6.0\phantom{1}& BSS& 11\cr
 46950&              & IC 2602             & 2--10\phantom{1}& BSS& 12\cr
 48943&              & LCC                 & 4.0\phantom{1}& BSS& 13\cr
 49934&              & IC 2391             & 3.0\phantom{1}& BSS& 14\cr
      &              & IC 2602             & 6.0\phantom{1}& BSS& 14\cr
 57669&              & IC 2602             & 3.0\phantom{1}& BSS& 15\cr
 69491&              & UCL(?)              & 3.0\phantom{1}&   ?& 16\cr
      &              & Cepheus OB6(?)      &10.0\phantom{1}&   ?& 16\cr
 76013&$\kappa^1$~Aps& LCC                 & 2.5\phantom{1}& BSS& 17\cr
 81377&$\zeta$~Oph   & US                  & 1.0\phantom{1}& BSS& 18\cr
 82868&              & IC 2602             & 6.0\phantom{1}& BSS& 19\cr
 91599&              & Perseus OB2         & 8.0\phantom{1}& DES& 20\cr
      &              & Perseus OB3         & 6.0\phantom{1}& BSS& 20\cr
102274&              & Cepheus OB2         & 2.5\phantom{1}& BSS& 21\cr
109556&$\lambda$~Cep & Cepheus OB3         & 4.5\phantom{1}& BSS& 22\cr
      &              & & & & \cr	     
      &J0826$+$2637  & Perseus OB3         & 1.0\phantom{1}&   ?&  1\cr
      &J0835$-$4510  & Vela OB2            &           0.01&   ?&  2\cr
      &J1115$+$5030  & Perseus OB3         & 1.5\phantom{1}&   ?&  4\cr
      &J1932$+$1059  & US                  & 1.0\phantom{1}& BSS&  8\cr
      &Geminga       & $\lambda$~Orionis   &           0.35&   ?&  9\cr\hline\hline
  \end{tabular}
  \relsize{+2}
\end{table}
}